%% file: paper.tex
\documentclass[11pt,twoside]{article}
\usepackage{authblk}
\usepackage{amsmath}
\usepackage{graphicx}
\usepackage{booktabs}
\usepackage{csquotes}
\usepackage{array}
\usepackage{multirow}
\usepackage{fancyhdr}
\usepackage{geometry}
\usepackage[hang,ragged,bottomfloats,belowfloats]{footmisc} 
\usepackage{hyperref}
\usepackage{url}
\usepackage[style=authoryear,maxbibnames=10,uniquename=false,uniquelist=false]{biblatex}
\usepackage{tabulary}
\usepackage{tabularx}
\usepackage{caption}
\usepackage{tcolorbox}
\usepackage[nobottomtitles*]{titlesec}
\usepackage[protrusion=true,expansion=false,tracking=true,nopatch=footnote]{microtype}
\usepackage[oldstyle,proportional]{libertinus}
\usepackage[ttdefault]{sourcecodepro}
\usepackage[semibold]{FiraSans}
\usepackage{rhtextmath}
\usepackage{wrapfig}

\newlength{\gridunit}
\normalsize
\newcount\gridlines
\gridlines=\dimexpr\textheight-\topskip+0.5\gridunit\relax
\divide\gridlines by \gridunit
\titleformat*{\section}{\sffamily\bfseries\large\raggedright\lineskiplimit=-0.5\gridunit\setlength{\baselineskip}{\gridunit}}
\titleformat*{\subsection}{\sffamily\bfseries\normalsize\raggedright\lineskiplimit=-0.5\gridunit\setlength{\baselineskip}{\gridunit}}
\titlelabel{\thetitle.\enspace}
\apptocmd{\appendix}{\titleformat{\section}{\sffamily\bfseries\large\raggedright\lineskiplimit=-0.5\gridunit\setlength{\baselineskip}{\gridunit}}{\appendixname~\thesection.}{.5em}{}}{}{\errmessage{could not patch \string\appendix}}

\titlespacing*{\section}{0pt}{\gridunit}{\gridunit}
\titlespacing*{\subsection}{0pt}{\gridunit}{\gridunit}

\apptocmd{\normalsize}{%
  \abovedisplayskip=0.75\gridunit plus 0.25\gridunit minus 0.1\gridunit\relax
  \belowdisplayskip=0.75\gridunit plus 0.25\gridunit minus 0.1\gridunit\relax
  \abovedisplayshortskip=0.25\gridunit plus 0.25\gridunit\relax
  \belowdisplayshortskip=0.5\gridunit plus 0.25\gridunit minus 0.1\gridunit\relax
}{}{\PackageError{paper}{could not patch \string\normalsize}{}}

\newcommand{\corpus}[1]{\emph{#1}}
\newcommand{\tabterm}[1]{\emph{#1}}

\fancypagestyle{firststyle}{
  \fancyhf{}
  \fancyfoot[C]{\sffamily\scriptsize \thepage}
}

\DeclareCaptionFormat{base}{\scriptsize\sffamily\ \textbf{#1#2}{#3}}
\newtcolorbox{protocolbox}[1]{
  colback=white, colframe=black, boxrule=0.8pt, arc=0.6pt,
  left=10pt, right=10pt, top=6pt, bottom=8pt,
  title={#1}, coltitle=black, colbacktitle=white, titlerule=0.8pt,
  toptitle=4pt,bottomtitle=4pt,
  fonttitle=\sffamily\bfseries\small,
  fontupper=\small,
  before skip=\gridunit, after skip=\gridunit
}

\newlength{\floatgutter}
\newlength{\floatcolumn}
\makeatletter
\newcommand{\WF@pendingtofloat}{%
  \ifvoid\WF@box\else
  \def\@tempa{wraptable }%
  \ifx\WF@wfname\@tempa\def\WF@type{table}\else\def\WF@type{figure}\fi
  \begingroup\@float{\WF@type}[tbp]\centering\noindent\box\WF@box\par\end@float\endgroup 
  \global\setbox\WF@box\box\voidb@x
  \global\c@WF@wrappedlines\z@
\fi}
\pretocmd{\ttl@straight@i}{\WF@pendingtofloat}{}{\PackageError{paper}{could not patch \string\ttl@straight@i}{}}
\newcommand{\WF@headingskips}{%
  \ifx\parshape\WF@fudgeparshape\ifnum\c@WF@wrappedlines>\@ne
  \global\advance\c@WF@wrappedlines-\tw@
\fi\fi}
\pretocmd{\ttl@straight@i}{\WF@headingskips}{}{\PackageError{paper}{could not patch \string\ttl@straight@i}{}}
\patchcmd{\WF@putfigmaybe}{\setbox\WF@box\hbox{\lower\intextsep\box\WF@box}}{\global\advance\WF@size-\intextsep}{}%
{\PackageError{paper}{could not patch \string\WF@putfigmaybe}{}}
\newlength{\wrapdrop}
\patchcmd{\WF@putfigmaybe}{\dp\WF@box\z@\box\WF@box}%
{\setbox\WF@box\hbox{\lower\wrapdrop\box\WF@box}\global\advance\WF@size\wrapdrop\dp\WF@box\z@\box\WF@box}{}%
{\PackageError{paper}{could not patch \string\WF@putfigmaybe}{}}
\makeatother

\AtBeginBibliography{\small}
\DefineBibliographyStrings{english}{page={},pages={}}
\appto{\bibsetup}{\emergencystretch=2em}
\makeatletter
\DeclareFieldFormat{eprint:arxiv}{%
  arXiv\addcolon\space
  \ifhyperref
    {\href{https://arxiv.org/\abx@arxivpath/#1}{\mbox{\nolinkurl{#1}}}}
    {\mbox{\nolinkurl{#1}}}}
\makeatother
\ExplSyntaxOn
\tl_new:N \l__pinpoint_postnote_tl
\cs_new_protected:Npn \__pinpoint_if_page_first:TF #1#2
{
  \savefield{postnote}{\l__pinpoint_postnote_tl}
  \exp_args:NnV \regex_match:nnTF { \A \d } \l__pinpoint_postnote_tl {#1} {#2}
}

\ExplSyntaxOff
\DeclareFieldFormat{postnote}{#1}

\DeclareBibliographyCategory{needsurl}

\renewbibmacro*{url+urldate}{%
  \ifcategory{needsurl}
  {\printfield{url}%
    \iffieldundef{urlyear}
    {}
    {\setunit*{\addspace}%
  \printurldate}}
{}}

\hypersetup{
  colorlinks=true,
  linkcolor=black,
  filecolor=black,
  urlcolor=black,
  citecolor=black,
}
\title{Quality, Stability, and Selection: \\Testing the Criteria Used to Evaluate \\Overlapping Community Detection\vspace{25pt}}
\author{Gabriel H. Pizzorno\textsuperscript{*}}
\affil{Harvard University\vspace{-20pt}}
\date{}

\begin{document}

\maketitle
\thispagestyle{firststyle}

\begin{abstract}
  \itshape
  Methods for detecting overlapping communities return covers, in which communities may share members. They are evaluated by three criteria: quality measures, which score a cover without a reference; stability, the agreement amongst covers from repeated runs; and accuracy against a reference cover. The methods have been tested at length, but the criteria far less so. This paper tests all three on benchmark graphs whose correct cover is known. Four quality measures were scored on the known cover and on versions degraded by merging communities, splitting them, or reassigning members at random. None could reliably rank covers of different granularity. For some methods, stability tracked accuracy while the communities remained detectable. After detection failed, it could remain high, because a method that finds nothing can return the same structureless cover on every run. Accuracy depended on the selection protocol, the rule that reduces a method's many runs to one score per graph. On the same runs, the protocol alone could decide which of two methods scored higher. Without a reference cover, agreement amongst quality measures has been taken as evidence of a good result. The tests show that such agreement is not independent evidence, and that adding stability does not make it so. Quality measures and stability both respond to the shape of a cover, and can agree for reasons unrelated to its correctness. Four safeguards are proposed. A quality measure should be validated against a trusted reference cover before adoption, and the four tested here should not be used to compare covers of different granularity. Stability should be reported only with a description of the cover's shape, and methods should be compared under a single selection protocol.
  \vspace{10pt}
\end{abstract}

\noindent\textbf{\footnotesize Keywords:}~{\footnotesize overlapping community detection, evaluation criteria, quality measures, overlapping modularity, run-to-run stability, selection bias, benchmark graphs}
\vspace{10pt}

\renewcommand{\thefootnote}{*}
\footnotetext{Department of History | Robinson Hall, Office 210 | 35 Quincy Street, Cambridge, MA 02138 | \href{mailto:pizzorno@fas.harvard.edu}{pizzorno@fas.harvard.edu}}
\renewcommand{\thefootnote}{\arabic{footnote}}

\section{Introduction}

Methods for detecting overlapping communities return \emph{covers}, in which communities may share members rather than forming a partition, whose groups are disjoint. Three criteria are commonly used to evaluate them. The first comprises \emph{internal quality measures}, which score a cover without comparing it with a reference. The second is \emph{stability}, measured as the agreement amongst covers returned by a method across random seeds. When a reference cover is unavailable, this agreement is treated as evidence that the result is reliable.\footnote{\emph{Stability} is used throughout this paper in this sense alone, as a property of a procedure. It is distinct from the quality function of the same name introduced by \textcite[12755]{delvenneStabilityGraphCommunities2010}, a property of a single partition (Section~\ref{sec:stability}).} The third criterion applies when a reference cover is available. Methods are then compared by their accuracy against it. This comparison depends on two choices: a \emph{selection protocol} reduces each method's runs to a score for each graph, while an \emph{index} measures that score against the reference cover. On benchmarks with known community structure, internal quality measures and stability can be assessed against accuracy. An internal measure is assessed by whether it ranks the planted cover above alternatives. Stability is assessed by whether greater agreement tracks higher accuracy. Assessments of measures that score a whole cover rather than a single community are few in the literature, and none extends to directed graphs with overlap (Section~\ref{sec:gap}). For stability on such graphs, no assessment appears to exist. This paper addresses that gap in four ways: internal quality measures are tested against known reference covers; stability is tested against accuracy on planted graphs; methods are compared under alternative selection rules; and paired comparisons of weighted and unweighted configurations are used to examine the limits of reference-free diagnostics on real networks.

The first test is applied to four internal quality measures (Section~\ref{sec:peak}). Each was scored on a benchmark's planted cover and on alternatives derived from it by merging communities, splitting them, or randomly reassigning memberships. A measure that behaves as intended should attain its maximum at the planted cover. Against splitting and reassignment, the measures behaved largely as intended. All four scored the planted cover above every randomly split cover, and their mean scores never rose under random reassignment of memberships. The clearest failure occurred under merging. The overlapping modularity \(Q_{ov}\) of \textcite{nicosiaExtendingDefinitionModularity2009} never attained its maximum at the planted cover on any planted graph, and a cover with fewer communities always scored higher. The failure recurred on subsamples of a real co-authorship network, and it was traced to one term in the definition of the measure.

The second test asks whether stability tracks accuracy on planted graphs (Section~\ref{sec:stability}). The answer depended on the method and on whether the planted communities remained detectable. For three of the four overlapping methods examined in this paper, stability generally tracked accuracy while they did. For the fourth, more stable settings were less accurate even then. Where communities were no longer detectable, one method's stability returned almost to its maximum at some settings while its accuracy fell to zero, because repeated runs usually produced the same degenerate cover. High stability was therefore no assurance of an accurate cover.

The third test examines how the selection protocol can change the ranking of methods by accuracy (Sections~\ref{sec:comparison} and~\ref{sec:discussion-index}). The four overlapping methods were compared with a partition baseline, which assigns each node to a single community. Three of them scored below the baseline even on graphs with planted overlap \autocite{pizzornoCoversPartitionsOverlapping2026}. The fourth scored above it on average, but the size and significance of its lead depended on the protocol. The lead was significant when every method was run at its documented default. It narrowed but remained significant when each method's setting was chosen graph by graph, using each graph's own planted cover. It was no longer significant when the setting was chosen on other graphs. On a suite in which overlapping nodes belong to three communities, the same runs placed the fourth method below the baseline, level with it, or above it, according to the protocol. The lead also varied with the benchmark and the index.

The fourth test considers the reference-free diagnostics together (Section~\ref{sec:reference-free}). In the absence of a reference cover, agreement amongst quality measures has been taken to strengthen the claim that a detected structure is good \autocite[26]{chakrabortyMetricsCommunityAnalysis2017}. Such agreement, however, can reflect shared responses to the shape and granularity of a cover rather than the correctness of its communities. It cannot establish accuracy, since several diagnostics may favour the same incorrect result. One method was run with and without each graph's edge weights. On planted graphs, where the more accurate configuration is known, the quality measures favoured the weighted configuration at similar rates whether it was the more accurate or the less accurate of the two. On a real email network, two quality measures and two stability indices all favoured the same configuration. Because that network has no reference cover, its accuracy ranking can be neither confirmed nor overturned.

The findings of the four tests have consequences for practice, and they expose a gap in evaluation. A quality measure should be tested against a trusted reference cover before it is adopted, and scores from the four measures examined should not be compared across covers that differ in granularity. Stability should not be reported alone. Like the quality measures, it responds to the shape of a cover, so a stability figure should be accompanied by a plain description of that shape in terms of community sizes and overlap. Method comparisons should use a matched selection protocol and identify scores selected with the scored graph's own reference cover. As for the gap, the available datasets may help explain why reference-free diagnostics have rarely been assessed against accuracy on the directed, weighted networks these methods target. Such networks generally lack an overlapping reference cover defined independently of node attributes, and the real datasets with reference covers are unweighted or use node metadata. This leaves few opportunities to make that assessment on real graphs (Section~\ref{sec:discussion-gap}). The protocol of the first test requires only a graph with a trusted reference cover and the measure under test. It can evaluate a proposed measure before adoption, but it cannot supply the missing real benchmark.

\section{Background and Related Work}
\label{sec:background}

\subsection{Overlapping Community Detection in Directed Graphs}
\label{sec:related-detection}

Community detection is ordinarily posed as a partition problem. Each node is assigned to exactly one group, and the grouping is chosen to make connections dense within groups and sparse between them \autocite{fortunatoCommunityDetectionGraphs2010,fortunatoCommunityDetectionNetworks2016}. In social and knowledge networks, however, individuals and entities may belong to several groups at once. \citeauthor{pallaUncoveringOverlappingCommunity2005} named this \emph{overlapping community structure} and observed that \enquote{most of the actual networks are made of highly overlapping cohesive groups of nodes} (\citeyear[814]{pallaUncoveringOverlappingCommunity2005}). A substantial literature has since followed \autocite{xieOverlappingCommunityDetection2013}.

This work was initially motivated by one use of community detection, on knowledge graphs built from archives of historical correspondence. In graph-based retrieval-augmented generation, a large language model extracts entities and their relations from a text corpus, the resulting graph is divided into communities, and summaries of those communities are used to answer queries \autocite[2]{edgeLocalGlobalGraph2024}. \citeauthor{edgeLocalGlobalGraph2024} detected the communities with Leiden \autocite{traagLouvainLeidenGuaranteeing2019}, applied recursively, so that each level of the resulting hierarchy is a partition that \enquote{covers the nodes of the graph in a mutually exclusive, collectively exhaustive way} (\citeyear[5--6]{edgeLocalGlobalGraph2024}).

In a correspondence network, however, a correspondent may belong to several circles at once. Its communities are therefore expected to overlap, and its edges are directed and weighted. An edge may encode a relation such as \emph{corresponded with} or \emph{influenced by}. The edge's direction distinguishes the initiator of a relation from its recipient. A method or measure that symmetrizes the graph discards that distinction. Few methods and measures address directed edges, weights, and overlap together.

Four methods are compared in this paper, out of the seven that a companion paper, \emph{Covers and Partitions} \autocite{pizzornoCoversPartitionsOverlapping2026}, studies in full. They are OSLOM \autocite{lancichinettiFindingStatisticallySignificant2011}, SLPA \autocite{xieSLPAUncoveringOverlapping2011}, ego-splitting \autocite{epastoEgoSplittingFrameworkNonOverlapping2017}, and CD-DAWN \autocite{kumarOverlappingCommunityDetection2025}, and each can produce an overlapping cover of a directed graph without a specified number of communities. Leiden, which optimizes modularity over partitions, serves as the \emph{partition baseline} \autocite{traagLouvainLeidenGuaranteeing2019}. Section~\ref{sec:methods} describes each method, the role of the baseline, and how each implementation was validated.

\subsection{Cover Quality Measures}
\label{sec:related-quality}

In the absence of a reference cover, quality is usually assessed with an internal measure. Most measures in current use descend from Newman's modularity, which compares the density of edges inside communities with the density that a degree-preserving null model would produce (\citeyear[8578]{newmanModularityCommunityStructure2006}). Later work has developed modularity in three directions. \citeauthor{leichtCommunityStructureDirected2008} adapted it to directed graphs by separating in-degree from out-degree in the null term (\citeyear[{2}]{leichtCommunityStructureDirected2008}). \citeauthor{kimFindingCommunitiesDirected2010} replaced edge counting with a random walk, comparing the probability that a walker is found inside a community with the product of the stationary probabilities of its members (\citeyear[{4}]{kimFindingCommunitiesDirected2010}). \citeauthor{nicosiaExtendingDefinitionModularity2009} extended the directed measure to overlapping covers by weighting each edge's contribution with belonging coefficients (\citeyear[6--7]{nicosiaExtendingDefinitionModularity2009}). By contrast, the measure that \citeauthor{kumarOverlappingCommunityDetection2025} proposed for covers of directed, weighted graphs departs from the modularity family, because it makes no comparison against a null model (\citeyear[18]{kumarOverlappingCommunityDetection2025}, see Section~\ref{sec:measures}).

In other overlapping measures, node pairs are weighted by the inverse of their membership counts \autocite[1708]{shenDetectOverlappingHierarchical2009}. \citeauthor{chenExtensionModularityDensity2014} extended modularity density to overlapping covers by multiplying the belonging coefficients of the two nodes (\citeyear[856]{chenExtensionModularityDensity2014}). \citeauthor{lazarModularityMeasureNetworks2010} defined a non-fuzzy measure \enquote{designed to rank the partitions of a network's nodes into overlapping communities} (\citeyear[{1}]{lazarModularityMeasureNetworks2010}). The broader family was surveyed by \textcite[{16--17}]{chakrabortyMetricsCommunityAnalysis2017}.

Three limitations of modularity are well established. First, it has a resolution limit. Communities below a scale set by the size of the graph cannot be resolved, because merging two of them raises the score even where they are cliques joined by a single edge \autocite[38--39]{fortunatoResolutionLimitCommunity2007}.

Second, the modularity function has a degenerate landscape. It \enquote{exhibits extreme degeneracies such that the globally maximum modularity partition is typically hidden among an exponential number of structurally dissimilar, high-modularity solutions} \autocite[{2}]{goodPerformanceModularityMaximization2010}. High-scoring solutions are therefore numerous, and the best is not distinguished from the rest. For three real metabolic networks, a reconstruction of the function showed \enquote{a broad and rugged region of high-modularity partitions with no clear global maximum} \autocite[{8}]{goodPerformanceModularityMaximization2010}.

Finally, the maximum need not lie at the planted structure. On benchmarks with heterogeneous community sizes, modularity with a tunable resolution \enquote{is not capable to recover the planted community structure, not even when it is pronounced and easily detectable by other methods, for any value of the resolution parameter} \autocite[{1}]{lancichinettiLimitsModularityMaximization2011}. These limitations affect the granularity of the maximum, the difficulty of finding it, and its agreement with the planted structure. Section~\ref{sec:peak} asks whether directed and overlapping measures themselves rank a known-correct cover above its rivals, independently of the procedure used to optimize them.

\subsection{Cover Comparison}
\label{sec:related-accuracy}

Where a reference cover is available, a detected cover is assessed by how closely it matches that reference. The partition-comparison indices in common use do not transfer directly to covers. The Rand index \autocite[847]{randObjectiveCriteriaEvaluation1971} and its chance-corrected form, the Adjusted Rand index \autocite[198]{hubertComparingPartitions1985}, classify pairs of nodes as either together or apart. This classification assumes that two nodes either share a community or do not. Applying these indices to a cover therefore requires collapsing it to a partition, which discards the multiple memberships that overlapping methods are meant to recover.

For covers, pair counting is generalized in the Omega index \autocite[238]{collinsOmegaGeneralFormulation1988}. Information-theoretic comparison has also been adapted to overlapping communities through normalized mutual information \autocites[14--15]{lancichinettiDetectingOverlappingHierarchical2009}[2]{mcdaidNormalizedMutualInformation2013}. \citeauthor{gatesElementcentricClusteringComparison2019} compared disjoint, overlapping, and hierarchical clusterings in a single element-centric framework (\citeyear[1]{gatesElementcentricClusteringComparison2019}).

Of these indices, normalized mutual information has a documented bias towards outputs with more communities \autocite[1584]{amelioNormalizedMutualInformation2015}. \citeauthor{jerdeeNormalizedMutualInformation2025} described it as having \enquote{a bias towards labelings with too many distinct label values} (\citeyear[1]{jerdeeNormalizedMutualInformation2025}). A correction for chance matters most \enquote{when the data size is small compared to the number of clusters present therein} \autocite[2837]{vinhInformationTheoreticMeasures2010}. \citeauthor{jebabliCommunityDetectionAlgorithm2018} described node-level comparison measures as \enquote{fairly insensitive to the variation of the overall community structure} and proposed instead the topology of community graphs, in which each community is represented as a node, as a basis for comparison (\citeyear[651]{jebabliCommunityDetectionAlgorithm2018}). Section~\ref{sec:accuracy} defines the three cover-aware indices used in this paper, and Section~\ref{sec:reporting} shows how stability comparisons change when covers are collapsed to partitions for the Adjusted Rand index.

\subsection{Benchmarks and \enquote{Ground Truth}}
\label{sec:related-benchmarks}

Comparison against a reference cover requires graphs that have one. Planted benchmarks supply it by construction. The Lancichinetti--Fortunato--Radicchi (LFR) family generates graphs whose degrees and community sizes follow power laws, with a tunable mixing parameter \autocite[{2}]{lancichinettiBenchmarkGraphsTesting2008}. It has been extended to directed and weighted graphs with overlapping communities \autocite[{1}]{lancichinettiBenchmarksTestingCommunity2009}. The limitation of any such benchmark is that the structure to be recovered is whatever the generator planted. A result obtained on such a benchmark therefore holds for the generator's graphs, and it transfers to a real network only insofar as that network resembles them.

Real networks distributed with \enquote{ground truth} almost always use node metadata as the reference, such as Wikipedia categories, publication venues, or declared group memberships. \citeauthor{peelGroundTruthMetadata2017} showed that treating metadata as a reference cover can mislead in both directions. A method may recover real structure that metadata do not encode. Conversely, metadata may encode distinctions that the structure does not contain (\citeyear[1--2]{peelGroundTruthMetadata2017}). A poor score against metadata is therefore not evidence against the method. Across many large networks, \citeauthor{hricCommunityDetectionNetworks2014} found \enquote{a marked separation between structural communities and metadata groups} (\citeyear[{1}]{hricCommunityDetectionNetworks2014}). The datasets used in this paper were selected with that caution in mind (Section~\ref{sec:data}).

The term \emph{reference cover} is therefore used throughout this paper for what a detected cover is scored against, without implying that it is the graph's true community structure. A cover assigned by a generator is called \emph{planted}, and for it the implication holds by construction. A cover derived from metadata is a reference only, since its relation to community structure cannot be assumed.\footnote{\emph{Ground truth} is retained only where a cited work uses it, or in the name of a dataset distributed under that description.}

\subsection{Validation Practice}
\label{sec:gap}

Quality measures such as those of Section~\ref{sec:related-quality} are typically introduced through axioms and worked examples. A measure is checked against properties such as vanishing on a structureless graph, and it is tested on a graph with obvious communities. Some quality measures have also been examined empirically. \citeauthor{almeidaThereBestQuality2011} studied five measures on clusters found in real graphs without a reference and found that they \enquote{have strong biases towards incorrectly awarding good scores to some kinds of clusters} (\citeyear[44]{almeidaThereBestQuality2011}). \citeauthor{yangDefiningEvaluatingNetwork2012} scored single communities defined by node metadata against randomized perturbations, on the premise that \enquote{an ideal community scoring function should give a maximal value when evaluated on the ground-truth community} (\citeyear[6]{yangDefiningEvaluatingNetwork2012}). The study found that \enquote{[m]odularity is not good at distinguishing true communities from randomized sets of nodes} (\citeyear[7]{yangDefiningEvaluatingNetwork2012}). On benchmark graphs with planted partitions, \citeauthor{emmonsAnalysisNetworkClustering2016} compared three quality measures with accuracy over the outputs of four detection methods and found \enquote{overall disagreement between stand-alone quality metrics and information recovery metrics} (\citeyear[15]{emmonsAnalysisNetworkClustering2016}). On real networks with metadata groups, \citeauthor{creusefondEvaluationPotentialQuality2016} asked whether ten quality functions \enquote{rank higher clusterings that are close to the ground-truth} (\citeyear[112]{creusefondEvaluationPotentialQuality2016}).

Three of these studies scored the outputs of detection methods, so the covers compared were whatever the methods happened to produce. Only \citeauthor{yangDefiningEvaluatingNetwork2012} constructed alternatives to a reference, by perturbing it, but for single communities. Such rivals can be constructed for a whole cover in the same way. A planted cover is correct by construction, and rivals are derived from it by merging its communities, splitting them, or randomizing their memberships. The test then asks whether a measure attains its maximum at the planted cover. Such a test does not appear to have been reported for measures that score a whole cover, and in particular not for directed graphs with overlap. Without one, a measure still returns well-formed scores for covers from any method on any graph, and those scores alone do not show whether its rankings reflect the quality of a cover or merely its granularity.

\section{Preliminaries}
\label{sec:preliminaries}

This section defines the notation and the evaluation measures, and then describes the datasets and the detection methods. The companion paper reports in full the method comparison, the validation of the implementations, and a study of OSLOM's use of edge weights and its other settings \autocite{pizzornoCoversPartitionsOverlapping2026}.

\subsection{Notation}
\label{sec:notation}

Let \(G = (V, E)\) be a directed graph with \(n = |V|\) nodes and \(m = |E|\) edges, with binary adjacency matrix \(A\) and, where the graph is weighted, edge weights \(W(u,v)\). Node \(i\) has out-degree \(k_i^{\text{out}}\) and in-degree \(k_i^{\text{in}}\) and, where weights are used, out-strength \(s_i^{\text{out}}\) and in-strength \(s_i^{\text{in}}\).

A \emph{cover} \(\mathcal{K} = \{C_1, \ldots, C_k\}\) is a set of subsets of \(V\), and a \emph{partition} is the special case in which the subsets are disjoint and their union is \(V\). The term \emph{cover} is taken from \citeauthor{lancichinettiDetectingOverlappingHierarchical2009}, who defined it as \enquote{a set of clusters such that each node is assigned to at least one cluster} and described it as \enquote{an extension of the traditional concept of graph partition} (\citeyear[4]{lancichinettiDetectingOverlappingHierarchical2009}). The definition used in this paper omits that requirement, because detected covers often leave nodes unplaced \autocite[4]{hricCommunityDetectionNetworks2014}.

Two sets describe how a cover treats the nodes. The set of communities that contain node \(v\) is written \(m_v\), and the set of nodes that the cover places in at least one community is \(V_{\mathcal{K}} = \bigcup_{C \in \mathcal{K}} C\). A node with no membership, for which \(m_v\) is empty, is \emph{unplaced}, and a node with more than one is an \emph{overlapping node}. A cover is therefore a partition exactly when \(|m_v| = 1\) for every node.

A reference cover places every node, whereas several of the methods compared in this paper leave nodes outside any community, so in general \(V_{\mathcal{K}} \subseteq V\) for a detected cover. Section~\ref{sec:accuracy} states how unplaced nodes enter each comparison between covers.

Overlapping modularity (Section~\ref{sec:measures}) weights each of a node's memberships. Following \citeauthor{nicosiaExtendingDefinitionModularity2009}, each node that a cover places carries \emph{belonging coefficients} \(a_{v,C} \in [0,1]\) that sum to one over the communities containing it (\citeyear[6]{nicosiaExtendingDefinitionModularity2009}). The coefficients are taken throughout to be uniform over a node's memberships, \(a_{v,C} = 1/|m_v|\) for \(C \in m_v\) and 0 otherwise, so every coefficient of an unplaced node is zero.\footnote{The uniform allocation is used because none of the methods compared reports graded memberships. Every measure that uses belonging coefficients therefore receives exactly the information that the methods emit, and no method is favoured. \citeauthor{gregoryFindingOverlappingCommunities2010} adopted the same allocation (\citeyear[11]{gregoryFindingOverlappingCommunities2010}).}

Covers of one graph also differ in \emph{granularity}. Of two covers, the one with fewer, larger communities is the \emph{coarser} and the one with more, smaller communities the \emph{finer}. The merge and split ladders of Section~\ref{sec:ladders} move a cover in these two directions.

Two conventions about community size are used throughout this paper. Communities of five or fewer members are termed \emph{very small}, and communities of 20 or more lie above the \emph{size floor} (Section~\ref{sec:data}).

\subsection{Quality Measures}
\label{sec:measures}

The four measures examined in Section~\ref{sec:peak} are set out below in a common notation. They differ in what they score and in their treatment of the null model. The first two are formulated for partitions, while the last two accommodate overlapping covers. Their original formulations can obscure these distinctions.

\emph{Directed modularity}, \(Q_{d}\), compares the observed number of within-community edges with the number expected under a degree-preserving null model \autocite[{2}]{leichtCommunityStructureDirected2008}. In-degree and out-degree are treated separately, and the measure is

\begin{equation}\label{eq:qd}
  Q_{d} = \frac{1}{m} \sum_{C} \left[ I_C - \frac{\kappa_C^{\text{out}} \kappa_C^{\text{in}}}{m} \right],
\end{equation}

\noindent where \(I_C\) counts directed edges with both endpoints in community \(C\), and \(\kappa_C^{\text{out}}\) and \(\kappa_C^{\text{in}}\) denote the total out-degree and in-degree of its members.

\emph{LinkRank modularity}, \(Q_{lr}\), replaces edge counts with the flow of a random walk \autocite[{3--4}]{kimFindingCommunitiesDirected2010}. Let \(\pi\) denote the PageRank vector and \(G_{ij}\) the Google matrix at damping \(\alpha\), so that \(\pi_i G_{ij}\) is the stationary probability that a walker follows the edge from \(i\) to \(j\). The measure sums that probability over the pairs of nodes inside each community and subtracts a null term,

\vspace{5pt}
\begin{equation}\label{eq:qlr}
  Q_{lr} = \sum_{C} \sum_{i,j \in C} \left[ \pi_i G_{ij} - \pi_i \pi_j \right],
\end{equation}
\vspace{5pt}

\noindent in which the null term \(\pi_i \pi_j\) treats the walker's presence at nodes \(i\) and \(j\) as independent.\footnote{The closed form used in this paper was checked against a dense computation built from the published equations alone \autocite[{3--4}]{kimFindingCommunitiesDirected2010}. The dense version reproduced the analytic modularity curves printed for the authors' model network \autocite[{8}]{kimFindingCommunitiesDirected2010} and agreed with the closed form to machine precision at a fixed PageRank vector. One convention departs from the published definition. A node of out-degree zero is given the teleportation row \((1 - \alpha)/n\) rather than the \(1/n\) of the published Google matrix, so a partition into one community scores \(-\alpha P_0\) rather than zero, where \(P_0\) is the PageRank mass on nodes of out-degree zero. The difference affects no score reported in this paper, because no graph on which \(Q_{lr}\) is scored has such a node. Those graphs are the nine partition graphs of the planted suite, the ten \corpus{DBLP} subsamples, and the 70 graphs of the \corpus{DBLP} granularity sweep, all described in Section~\ref{sec:data}, and every ring of cliques in Section~\ref{sec:resolution}.}

\emph{Overlapping modularity}, \(Q_{ov}\), weights each edge's contribution by applying a function \(F\) to its endpoints' belonging coefficients \autocite[7--8]{nicosiaExtendingDefinitionModularity2009}. The authors left \(F\) unspecified \autocite[6]{nicosiaExtendingDefinitionModularity2009}. This paper uses the product form \(F(a_{i,C}, a_{j,C}) = a_{i,C}\,a_{j,C}\).\footnote{In their experiments, \textcite[14]{nicosiaExtendingDefinitionModularity2009} used a logistic function of the belonging coefficients but did not specify its steepness. \textcite[11]{gregoryFindingOverlappingCommunities2010} set it to 30. The logistic function is not used in this paper because it sharply suppresses three-way overlap. For a node with three memberships, it assigns a factor of \(4.5 \times 10^{-5}\), whereas the product form assigns \(1/3\). An edge between two such nodes therefore contributes only \(2 \times 10^{-9}\) under the logistic form. The product form does not account for the failure reported in Section~\ref{sec:peak}. On a partition, each node's belonging coefficient is either one or zero. As a result, any \(F\) near one at \((1,1)\) and near zero elsewhere damps the null term by the same mean-belonging factor. When evaluated on the ladders of Section~\ref{sec:ladders}, the logistic form also failed the merge ladder on all 36 graphs of the planted suite and all ten co-authorship subsamples. It preferred a cover of the same size as the product form on all 46 graphs.} The null term differs from the observed-edge term because it averages belonging over the whole node set, through the factors

\begin{align}
  \beta^{\text{out}}_{i,C} &= \frac{1}{n} \sum_{j \in V} F(a_{i,C}, a_{j,C}), \label{eq:beta-out} \\[0.5em]
  \beta^{\text{in}}_{j,C} &= \frac{1}{n} \sum_{i \in V} F(a_{i,C}, a_{j,C}). \label{eq:beta-in}
\end{align}

\noindent Substituting these factors gives

\begin{equation}\label{eq:qov}
  Q_{ov} = \frac{1}{m} \sum_{C} \sum_{i,j \in V} \Bigl[ F(a_{i,C}, a_{j,C}) A_{ij} - \frac{\beta^{\text{out}}_{i,C} k_i^{\text{out}} \beta^{\text{in}}_{j,C} k_j^{\text{in}}}{m} \Bigr].
\end{equation}
\vspace{5pt}

Because \(F\) factorizes into per-node terms, each factor is a node's own coefficient multiplied by the community's mean belonging, \(\beta^{\text{out}}_{i,C} = a_{i,C}\,\bar{a}_C\), where \(\bar{a}_C = n^{-1}\sum_j a_{j,C}\) is the mean of the coefficients over all nodes. The null term is therefore Newman's term scaled down by \(\bar{a}_C^{\,2}\), which damps its contribution. Section~\ref{sec:mechanism} tests whether this damping explains the measure's behaviour.\footnote{The null term is easy to implement incorrectly, and no public implementation examined computes this form. In \emph{cdlib} version 0.4.1 \autocite{rossettiCDLIBPythonLibrary2019}, it is summed over adjacent pairs only and every graph is treated as undirected. That implementation assigns 0.1447 to the single-community cover on each of 20 structureless random graphs, although the definition requires zero \autocite[8]{nicosiaExtendingDefinitionModularity2009}, and it scores that cover above the planted partition on each of the nine partition graphs of the planted suite (Section~\ref{sec:data}). \emph{Graph-tool} version 3.9 \autocites{peixotoGraphtoolPythonLibrary2014}[2]{peixotoDescriptiveVsInferential2023} accepts an overlapping cover but returns Newman's modularity for the partition that groups nodes with identical memberships, which is not an overlapping modularity. Neither \emph{networkx} 3.6.1 \autocite{hagbergExploringNetworkStructure2008} nor \emph{igraph} 1.0.0 \autocite{antonovIgraphEnablesFast2023} scores overlapping covers. Every reported \(Q_{ov}\) value uses the form in Equation~\eqref{eq:qov}.}

\emph{Directed weighted overlapping modularity}, \(Q_{dwo}\), differs in form from the preceding measures \autocite[18]{kumarOverlappingCommunityDetection2025}. It is the mean over communities of a cohesion term \(\lambda_C\), reduced according to the memberships that the community's nodes hold,

\vspace{-5pt}
\begin{equation}\label{eq:qdwo}
  Q_{dwo} = \frac{1}{|\mathcal{K}|} \sum_{C} \left[ 1 - \frac{\sum_{v \in C} |m_v|}{|C|\,|\mathcal{K}|} \right] \lambda_C,
\end{equation}

\noindent with

\vspace{-5pt}
\begin{equation}\label{eq:lambda}
  \lambda_C = \frac{1}{2}\left( \frac{1}{s^{-}(C)} + \frac{1}{s^{+}(C)} \right) \sum_{u,v \in C} W(u,v),
\end{equation}
\vspace{5pt}

\noindent where \(s^{-}(C)\) and \(s^{+}(C)\) are the total in- and out-strength of \(C\)'s members over the whole graph. The term \(\lambda_C\) is the fraction of the community's strength that stays inside it, and the factor that precedes it penalizes multiple membership.\footnote{When every edge has a reciprocal of equal weight, \(s^{-}(C) = s^{+}(C)\) and \(\lambda_C\) reduces to \(d_{int}(C)/d(C)\). Thus, \(Q_{dwo}\) coincides with the weighted overlapping modularity \(Q_{wo}\) defined for undirected graphs \autocite[1355]{kumarNeighborhoodProximityBased2019}. Both measures are implemented in the package that accompanies this paper. The \(Q_{dwo}\) implementation is tested against this identity. Transposing in- and out-strength, dropping the factor of two, or using an incorrect normalizer would violate it.} Because \emph{the measure has no null model}, it does not compare a community's density with what chance would produce. It scores cohesion while penalizing overlap. The consequences of this design are reported in Sections~\ref{sec:peak-results} and~\ref{sec:battery}.

\subsection{Comparative Measures}
\label{sec:accuracy}

Accuracy against a reference cover is assessed with three cover-aware indices: the Omega index, an overlapping form of normalized mutual information, and the average best-match \(F_1\). The first two are also used to measure run-to-run stability, while the third is used only for accuracy. A separate stability comparison uses the Adjusted Rand index after each cover is collapsed to a partition. Section~\ref{sec:reporting} reports how the indices behave as measures of stability.

The \emph{Omega index}, \(\Omega\), generalizes the Adjusted Rand index to covers \autocite[238--239]{collinsOmegaGeneralFormulation1988}. The latter requires each node to be assigned to a single community. By contrast, \(\Omega\) compares the number of communities shared by each pair of nodes in the two covers, and it corrects the observed agreement for the agreement expected by chance. Every node of \(V\) enters the comparison. An unplaced node shares no community with any other, so it is compared as a node with no memberships, and its omission is not removed from the comparison. For each cover \(\mathcal{K}\), let \(j_{\mathcal{K}}(u,v) = |\{C \in \mathcal{K} : u \in C,\; v \in C\}|\) denote the number of communities containing both \(u\) and \(v\), and let \(t_j(\mathcal{K})\) denote the number of node pairs that share \(j\) communities. Across the \(P = n(n-1)/2\) unordered node pairs in \(V\), the observed and expected agreement are

\vspace{-5pt}
\begin{equation}\label{eq:omega-terms}
  \omega_{\text{obs}} = \frac{1}{P} \bigl|\{\{u,v\} : j_{\mathcal{K}}(u,v) = j_{\mathcal{K}'}(u,v)\}\bigr|, \qquad \omega_{\text{exp}} = \frac{1}{P^{2}} \sum_{j \ge 0} t_j(\mathcal{K})\, t_j(\mathcal{K}'),
\end{equation}

\noindent and \(\Omega(\mathcal{K}, \mathcal{K}') = (\omega_{\text{obs}} - \omega_{\text{exp}}) / (1 - \omega_{\text{exp}})\). The Omega index is the primary index used throughout this paper because it is chance-corrected and does not reward degenerate covers. One degenerate case remains. If every pair of nodes shares the same number of communities in one cover, and every pair shares that same number in the other, then all pairs agree and chance predicts as much, so \(\omega_{\text{obs}} = \omega_{\text{exp}} = 1\) and the correction reduces to \(0/0\). This happens, for example, when both covers are the all-singleton partition, or when both are the single community of all nodes. The implementation used in this paper sets \(\Omega\) to one in that case, which is correct where the two covers are the same degenerate cover (Section~\ref{sec:stability-planted}). It rejects a cover that holds a node outside the set over which the comparison is made, so that two unrelated covers cannot be scored as identical by that route.

Computing \(\Omega\) by enumeration holds a count for every pair of nodes that shares a community, so the memory it needs grows with the sum of the squared community sizes, reaching tens to hundreds of gigabytes on degenerate covers (Section~\ref{sec:reporting}). Both terms of \(\Omega\) can instead be computed without enumerating node pairs, by counting them through the nodes that several communities have in common. Let \(N_{kj}\) be the number of node pairs that share \(k\) communities in \(\mathcal{K}\) and \(j\) in \(\mathcal{K}'\). These counts give both terms, since \(\omega_{\text{obs}} = P^{-1} \sum_{k} N_{kk}\) and the \(t_j\) of the two covers are the row and column sums of \(N\). They are obtained indirectly. For a set \(\mathcal{R}\) of \(r\) communities of \(\mathcal{K}\) and a set \(\mathcal{S}\) of \(s\) communities of \(\mathcal{K}'\), let \(n_{\mathcal{R}\mathcal{S}}\) be the number of nodes that belong to every community in \(\mathcal{R}\) and in \(\mathcal{S}\). Any two of those nodes share all the communities chosen, so they contribute \(\binom{n_{\mathcal{R}\mathcal{S}}}{2}\) pairs. Conversely, a pair that shares \(k\) and \(j\) communities is counted once for each of the \(\binom{k}{r}\binom{j}{s}\) choices of \(\mathcal{R}\) and \(\mathcal{S}\) from amongst them. Summing over the choices of each size therefore gives the binomial moments

\vspace{-5pt}
\begin{equation}\label{eq:omega-moments}
  M_{rs} = \sum_{|\mathcal{R}| = r,\, |\mathcal{S}| = s} \binom{n_{\mathcal{R}\mathcal{S}}}{2} = \sum_{k \ge r,\, j \ge s} \binom{k}{r} \binom{j}{s} N_{kj},
\end{equation}
\vspace{5pt}

\noindent from which the counts follow by binomial inversion,

\begin{equation}\label{eq:omega-inversion}
  N_{kj} = \sum_{r \ge k,\, s \ge j} (-1)^{(r - k) + (s - j)} \binom{r}{k} \binom{s}{j} M_{rs}.
\end{equation}
\vspace{1pt}

\noindent Only sets to which at least one node belongs contribute to \(M_{rs}\), and every such set is a subset of one node's memberships. The moments can therefore be accumulated node by node, at a cost of \(2^{|m_v| + |m'_v|}\) for a node \(v\) with \(|m_v|\) memberships in \(\mathcal{K}\) and \(|m'_v|\) in \(\mathcal{K}'\). The arithmetic is exact in integers, and on the 10,830 pairs of covers scored both ways the result matched enumeration exactly. The cost is independent of community size but exponential in the memberships per node, which rules out a cover in which one node holds 110 memberships (Section~\ref{sec:reporting}).

\emph{Overlapping normalized mutual information} represents each community as a binary random variable over the nodes. For community \(C\), the variable \(X_C\) indicates whether a node of \(V\) belongs to \(C\), an unplaced node belonging to none, so its entropy \(H(X_C)\) is determined by the share \(|C|/n\). Following \textcite[14--16]{lancichinettiDetectingOverlappingHierarchical2009}, abbreviated LFK below, the conditional entropy of \(X_C\) given the other cover is the smallest of its conditional entropies given the communities of that cover,

\vspace{1pt}
\begin{equation}\label{eq:conditional-entropy}
  H(X_C \mid \mathcal{K}') = \min_{C' \in \mathcal{K}'} H(X_C \mid X_{C'}).
\end{equation}
\vspace{1pt}

\noindent The minimum is taken only over the communities \(C'\) that satisfy an admissibility constraint, \(h(p_{11}) + h(p_{00}) > h(p_{01}) + h(p_{10})\), where \(h(p) = -p \log p\) and \(p_{11}\), \(p_{10}\), \(p_{01}\), and \(p_{00}\) are the shares of nodes in both communities, only in \(C\), only in \(C'\), and in neither. If no \(C'\) qualifies, the conditional entropy is set to \(H(X_C)\), as it would be if the other cover carried no information about \(C\). The entropy of a cover is the sum of its communities' entropies, \(H(\mathcal{K}) = \sum_{C \in \mathcal{K}} H(X_C)\), and its conditional entropy is \(H(\mathcal{K} \mid \mathcal{K}') = \sum_{C \in \mathcal{K}} H(X_C \mid \mathcal{K}')\). Two normalizations of the information that the covers share are in use,

\begin{align}
  \text{NMI}_{\max} &= \frac{\tfrac{1}{2}\bigl[\left(H(\mathcal{K}) - H(\mathcal{K} \mid \mathcal{K}')\right) + \left(H(\mathcal{K}') - H(\mathcal{K}' \mid \mathcal{K})\right)\bigr]}{\max\{H(\mathcal{K}),\, H(\mathcal{K}')\}}, \label{eq:nmi-max} \\[1.5em]
  \text{NMI}_{\text{LFK}} &= 1 - \frac{1}{2}\left[ \frac{1}{|\mathcal{K}^{+}|} \sum_{C \in \mathcal{K}^{+}} \frac{H(X_C \mid \mathcal{K}')}{H(X_C)} + \frac{1}{|\mathcal{K}'^{+}|} \sum_{C' \in \mathcal{K}'^{+}} \frac{H(X_{C'} \mid \mathcal{K})}{H(X_{C'})} \right], \label{eq:nmi-lfk}
\end{align}
\vspace{10pt}

\noindent where \(\mathcal{K}^{+}\) contains the communities with non-zero entropy, the only ones for which the LFK ratio is defined. This paper uses \(\text{NMI}_{\max}\) \autocite[2--3]{mcdaidNormalizedMutualInformation2013}, because the LFK form does not reach zero for unrelated covers.\footnote{On one of the \corpus{wiki-topcats} subsamples of Section~\ref{sec:data}, LFK scored a single all-node community 0.5000 and a cover detected on the same graph 0.0171, rating the degenerate cover more than an order of magnitude above the detected one. \(\text{NMI}_{\max}\) scored the same degenerate cover zero, as a normalization should.}

\emph{Average best-match \(F_1\)} \autocite[594]{yangOverlappingCommunityDetection2013} scores each detected community against its best-matching reference community, where the match \(f\) between two communities is the harmonic mean of the precision and recall of one against the other. The calculation is repeated in the reverse direction, and the two mean scores are averaged, giving

\vspace{5pt}
\begin{equation}\label{eq:f1}
  f(C, C') = \frac{2\,|C \cap C'|}{|C| + |C'|}, \qquad F_1 = \frac{1}{2}\left[ \frac{1}{|\mathcal{K}|} \sum_{C \in \mathcal{K}} \max_{C' \in \mathcal{K}'} f(C, C') + \frac{1}{|\mathcal{K}'|} \sum_{C' \in \mathcal{K}'} \max_{C \in \mathcal{K}} f(C, C') \right].
\end{equation}
\vspace{5pt}

\noindent Although it is not chance-corrected, average best-match \(F_1\) is the most sensitive of the three indices to whether communities are the right size. It is therefore read alongside the other two in Section~\ref{sec:discussion-index}.

When \(\Omega\) and \(\text{NMI}_{\max}\) measure stability, the covers compared are those that different seeds return, and the comparison runs over the nodes that at least one of them places. A node that no seed places does not enter it, which is why the share of nodes placed is reported beside a stability figure (Section~\ref{sec:reporting}). The consensus streams of Section~\ref{sec:consensus} are the exception, and their agreement is computed over all of \(V\).

\subsection{Datasets}
\label{sec:data}

This paper draws on 16 datasets, summarized in Table~\ref{tab:corpora} alongside their structure and their roles. The provenance of a reference cover determines the scope of any conclusion about accuracy. A planted cover is correct by construction, so it supports claims about whether a measure ranks the known structure highest and whether a method recovers it. A cover derived from metadata supports less. Scores against such covers are read with the caution of Section~\ref{sec:related-benchmarks} and are not taken as absolute accuracy. Datasets without reference covers are assessed only with diagnostics that do not require them.

\input{tables/corpora}

The primary synthetic family is a directed LFR suite with planted overlap, the \emph{main suite}. It follows the original construction \autocite[{2}]{lancichinettiBenchmarkGraphsTesting2008} and its extension to directed graphs and overlapping communities \autocite[{1}]{lancichinettiBenchmarksTestingCommunity2009}. The \emph{mixing parameter} \(\mu\) is the fraction of each node's edges that join it to nodes with which it shares no community \autocite[{2}]{lancichinettiBenchmarksTestingCommunity2009}, and it is applied to outgoing and incoming edges alike. The base suite contains 36 graphs of 1,000 nodes. It covers nine requested mixing parameters, from \(\mu = 0.1\) to \(\mu = 0.8\), and four levels of planted overlap: none, ten, 30, and 50 percent of nodes. Each combination of a mixing parameter and an overlap level is a \emph{parameter set}, and pooled analyses use ten realizations of each, including the base graph. At the three non-zero levels, each overlapping node belongs to exactly two communities. This two-membership cap is imposed by the configuration rather than by the generator. A variant with three memberships per overlapping node, the \emph{three-membership suite}, uses the same grid and serves as a control in Sections~\ref{sec:stability-planted} and~\ref{sec:generalisation}. The four-membership and five-membership suites serve the same role in Section~\ref{sec:generalisation}.

The suite was generated by a reimplementation, written so that direction, edge weights, and planted overlap could be combined in one generator. Edges are wired using a directed configuration model with stub matching rather than the rewiring procedure of the LFR authors' reference implementation. The generator was tested for the properties its use depends on, namely planted overlap exactly as requested, community sizes within their bounds, realized mixing that tracks the request, and direction and edge weights as specified. Its graphs were not compared with those of the reference implementation one by one, since the two generators realize different mixing and degree from the same request. As a check on the main result of Section~\ref{sec:peak}, the base suite's parameter sets were also generated with the reference implementation (the outcome is reported in Section~\ref{sec:peak-results}). The realized mixing parameter is the fraction of a graph's edges whose endpoints share no planted community, \(|\{(u,v) \in E : m_u \cap m_v = \emptyset\}|/m\) with \(m_v\) read from the planted cover, and it differs from the requested value. At a requested value of \(\mu = 0.5\), it is higher on every base graph of the two LFR suites and of the block model introduced below, and ranges from 0.508 to 0.547. Graphs are labelled throughout by the requested value.\footnote{The realized value is recorded for each graph. Relabelling the graphs by it removes the highest level from every subset defined by \(\mu \le 0.5\). Under that relabelling, no result reported in this paper changes its sign or its significance at the five percent level, except SLPA's place above ego-splitting under \(\Omega\) on the graphs with planted overlap (Section~\ref{sec:discussion-index}), which the realized labels reverse, leaving ego-splitting ahead by 0.0006.} Because overlap and mixing are varied independently, the suite separates overlap from the strength of community structure. Its nine graphs with no planted overlap have planted partitions, on which all four quality measures are defined, and are called its \emph{partition graphs}.

The same family was also built at 10,000 and 50,000 nodes. Community sizes and density were held constant, so the number of communities, not their size, grows with the graph. This choice reflects findings for large real networks, in which the tightest communities extend to about 100 nodes, while larger communities are progressively less separated from the rest of the network \autocite[29]{leskovecCommunityStructureLarge2009}. The \emph{10,000-node suite} covers the same nine mixing parameters, and the \emph{50,000-node suite} was built only up to \(\mu = 0.6\). Both are used in the ladders of Section~\ref{sec:peak}, and the larger is also used in the comparison of grid scaling in Section~\ref{sec:matched}.

A second planted family tests whether findings from the main suite depend on its generator. It is a directed, overlapping, degree-corrected stochastic block model with 1,000 nodes, the \emph{block-model suite}. Its base suite covers the same 36 parameter sets, and pooled analyses use ten realizations of each, permitting comparisons at matched requested mixing and overlap levels. The degree correction follows the model of \textcite[3--4]{karrerStochasticBlockmodelsCommunity2011}, in which a block-level rate is multiplied by per-vertex parameters to produce a heterogeneous degree sequence. The overlap formulation follows \textcite[2--3]{ballEfficientPrincipledMethod2011}, in which each vertex carries one propensity per community and communities contribute edges to it independently.\footnote{In the overlap model of \textcite[2]{ballEfficientPrincipledMethod2011}, links are coloured and each vertex's memberships are inferred from the colours it carries. In this paper, the cover is planted first and edges are then drawn conditional on it. Accuracy must be measured against a planted cover. Inferring memberships from the realized colouring would make the reference depend on the draw. The model's overlap formulation is used, but its inference procedure is not reproduced.}

Four deliberate design choices distinguish the block model from LFR and make the comparison informative. LFR draws community sizes and degrees from power-law distributions, wires edges by matching configuration-model stubs, and enforces the mixing parameter separately at each node. By contrast, the block model draws community sizes uniformly and generates degree heterogeneity from log-normal per-vertex propensities. It samples a Poisson count independently for each ordered node pair and includes an edge whenever that count is non-zero. Its mixing parameter is a single global target, the expected fraction of edges between nodes sharing no community. The block-model graphs contain between 13 and 18 planted communities, compared with 16 to 30 in the main suite.\footnote{The directed extension of the block model is this paper's own. The degree-corrected block model of \textcite[3]{karrerStochasticBlockmodelsCommunity2011} and the overlap model of \textcite[2]{ballEfficientPrincipledMethod2011} are formulated for undirected graphs. In this construction, in- and out-degree are corrected separately. The suite is therefore reported as a construction of this paper, without a claim that it reproduces the published behaviour of either model.}

The two families were not calibrated to be equally difficult. A partition cannot hold a node's second membership, so the score of a planted cover collapsed to a partition falls as planted overlap rises, by amounts that differ between families \autocite{pizzornoCoversPartitionsOverlapping2026}. Raw accuracies therefore cannot be compared directly across overlap levels or families. Instead, the margin is calculated for each graph by subtracting the partition baseline's score from the method's score on that same graph. These paired differences are then summarized within each family and overlap level.

The \emph{weighted suite} is used only to study edge weights (Section~\ref{sec:weights}). It applies the LFR construction with planted weights to 16 combinations of four mixing parameters, from \(\mu = 0.3\) to \(\mu = 0.6\), and the same four overlap levels. Each combination has ten realizations of a 1,000-node graph. The weights are drawn at one setting across all graphs, independently of \(\mu\). Internal edges have mean weights between 2.17 and 2.52 times those of external edges. As a result, between 17 and 43 percent of total edge weight crosses a community boundary, compared with 33 to 64 percent of edges. The weights therefore distinguish planted communities more clearly than the topology does on every graph. Three additional weighted suites are used only to bound the conclusions about weights and are described in Section~\ref{sec:weights}. The \emph{contrast suite} varies the contrast between internal and external edge weights. The \emph{weights-only suite} removes planted structure from the topology while retaining it in the weights. The \emph{real-weight suite} reuses the same planted graph realizations but draws their edge weights from the distributions observed in two real networks described below, \corpus{Enron} and \corpus{Advogato}.

For the quality-measure analysis, a further labelled family was drawn from the \corpus{DBLP} co-authorship network, where publication venues supply the reference communities \autocite[3]{yangDefiningEvaluatingNetwork2012}. It comprises ten distinct subsamples, each seeded on a different venue and containing 60 reference communities. Across the subsamples, between 27 and 35 percent of nodes overlap. Each subsample was also scored against a partition that assigns every author to the venue containing most of that author's edges. This partition allows the partition-based measures \(Q_d\) and \(Q_{lr}\) to be applied. The \corpus{DBLP} subsamples test whether the quality-measure findings are specific to the LFR benchmark. LFR and \corpus{DBLP} differ in construction, degree distribution, and community definition, so a finding replicated across both is not confined to the main suite.\footnote{The same \corpus{DBLP} construction is also swept across seven granularities, from five to 150 reference communities, to test whether a measure responds to granularity. The results are reported in Section~\ref{sec:mechanism}.}

The \corpus{wiki-topcats} subsamples use editorial categories as reference covers \autocite{klymkoUsingTrianglesImprove2014,yinLocalHigherOrderGraph2017}. In this paper they serve the paired-interval example of Section~\ref{sec:matched}, and the comparison of methods on them is reported in the companion paper \autocite{pizzornoCoversPartitionsOverlapping2026}. That comparison also gives the basis for the size floor of Section~\ref{sec:notation}, the size at which OSLOM's community counts on these subsamples approach those of the reference, and another graph may call for another floor.

The reference-free analysis uses the \corpus{Enron} email network \autocite{Klimt2004IntroducingTE} and the smaller \corpus{Advogato} developer trust network \autocite[658]{massaBowlingAloneTrust2009}. Neither has a reference cover. \corpus{Enron} job titles are not used as one because they encode seniority rather than community membership. For example, two vice presidents in different business units need not belong to the same community. Scoring against those titles would make the substitution that \citeauthor{peelGroundTruthMetadata2017} warned against (\citeyear[3]{peelGroundTruthMetadata2017}). \corpus{Advogato} represents developer trust rather than email correspondence, providing a contrast to \corpus{Enron}.\footnote{All graphs are reported as loaded after self-loops are discarded, so they may differ from their source catalogues. For \corpus{Advogato}, 3,992 catalogue edges are self-loops and 1,386 catalogue nodes occur in no edge. The loaded graph therefore contains the 5,155 nodes and 47,135 edges reported in Table~\ref{tab:corpora}.} Both networks are directed and weighted, matching the methods' target setting. Reference-cover benchmarks for this setting remain scarce.

Unless stated otherwise, \corpus{Enron} refers to the internal network of correspondents within the organization. The unrestricted graph also includes an external tail composed largely of one-off and automated senders. Because the tail changes the graph's component structure, the unrestricted graph is used to examine the effect of including external senders. Without reference covers, the analysis can describe community counts, the number of very small communities, the extent of detected overlap, and run-to-run variability, but cannot establish accuracy.

\subsection{Detection Methods}
\label{sec:methods}

Four overlapping methods are compared in Section~\ref{sec:comparison}. OSLOM, the Order Statistics Local Optimization Method, tests each community for statistical significance against a configuration-model null \autocite{lancichinettiFindingStatisticallySignificant2011}. SLPA, the Speaker--listener Label Propagation Algorithm, has each node accumulate the labels it hears in repeated exchanges with its neighbours \autocite{xieSLPAUncoveringOverlapping2011}. Ego-splitting resolves each node's ego-network into personas and partitions the resulting graph \autocite{epastoEgoSplittingFrameworkNonOverlapping2017}. CD-DAWN, for Community Detection in Directed And Weighted Networks, expands communities from seeds \autocite{kumarOverlappingCommunityDetection2025}. All four are also examined in the stability study (Section~\ref{sec:stability}), and OSLOM and SLPA in the reference-free diagnostics of Section~\ref{sec:battery}.

Leiden \autocite{traagLouvainLeidenGuaranteeing2019} serves as the partition baseline. Because it returns a partition, it cannot recover a planted overlapping cover. An overlapping method that does not outperform it on graphs with planted overlap has therefore shown no gain in accuracy from its additional machinery. The baseline optimizes the objective of \textcite[{2--3}]{reichardtStatisticalMechanicsCommunity2006}, whose configuration-model null term is scaled by a resolution parameter, and at resolution one that objective is the directed modularity \(Q_{d}\) of Section~\ref{sec:measures}. The directed implementation in the \emph{leidenalg} package was run at nine resolutions: 0.5, 1.0, 1.5, 2.0, 2.5, 3, 4, 6, and 8.

The package released with this paper implements SLPA, ego-splitting, and CD-DAWN, and each implementation departs from its method's published description \autocite{pizzornoCoversPartitionsOverlapping2026}. Ego-splitting was extended to directed and weighted graphs. CD-DAWN follows its authors' code in three respects where that differs from their pseudocode. SLPA departs furthest, and in this paper the name refers to the implementation documented in the companion paper. After propagation, that implementation applies a size-shaping stage, which merges communities below a lower size threshold and splits those above an upper one. The stage's parameters are swept when SLPA is evaluated against a reference cover. The stage is disabled in the reference-free analyses, including the comparison with OSLOM in Section~\ref{sec:battery}, where cover shape is assessed without post-processing.

All three implementations were validated \autocite{pizzornoCoversPartitionsOverlapping2026}: CD-DAWN against its authors' compiled binary; SLPA against the scores of the implementation in \emph{cdlib} \autocite{rossettiCDLIBPythonLibrary2019}, which runs only on the undirected projection of a graph; and ego-splitting's extension with unit tests on constructed graphs.

OSLOM was run without modification, compiled from its authors' source distribution (version 2.5). Its directed binary was used on directed graphs and its undirected binary wherever an undirected graph was clustered, including in the consensus procedure of Section~\ref{sec:consensus}. OSLOM's search depth is the number of optimization runs it makes at each level of its hierarchy. It is a free parameter, and Section~\ref{sec:discussion-shared} measures its effect. The planted suites and the \corpus{wiki-topcats} subsamples were run at the authors' default depth, ten runs at the first level and 50 at higher levels (10/50). The graphs without reference covers and every base run of the consensus sweep were run at 3/3, because the default was unaffordable for the unrestricted graph and for the sweep. The internal \corpus{Enron} graph was rerun at the default depth as a control (Sections~\ref{sec:inversion} and~\ref{sec:discussion-shared}). Within each group, every configuration compared was run at the same depth. OSLOM reads edge weights only under its \texttt{\footnotesize{-w}} option and treats a graph as unweighted under \texttt{\footnotesize{-uw}}. The tables use the two options as labels for the weighted and unweighted configurations of every method.

\section{Behaviour of the Quality Measures at the Reference Cover}
\label{sec:peak}

Validation begins with a direct test of whether each quality measure attains its maximum at the planted cover. A measure that scores any rival above the planted cover fails this test, regardless of its agreement with other measures. Without a trusted reference cover, this failure cannot be identified on the graphs where the measures are ordinarily applied. The section sets out the protocol, reports its results, traces the main failure to its mechanism, tests the four measures for the resolution limit, and states what the two failures imply for the comparison of covers.

\subsection{The Protocol}
\label{sec:ladders}

The protocol scores each measure on three one-parameter ladders of rival covers, each starting at the planted reference: the \emph{merge}, \emph{split}, and \emph{noise} ladders. It also scores two degenerate covers, the one-community cover and the all-singleton partition. The box below specifies the protocol so that it can be applied to measures not examined in this paper.

The ladders test distinct failure modes. The noise ladder tests sensitivity to membership structure at fixed granularity, while the merge and split ladders test preferences for coarser and finer covers, respectively. A maximum away from the reference on the merge ladder signals a resolution-limit bias, and on the split ladder it signals a fine-cover bias.

\vspace{5pt}
\begin{protocolbox}{Specification of the Validation Protocol}
  \textbf{Inputs.} A benchmark graph with a planted or otherwise trusted reference cover and the measure under test. The protocol requires no community-detection run. Its random-split and noise ladders are random constructions. In this paper the random split is drawn once and each level of the noise ladder 30 times.

  \smallskip
  \textbf{Rival covers.} Three one-parameter ladders begin at the reference cover and move away from it one rung at a time. Two degenerate covers are scored separately.
  \emph{Merge}: fuse the two smallest communities at each rung until one remains.
  \emph{Split}: divide the largest community at each rung, either by halving it at random (the \emph{random split}) or by cutting it along a sparse internal cut (the \emph{structural split}). The cut is found on the community's induced subgraph, treated as undirected. A disconnected community is split into its largest component and the remainder. For a connected community, the initial split follows the sign of its Fiedler vector \autocites{fiedlerAlgebraicConnectivityGraphs1973}{fiedlerPropertyEigenvectorsNonnegative1975} and is refined by the Kernighan--Lin heuristic \autocite{kernighanEfficientHeuristicProcedure1970}, which preserves the sizes of the two parts while reducing the number of edges between them. Each split ladder has eight rungs in this paper, and Section~\ref{sec:resolution} extends the structural one to 120.
  \emph{Noise}: at each rung, select each membership of the reference cover independently for reassignment with probability equal to the rung's level, then assign each selected membership to a uniformly chosen community. The expected fraction of memberships selected is therefore the rung's level. No community is created, and one is lost only if reassignment leaves it empty. The levels in this paper are 0.1, 0.2, 0.4, 0.6, 0.8, and 1. Each level is drawn repeatedly, and the measure is read from its mean score at each level and the standard error of that mean.
  \emph{Degenerate covers}: the whole graph as one community, and the all-singleton partition.

  \smallskip
  \textbf{Invariants, asserted at every rung.} Every node in the reference cover remains assigned to at least one community, so that a deleted community cannot be mistaken for a merge. For partition references, the total number of node--community memberships is also conserved. The count may legitimately fall on an overlapping reference, since merging two communities that share a node removes one membership, as does reassignment on the noise ladder to a community that already contains the node.

  \smallskip
  \textbf{Pass.} The measure attains its maximum at the reference cover on the merge and split ladders, both of which start there. Its mean score on the noise ladder never rises from one level to the next by more than two standard errors of the difference. The reference scores above both degenerate covers.

  \smallskip
  \textbf{Fail.} A maximum away from the reference locates the bias. On the merge ladder it indicates a preference for coarse covers, the signature of a resolution limit, and on the split ladder a preference for fine covers. A mean score that rises as memberships are randomized indicates insensitivity to structure. A score for either degenerate cover at or above the reference is an outright failure.

  \smallskip
  \textbf{Cost.} One evaluation per rung of each ladder and per degenerate cover, for each draw of a random ladder.
\end{protocolbox}

\subsection{Results}
\label{sec:peak-results}

The protocol's three simplest tests ask whether a measure scores the planted cover above every randomly split cover, whether it scores that cover above both degenerate covers, and whether its mean score never rises as the share of randomized memberships increases. They were applied over ten realizations of each planted family. \(Q_{d}\) and \(Q_{lr}\) are defined on the 180 partition graphs, and \(Q_{ov}\) and \(Q_{dwo}\), the two measures that accommodate overlapping covers, on all 720 graphs. All four measures passed the three tests on every graph. On the noise ladder, no measure's mean score rose from one level to the next on any graph, and the smallest fall between two adjacent levels was 1.4 standard errors.\footnote{One draw of each level does not suffice. Where the scores at two levels are close, a single draw decides their order. Read as 30 separate ladders, the draws showed the score of \(Q_{ov}\) rising between two levels on five to 28 of the 720 graphs, depending on the draw, that of \(Q_{dwo}\) on seven to 26, and those of \(Q_{d}\) and \(Q_{lr}\) on up to four and 12 of the 180 partition graphs. A noise-degraded cover scored above the planted cover in one draw, for both \(Q_{ov}\) and \(Q_{dwo}\), at the lowest noise level on a block-model graph at \(\mu = 0.8\) with 50 percent planted overlap.} The structural-split test was harder, and three of the measures failed it on some graphs, but only by small margins. \(Q_{ov}\) scored a structurally split cover above the planted cover on 146 of the 720 graphs, by less than 0.00003 in each case, and \(Q_{d}\) and \(Q_{lr}\) did so on 14 and 17 of the 180 partition graphs, by at most 0.0005 and 0.0012. \(Q_{dwo}\) did so on none. Taken alone, the simpler tests would have made the measures appear to behave sensibly.

\input{tables/peaks-at-truth}

The merge ladder revealed the failure that the simpler tests missed. Table~\ref{tab:peaks-at-truth} reports the merge-ladder results on the nine partition graphs, where all four measures are defined.\footnote{\(Q_{d}\) and \(Q_{lr}\) are partition measures. On an overlapping cover, the directed modularity counts a shared node's degree once for each community containing it, so the quantity returned has no principled definition. Therefore, their scores on overlapping covers, planted or detected, are not reported anywhere in this paper.} On every graph of the suite, across all mixing parameters and overlap levels, \(Q_{ov}\) failed to attain its maximum at the planted cover. It systematically preferred coarser covers. \citeauthor{kumarNeighborhoodProximityBased2019} noted that the measure \enquote{does not punish the existence of \enquote{giant} communities} (\citeyear[1355]{kumarNeighborhoodProximityBased2019}). Across the nine partition graphs, the highest-scoring merge-ladder covers had a median of two communities, while the planted covers had a median of 21. The median score gap, the amount by which the best rival outscored the planted cover, was 0.1197.\footnote{The merge ladder was run to \(k = 1\), so it spanned its whole domain, and its last rung is the one-community cover that the protocol also scores separately. The random-split ladder was not run to its own endpoint, and the effect of that asymmetry was checked. No measure took its maximum at the ladder's last rung on any graph, and every measure already scored that rung below the planted cover, so extending the ladder was unlikely to change the result. The structural-split ladder is different. Some of its maxima fell on its last rung, so an advantage measured on it is a lower bound (Section~\ref{sec:mechanism}).} That median reflected the harder half of the mixing range. On the single graph at each mixing level, the gap rose from 0.0105 at \(\mu = 0.1\), where the preferred cover contained five communities, to 0.2600 at \(\mu = 0.8\), where it contained two. Over ten realizations of each planted family, \(Q_{ov}\) peaked at the planted cover on none of the 360 graphs in either family. On the 36 graphs generated with the LFR authors' reference implementation (Section~\ref{sec:data}), scored on the merge and random-split ladders, \(Q_{ov}\) likewise attained its maximum at the planted cover on none, with a median score gap of 0.1575, against 0.1695 on the suite's base graphs.

The degree of coarsening also depended on the merge rule. Fusing the two smallest communities, as above, produced the more extreme result. When a uniformly random pair was fused instead, the maximizing covers contained between two and ten communities across the nine graphs and three seeds, compared with 21 in the planted covers. The preference for coarser covers was a property of the measure, but the ladder path affected the exact granularity.

\(Q_{dwo}\) also failed on the merge ladder, though less severely. It peaked at the planted cover on two of the nine graphs. Its preferred cover contained a median of 17 communities, and on the other seven graphs its best rival exceeded the planted score by up to 0.1142, at \(\mu = 0.8\). \(Q_{d}\) and \(Q_{lr}\) each failed on two of the nine graphs, at \(\mu = 0.75\) and \(0.8\), where no method recovered the planted cover (Section~\ref{sec:comparison-results}) and a higher score for another cover is weaker evidence against a measure.

The failure was not specific to the LFR generator. The ten undirected \corpus{DBLP} co-authorship subsamples used publication venues as reference communities. On every subsample, \(Q_{ov}\) failed to peak at either reference: the venue cover or the partition assigning each author to the venue containing most of that author's edges. For the venue cover and the author-to-venue partition, respectively, the best rival exceeded the reference by median score gaps of 0.1716 and 0.1292. All four measures were defined for the author-to-venue partition, but none peaked at the reference on any subsample. The median score gaps for \(Q_{d}\) and \(Q_{lr}\), at 0.0022 and 0.0014, were much smaller than the 0.1292 for \(Q_{ov}\). A metadata reference need not be the structural optimum (Section~\ref{sec:related-benchmarks}), so a failure to peak there is weak evidence when taken alone. The evidence against \(Q_{ov}\) is the size of its gap, more than 50 times that of the undamped \(Q_{d}\) on the same partitions.

The failure also persisted at larger graph sizes. On the LFR family generated at 10,000 and 50,000 nodes, \(Q_{ov}\) attained its maximum at the planted cover on none of the 180 graphs at the smaller size and none of the 72 at the larger. At both sizes, the maximizing covers contained between two and six communities, while the planted covers contained between 178 and 274 communities at 10,000 nodes and between 899 and 1,372 at 50,000 nodes.

\subsection{The Damping Mechanism}
\label{sec:mechanism}

The preference of \(Q_{ov}\) for coarse covers arose from the damping in its null term (Section~\ref{sec:measures}). On a partition, the mean belonging \(\bar{a}_C\) of community \(C\) is its share of the nodes, \(|C|/|V|\). The null term in \(Q_{ov}\) is therefore the corresponding term in \(Q_{d}\), multiplied by \(\bar{a}_C^{\,2}\). For \(k\) equally sized communities, this factor is \(1/k^2\), so the finer the cover, the less of the null term remains. In the 1,000-node graphs, where the planted communities contained a few dozen nodes, almost none remained. At the planted covers of the nine partition graphs, the damped null term was between 0.34 and 0.35 percent of the undamped null term in \(Q_{d}\). With the null term nearly absent, the score of \(Q_{ov}\) was close to the fraction of edges within communities, which cannot decrease as communities are merged. The merge ladder therefore exposed the resulting preference for coarse covers.

On a partition, the effect of the damping can be isolated exactly, because \(Q_{ov}\) and \(Q_{d}\) then have identical observed-edge terms and only \(Q_{ov}\) damps the null term by \(\bar{a}_C^{\,2}\). At every rung of every ladder, therefore,

\vspace{5pt}
\begin{equation}\label{eq:damping}
  Q_{ov} - Q_{d} = m^{-1}\sum_C (1 - \bar{a}_C^{\,2})\,\kappa_C^{\text{out}} \kappa_C^{\text{in}} / m.
\end{equation}
\vspace{5pt}

\noindent This difference alone moved the maximum. Across the nine partition graphs, the maxima of the two measures were separated by a median of 18 communities. \(Q_{d}\) peaked at 19 to 21 communities, against a planted median of 21, whereas \(Q_{ov}\) peaked at two to five. As communities were merged, both the within-community edge fraction and the damped null term rose. The \(Q_{ov}\) score stopped increasing once the rise in the null term outweighed the gain in within-community edges. At its maximizing cover, the damped null term had recovered to between 4.9 and 27 percent of the undamped term in \(Q_{d}\). Changing only the damping factor therefore shifted the maximum to a much coarser cover.

\input{tables/granularity-dose}

If damping is the mechanism, the failure should worsen as the reference grows finer and the damping stronger. This was tested with a sweep of reference granularity on \corpus{DBLP} (Section~\ref{sec:data}), in which the same subsampling was repeated with references of five to 150 communities. Across this range, the median null ratio at the venue cover, the ratio of the damped null term to the undamped term, spanned more than three orders of magnitude (Table~\ref{tab:granularity-dose}). \(Q_{ov}\) did not peak at the venue reference on any of the 70 graphs and peaked at the partition reference on only one, at five communities, the least-damped setting. As the reference grew finer, damping strengthened and the median score gap grew at every step, from 0.0299 to 0.2098 against the venues, and from 0.0197 to 0.1459 against the partition. By contrast, the median score gap for \(Q_{d}\), which uses the same expression without damping, remained within 0.01 of zero at every partition granularity. The partition sweep also provided a control. At five communities, \(Q_{d}\), \(Q_{lr}\), and \(Q_{dwo}\) peaked at the reference on most graphs, but from 60 communities onwards, none did. This showed that the sweep registered granularity-dependent behaviour in measures other than \(Q_{ov}\).

A further comparison separated damping from overlap, using an overlapping modularity whose null term is not damped. The \(EQ\) measure of \textcite[1708--1709]{shenDetectOverlappingHierarchical2009} is defined on a graph's undirected view as

\vspace{-5pt}
\begin{equation}\label{eq:eq}
  EQ = \frac{1}{2|E^{u}|} \sum_{C} \sum_{i,j \in C} \frac{1}{|m_i|\,|m_j|} \left[ A^{u}_{ij} - \frac{k^{u}_i k^{u}_j}{2|E^{u}|} \right],
\end{equation}
\vspace{5pt}

\noindent where \(E^{u}\), \(A^{u}\), and \(k^{u}_i\) are the edge set, the adjacency matrix, and the degrees of that view. Each pair of nodes is weighted by the inverse product of its membership counts, which is the product form of Section~\ref{sec:measures} under uniform belonging coefficients. On an undirected graph, \(EQ\) therefore has the observed-edge term of \(Q_{ov}\) in Equation~\eqref{eq:qov}, and Newman's null term where \(Q_{ov}\) has that term damped by \(\bar{a}_C^{\,2}\). It reduces to Newman's modularity on a partition. Scored on the merge and random-split ladders (Table~\ref{tab:eq-checks}), \(EQ\) peaked at the planted cover on 32 of the 36 LFR graphs, while its maximum exceeded the planted score by at most 0.0003 on the other four. On the structural-split ladder, it scored a split cover above the planted cover on nine of the 27 overlapping graphs. \(EQ\), \(Q_{ov}\), and \(Q_{dwo}\) were also scored on the ten \corpus{DBLP} subsamples, all three on the same undirected graphs. Along the merge ladder, the maximizing cover for \(EQ\) contained a median of 58.5 communities, against 60 venues in the reference, whereas the maximizing cover for \(Q_{ov}\) contained two. When a uniformly random pair of communities was merged at each rung instead, over three seeds, \(EQ\) scored some merged cover above the reference on 11 of the 27 overlapping LFR graphs and on nine of the ten \corpus{DBLP} subsamples, by at most 0.0049 and 0.0079, respectively. \(Q_{ov}\) did so on every one of those graphs, by up to 0.2529 and 0.1674. The coarse-cover preference of \(Q_{ov}\) was therefore not a general property of overlapping modularity.

\(EQ\) had limitations of its own. For most \corpus{DBLP} graphs, both \(EQ\) and \(Q_{dwo}\) gave a structurally split cover a higher score than the reference, by at most 0.0010 for \(EQ\) and 0.0141 for \(Q_{dwo}\) within the eight rungs of that ladder. The best rung for \(EQ\) was the last on five of the ten, so the ladder does not bound how much finer than the reference its preferred cover is. Its authors also noted that \(EQ\) shares the resolution limit of modularity \autocite[1709]{shenDetectOverlappingHierarchical2009}.

\input{tables/eq-checks}

Together, these comparisons identified damping of the null term as the mechanism behind the coarse-cover preference of \(Q_{ov}\). Within the tested range, stronger damping accompanied larger score gaps. The sweep itself did not reach the undamped end, since even at its least-damped setting the null ratio at the reference never exceeded 0.11, and that end is supplied by the two undamped measures. On partitions the undamped form is \(Q_{d}\), which peaked at the planted cover on seven of the nine LFR partition graphs and stayed within 0.01 of the reference on the co-authorship partitions without peaking there. On overlapping covers it is \(EQ\), which showed no comparable coarse-cover preference, though it retained a structural-split bias.

The larger LFR graphs extended the same pattern. At the planted partitions, the damped null term was between 0.0049 and 0.0056 percent of the undamped null term in \(Q_{d}\) at 10,000 nodes, and was 0.0002 percent at 50,000 nodes. The score gap changed little with graph size. At each mixing parameter, the median score gap, over every overlap level and realization, was larger by 0.005 to 0.017 at 10,000 nodes than at 1,000, and at 50,000 nodes it was within 0.003 of its value at 10,000. With the null term negligible, \(Q_{ov}\) again approached the fraction of edges within communities. The score gap was therefore governed by the edges left between communities in the planted cover, and it increased with mixing, from 0.069 at \(\mu = 0.1\) to 0.290 at \(\mu = 0.8\) at 10,000 nodes, against 0.058 and 0.274 at 1,000 nodes. The failure thus lies in the ranking, and no rescaling of the scores can correct it, since changing their magnitudes cannot move the maximum from a coarse cover to the planted one.

\subsection{Resolution Limits and Two Proposed Remedies}
\label{sec:resolution}

The damping is specific to \(Q_{ov}\). A second preference for coarser covers, the resolution limit, is an established failure of modularity, and it was present in all four measures examined in this paper. The test for it is a ring of cliques, a deterministic construction that makes the intended communities unambiguous without a random graph generator or a judgement about communities in empirical data. It adapts the ring of \textcite[39]{fortunatoResolutionLimitCommunity2007} to directed graphs, with identical four-node cliques arranged in a ring and neighbouring cliques joined by one reciprocal pair of directed edges. The planted cover assigns each clique to its own community, while the rival merges adjacent pairs. A measure exhibits the resolution limit when it scores the merged cover above the planted one.

\input{tables/resolution-limits}

Table~\ref{tab:resolution-limits} shows that every measure scored the pairwise-merged cover above the planted cover once the ring was large enough. \(Q_{ov}\) first did so at six cliques, while \(Q_{dwo}\), \(Q_{d}\), and \(Q_{lr}\) first did so at 16. From that point, the merged cover's advantage grew with ring size. This pattern was consistent with the merge-ladder results and came from an independent test.

One entry of the table has an exact explanation. Under \(Q_{dwo}\) and \(Q_{d}\), the two covers tied at 14 cliques, and the tie follows from an identity between the two measures. Let \(q\) be the clique size, \(K = q(q-1)+2\) the total degree of each clique, and \(r\) the number of cliques. Under both, the pairwise-merged cover outscores the planted cover by \(1/K - 1/r\), although the two measures' absolute scores differ. The covers therefore tie at \(r = K\), which is 14 cliques for \(q = 4\), and eight, 22, and 32 cliques for cliques of three, five, and six nodes.\footnote{For \(q = 4\), with \(m = rK\) edges, the planted cover scored \(6/7 - 1/r\) under \(Q_{d}\) and \(6/7 - 6/(7r)\) under \(Q_{dwo}\), while the pairwise-merged cover scored \(13/14 - 2/r\) and \(13/14 - 13/(7r)\).} The score-difference identity is a consequence of the ring's uniform structure. Each clique has total degree \(K\), so the null term in \(Q_{d}\) and the multiple-membership factor in \(Q_{dwo}\) reduce to the same function of the number of communities. This identity need not hold on irregular graphs.

Because all four measures exhibited this limit, none can be relied on to compare covers of different granularity. The constraint in Section~\ref{sec:granularity-constraint} therefore applies to all four. Nor does the bias always favour merging. Partition density, used in the link-community framework, has an inverse resolution limit and strongly favours separating a triangle from a neighbouring link community, even when the two share no node \autocite[1--3]{leeInverseResolutionLimit2017}. Scale bias can therefore favour covers that are too coarse or too fine.

The protocol detected such a fine-scale bias in \(Q_{ds}^{ov}\), the overlapping extension of modularity density proposed by \textcite[858--859]{chenExtensionModularityDensity2014}, an objective designed to address scale bias in modularity. It adds a density weighting and a penalty on edges between communities, and their Equation 19 gives it in full. In the configuration recommended by its authors, each node's belonging coefficient is the reciprocal of its membership count, and each edge contribution is weighted by the product of the coefficients at its endpoints \autocite[861]{chenExtensionModularityDensity2014}.

\(Q_{ds}^{ov}\) was tested on 72 graphs from the two planted families, using the first realization for each setting. For \(\mu \le 0.5\), it scored the planted cover at least as highly as every cover on the merge ladder for 37 of the 40 graphs, while \(Q_{ov}\) did so for none. Across all 72 graphs, it also scored the planted cover at least as highly as every randomly split or noise-degraded cover. On the structural-split ladder, however, its highest-scoring cover was finer than the planted cover on 32 of the 40 lower-mixing graphs. These covers contained between one and 20 more communities than the planted cover, with a median of three, and the largest score advantage was 0.0032, well above the structural-split exceptions reported in Section~\ref{sec:peak-results}. Identifying these maxima required extending the structural-split ladder to 120 rungs. Every score profile reached its maximum before the final rung, and across all 72 graphs the deepest maximum was 37 communities finer than the planted cover. At \(\mu > 0.5\), a structurally split cover still outscored the planted cover on 24 of the 32 graphs, and a coarse-side failure became the rule. The single-community cover scored at least as highly as the planted cover on 27 of 32 graphs. None of the four measures in Section~\ref{sec:peak-results} showed this single-community preference. These results showed that a measure designed to avoid a coarse resolution limit could nevertheless favour covers that were too fine or too coarse, depending on the mixing level. The ladders revealed which failure occurred.

A measure can be free of the limit, and the constant Potts model (CPM) of \textcite[{2}]{traagNarrowScopeResolutionlimitfree2011} is an example. Its authors called an objective resolution-limit-free when a partition that is optimal for a graph remains optimal for every subgraph induced by a subset of its communities, and showed that CPM has this property \autocite[{4--5}]{traagNarrowScopeResolutionlimitfree2011}. It compares each community's internal-edge count with a fixed resolution parameter times the number of node pairs in that community, rather than with a null model scaled to the whole graph. On the ring, the planted clique cover scored higher than the pairwise-merged cover at every size once the resolution exceeded \(1/16\). CPM was not amongst the four measures examined because it requires a chosen resolution. It was nevertheless tested on the partition ladders, by computing the interval of resolutions at which the reference scores at least as highly as every rival. For all 90 LFR graphs with planted partitions, some CPM resolution scored the planted cover at least as highly as every merged and randomly split rival. The harder test replaced the random-split rivals with structurally split rivals. Under this test, a CPM resolution scored the planted cover at least as highly as every merged and structurally split rival on 74 of the 90 LFR graphs, but on none of the ten \corpus{DBLP} graphs with reference partitions.

\subsection{The Granularity Constraint}
\label{sec:granularity-constraint}

Taken together, the validation results showed that the four measures' scores could not safely rank covers with different numbers of communities. Such differences in community count are common in method comparisons. In that setting, a score difference cannot be interpreted as evidence that one cover better captures the graph's structure. When a reference cover is available, detected covers should instead be compared by their accuracy under a chance-corrected index. Without a reference, each cover's shape should be reported directly (Section~\ref{sec:reporting}).

Nor is an equal number of communities sufficient. Covers with the same number of communities can have different community-size distributions, which can affect their quality scores independently of how well they capture the graph's structure. Score comparisons should therefore be restricted to covers that match in both count and shape.

These restrictions apply to the four measures tested, not to measure-based comparison in general. \(EQ\) came within 0.0003 of passing the merge ladder on the main suite, while CPM passed it at a suitable resolution on the LFR partition graphs. CPM could therefore support comparisons across community counts on partitions for which such a resolution exists, and \(EQ\) only where its resolution limit and its preference for split covers do not bind. Each would first have to pass the full protocol on the graphs under study.

\section{Run-to-Run Stability}
\label{sec:stability}

Run-to-run stability measures the agreement amongst covers produced by repeated runs of a method on the same graph. For \(R\) seeds yielding covers \(\mathcal{K}_1, \ldots, \mathcal{K}_R\), the stability reported under \(\Omega\) is the mean of that index over the \(R(R-1)/2\) unordered pairs of seeds,

\begin{equation}\label{eq:stability}
  \bar{\Omega} = \frac{2}{R(R-1)} \sum_{i < j} \Omega(\mathcal{K}_i, \mathcal{K}_j).
\end{equation}
\vspace{5pt}

\noindent \(\text{NMI}_{\max}\) is aggregated over the same seed pairs, and percentile-bootstrap intervals are obtained by resampling the seeds. Stability is attractive because it requires only the graph, and it is often used when no reference cover is available. In clustering, stability under data resampling has been proposed as a criterion for choosing the number of clusters \autocites[6]{ben-hurStabilityBasedMethod2002}[1299]{langeStabilitybasedValidationClustering2004}. In community detection, agreement across random seeds has likewise been used to choose a resolution \autocite[35]{fortunatoCommunityDetectionNetworks2016} and has been reported as evidence of a method's quality \autocite[27, 30]{kumarOverlappingCommunityDetection2025}.

Network science also uses \emph{stability} for a different quantity: \citeauthor{delvenneStabilityGraphCommunities2010} defined it as a partition-quality function based on the clustered autocovariance of a Markov process, evaluated across Markov time (\citeyear[12755]{delvenneStabilityGraphCommunities2010}). That function scores a single partition of a single graph. In this paper, by contrast, \emph{stability} means agreement amongst a method's covers across random seeds, which is a property of the procedure, not of an individual partition.

This section first relates stability to the structure of an objective, as earlier work has done. It then tests stability against accuracy on planted graphs and on the comparison of OSLOM's two weight configurations. From those tests it derives a reporting rule, which it then applies to consensus clustering.

\subsection{Stability and Objective Structure}
\label{sec:stability-objective}

Prior work has shown that stability need not indicate quality. For large samples, \textcite[5, 11--12]{ben-davidSoberLookClustering2006} showed that stability is determined by the structure of the objective. A unique global minimizer yields a stable algorithm, while symmetries in the data can create multiple minimizers and instability, regardless of whether the resulting clustering is correct. \Textcite[250]{luxburgClusteringStabilityOverview2010} made the implication explicit for \(K\)-means. At a fixed \(K\), stability depends on whether the objective has one or several global minima, not on whether \(K\) is correct. In network analysis, \citeauthor{peixotoRevealingConsensusDissensus2021} also showed that consensus summaries can mislead when the distribution of partitions is sufficiently heterogeneous (\citeyear[{1, 10}]{peixotoRevealingConsensusDissensus2021}).

\input{tables/cover-shape}

The large-sample result of \textcite[11--12]{ben-davidSoberLookClustering2006} rests on two conditions. The algorithm attains the optimum of its objective, and stability is measured across independent samples of the data. The procedures examined in this paper differ from that setting in both respects. They are rerun on one fixed graph rather than on independent samples, and they use local updates from random starting points without a guarantee of reaching an objective's optimum. Where an optimum exists, it may also have many close rivals. The modularity objective, for example, typically admits an exponential number of structurally diverse partitions with scores close to the optimum \autocite[{10}]{goodPerformanceModularityMaximization2010}. This degeneracy suggests a possible source of instability, but near-optimal alternatives alone do not determine how a particular randomized procedure behaves across runs.

High stability can therefore have another source, as an example from the real graphs shows. On the \corpus{Advogato} trust network, SLPA's unweighted configuration had \(\Omega = 0.888\), yet its cover was degenerate. Of its communities, 98 percent had five or fewer members, and one community contained 80 percent of the nodes (Table~\ref{tab:cover-shape}). The departures of this paper's SLPA from the published method (Section~\ref{sec:methods}) did not explain the degeneracy. With the published grouping and nesting rules, on the same 20 seeds, the largest community again contained 80 percent of the nodes, and agreement was \(\Omega = 0.889\). The example shows that the stability of a procedure rerun on one fixed graph is a different matter from the behaviour of objective minimizers that \textcite{ben-davidSoberLookClustering2006} characterized.

The frequency with which parts of a cover recur across a solution landscape has been used to assess their significance, including for overlapping and hierarchical covers \autocite[{3}]{calatayudExploringSolutionLandscape2019}. The example also qualifies estimates of that kind. Recurrence should be interpreted together with the procedure that generated the solutions, since the procedure itself may be the source of their agreement.

\subsection{Stability and Accuracy on Planted Graphs}
\label{sec:stability-planted}

\begin{figure}[tbp]
  \includegraphics[width=\textwidth,alt={Three panels, one per planted family: the main suite, the three-membership suite, and the block-model suite. Each plots, against the requested mixing parameter, the accuracy and the stability of OSLOM and of SLPA, and OSLOM's share of very small communities, averaged over settings and graphs. At mixing parameters up to 0.5 the more stable method is usually also the more accurate. Above it, accuracy falls towards zero for both methods, while OSLOM's stability falls and then rises again as its share of very small communities increases.}]{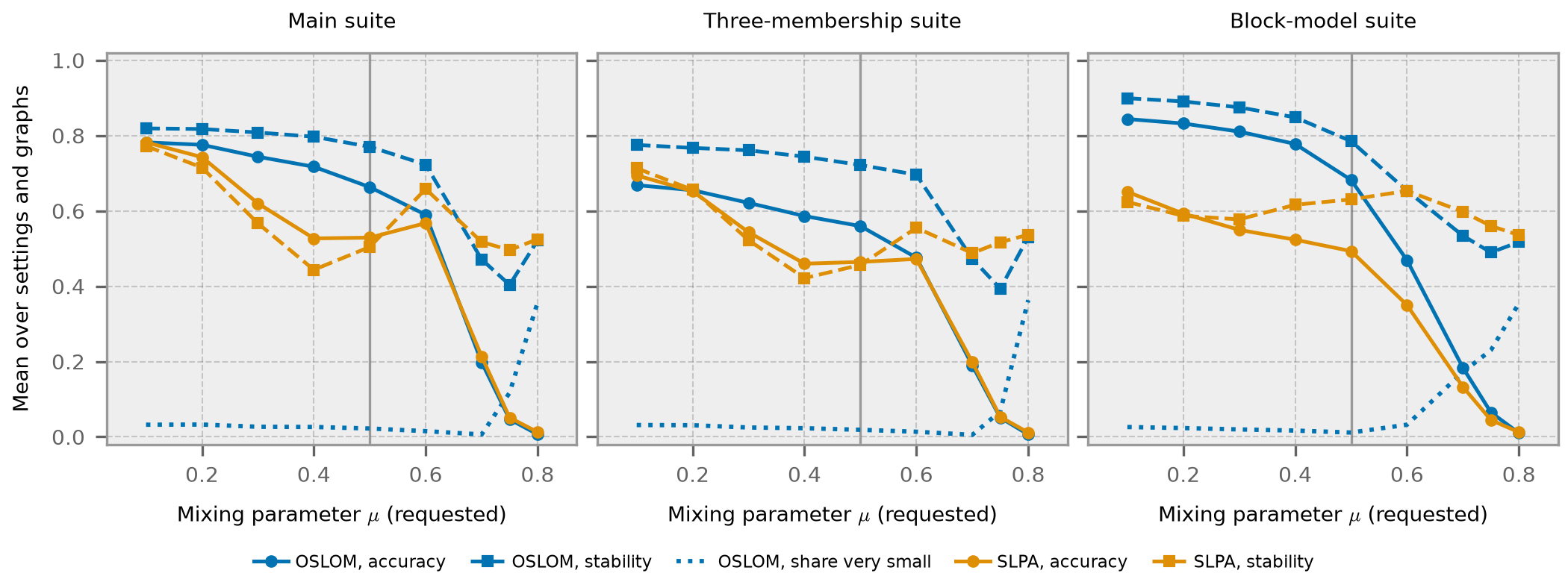}
  \caption{Accuracy against the planted cover (solid), stability (dashed), and, for OSLOM, the share of communities with five or fewer members (dotted), each averaged over every setting and graph at each requested mixing parameter, one panel per planted suite. The vertical rule marks \textit{μ}\tmthick{}=\tmthick{}0.5.}
  \label{fig:stability-regime}
\end{figure}

A planted cover makes it possible to test whether run-to-run stability tracks accuracy. The test was applied to the four methods of Section~\ref{sec:methods} on the main suite, the three-membership suite, and the block-model suite (Section~\ref{sec:data}). OSLOM and SLPA are reported first and shown in Figure~\ref{fig:stability-regime}, and ego-splitting and CD-DAWN follow. Each parameter set has ten graph realizations, and the plotted values average across each method's settings and the realizations at each mixing parameter.

At \(\mu \le 0.5\), stability tracked accuracy for OSLOM and SLPA. The more stable of the two methods was also the more accurate on 185 to 198 of the 200 graphs in each family, whether each method was evaluated at its most accurate setting or at the setting with median accuracy. Within each method, stability also ranked settings by accuracy. The median rank correlation was between 0.70 and 0.83 for SLPA and between 0.86 and 0.89 for OSLOM. The median loss in accuracy from choosing the most stable rather than the most accurate setting was about 0.009 at its largest.

Above \(\mu = 0.5\), the relationship weakened for both methods. When each was evaluated at its most accurate setting, the more stable method was also the more accurate on only 79 to 85 of the 160 graphs in each family, a rate indistinguishable from chance under a two-sided binomial test at the five percent level. At the setting with median accuracy, the count was 99 to 110, significantly above chance under the one-sided form of the same test.

OSLOM's stability initially fell with accuracy, then rose again as the share of very small communities increased. At \(\mu = 0.8\), the mean over its settings was between 0.517 and 0.531 for stability and between 0.006 and 0.010 for accuracy.\footnote{In 35.4 to 38.1 percent of the graph--setting combinations behind those figures, every seed returned one of the two degenerate covers of Section~\ref{sec:ladders}, the all-singleton partition or the single community containing every node. This is the degenerate case of Section~\ref{sec:accuracy}, in which the agreement expected by chance is total and \(\Omega\) is set to one. Those combinations therefore contributed perfect agreement to the stability reported, and outside them the mean of \(\Omega\) in the three families ranged from 0.242 to 0.253. The share of very small communities shows which covers were fragmented. The figure gives that share for OSLOM alone, because SLPA's settings impose a minimum community size of ten or 20 members, so its covers on the planted families contain no very small community. The SLPA configuration used on the real graphs has no such minimum and is evaluated on the planted families below. Mixing parameters are the requested values (Section~\ref{sec:data}).} That mean stability combined two kinds of cover, from the two ends of OSLOM's threshold grid. At thresholds of 0.2 and below, OSLOM returned the all-singleton partition for 92 to 96 percent of the combinations of graph and setting, and its covers agreed across seeds at 0.976 to 0.990 despite an accuracy of zero. At thresholds of 0.8 and above, its covers contained a median of 35 to 38 communities, almost none of them very small, and agreed at only 0.212 to 0.216.

The other two methods showed that the relationship also depended on the method. Up to \(\mu = 0.5\), stability ranked ego-splitting's settings by accuracy, as it did OSLOM's and SLPA's. The median rank correlation was between 0.69 and 0.75, and the median loss from choosing the most stable rather than the most accurate setting was about 0.014 at its largest. At \(\mu = 0.8\), ego-splitting's covers agreed across seeds at 0.348 to 0.355, well below perfect agreement. By contrast, CD-DAWN's correlation up to \(\mu = 0.5\) was negative, between −0.72 and −0.67. The median accuracy loss from choosing its most stable rather than its most accurate setting ranged from 0.062 to 0.472. At \(\mu = 0.8\), its covers agreed at 0.767 to 0.943 while scoring 0.001 on average against the planted cover.\footnote{Between 38.5 and 46.9 percent of CD-DAWN's graph--setting combinations at \(\mu = 0.8\) were in the degenerate case of the preceding footnote. That case is also reached when every seed returns a single community holding every node it places, since unplaced nodes do not enter a stability figure (Section~\ref{sec:accuracy}). Outside it, the mean of \(\Omega\) ranged from 0.562 to 0.907. Ego-splitting reached the degenerate case in only 1.7 to 2.4 percent of its combinations.} When CD-DAWN was one of the two methods compared, the more stable method was also the more accurate on as few as 17 of the 200 graphs at \(\mu \le 0.5\). Thus, CD-DAWN's more stable settings were less accurate even where detection succeeded.

For the real-graph analyses, OSLOM was run at search depth 3/3 and SLPA without its size-shaping stage (Section~\ref{sec:methods}). These configurations were also run on the planted families at the same settings (Table~\ref{tab:realdata-planted}). Without size shaping, 17.6 to 20.8 percent of SLPA's communities were very small in the two LFR suites and 72.4 to 87.6 percent in the block model. Within a graph, a larger share was associated with lower stability across SLPA's settings. The median rank correlation ranged from −1.00 to −0.70.

At \(\mu \le 0.5\), the more stable of the two configurations was also the more accurate on 237 of the 240 graphs on which they differed. Across these configurations, the rank correlation between the share of very small communities and stability was between −0.82 and −0.55 in each family. The positive association observed on the real graphs (Section~\ref{sec:reporting}) appeared only above \(\mu = 0.5\) on the block model, where the correlation was 0.74. In that regime, 87.6 percent of SLPA's communities were very small, its covers agreed across seeds at 0.880, and its accuracy against the planted cover was 0.000. Thus, for these real-graph settings, stability was positively associated with fragmentation only where detection had already failed.

Across the planted families, the relation of stability to accuracy depended on both the mixing level and the method. Stability tracked accuracy for some methods up to \(\mu = 0.5\), the relationship weakened above that boundary, and CD-DAWN's settings were ranked in the wrong direction even below it. The \corpus{Enron} comparisons that follow illustrate this ambiguity without establishing a general failure, because no planted reference is available. Section~\ref{sec:consensus} tests whether increasing agreement under consensus clustering preserves detected overlap.

\subsection{Stability Without a Reference Cover}
\label{sec:inversion}

Without a reference cover, a difference in stability is often the only evidence available for preferring one result to another. This subsection takes one such difference, between two configurations of OSLOM on \corpus{Enron}, and asks whether a difference of that kind marks the more accurate configuration on planted graphs, where accuracy is known.

The comparison is between OSLOM run with a graph's own edge weights, the \emph{weighted} configuration, and without them, the \emph{unweighted} configuration. Weights can affect a result in two ways, through the values they carry and through the fact that the method reads weights at all, and two controls separate the two. The \emph{constant} configuration assigns unit weight to every edge, so that weights are read but carry no information. The \emph{permuted} configurations shuffle the graph's weights across its edges, so that the values are kept but not their placement. Section~\ref{sec:reference-free} uses the same configurations, and Section~\ref{sec:weights} reports which was the more accurate on planted graphs.

On the internal \corpus{Enron} graph, OSLOM's weighted configuration was more stable than the unweighted one, with \(\Omega = 0.597\) compared with 0.500 (Table~\ref{tab:cover-shape}). The gain was 0.097, with a 95 percent bootstrap interval of 0.064 to 0.130 over the seeds shared by the two configurations. The weighted cover also had twice the share of very small communities, 36.1 percent against 16.9, though about as many of them, since it had half as many communities in all.

\input{tables/realdata-planted}

The two \corpus{Enron} graphs differed in the same direction. On the unrestricted graph, which also includes external senders (Section~\ref{sec:data}), OSLOM's unweighted configuration had \(\Omega = 0.626\), compared with 0.500 on the internal graph. The share of very small communities in its cover was also higher, at 74.1 percent against 16.9. In both comparisons, the higher stability accompanied a larger share of very small communities, although no planted reference was available to assess accuracy.

The weighted planted suite (Section~\ref{sec:data}) made it possible to compare stability with accuracy for the same method and the same pair of configurations. On that suite, OSLOM was more accurate without the graph's own edge weights (Section~\ref{sec:weights}, Table~\ref{tab:planted-battery}). Across the 160 graph realizations at the setting used on \corpus{Enron}, the unweighted configuration was more accurate on 124, and on 49 of those \(\Omega\) favoured the weighted configuration instead. Larger differences in \(\Omega\) favoured the configuration with higher accuracy more often.

\input{tables/planted-battery}

A \emph{planted calibration} then tested whether a gain of the size seen on \corpus{Enron} indicated greater accuracy. Three criteria were applied to the comparisons on the planted suite. Each comparison set the weighted configuration, the constant configuration, and five permuted configurations against the unweighted one. A comparison was classified as \corpus{Enron}-like if the weighted gain was at least as large as \corpus{Enron}'s, the constant-weight gain was less than half the weighted gain, and the mean gain under permuted weights was negative. Let \(g(a) = \bar{\Omega}(a) - \bar{\Omega}(\text{unweighted})\) denote configuration \(a\)'s stability gain over the unweighted configuration, and let \(g_{\text{Enron}}\) denote the measured \corpus{Enron} gain. The three criteria were

\vspace{5pt}
\begin{equation}\label{eq:enron-criteria}
  g(\text{weighted}) \ge g_{\text{Enron}}, \qquad g(\text{constant}) < \tfrac{1}{2}\,g(\text{weighted}), \qquad g(\text{permuted}) < 0,
\end{equation}
\vspace{5pt}

\noindent where \(g(\text{permuted})\) is the mean gain over the five permutations.

On \corpus{Enron} itself, the constant and permuted configurations served as controls, and they suggested that the stability gain was not simply a by-product of reduced overlap. The weighted configuration placed 2.57 percent of nodes in more than one community, compared with 6.59 percent for the unweighted one. The constant configuration reduced overlap almost as much, to 3.33 percent, but produced no stability gain (difference 0.000, interval −0.053 to 0.049). Each of the five weight permutations lowered stability relative to the unweighted configuration, by 0.039 to 0.091.\footnote{The fall was resolved under four of the five permutations. At OSLOM's default depth of 10/50, the gain under the graph's own weights was smaller, at 0.059 (0.037 to 0.080), and constant and permuted weights alike made the covers less stable than they were without weights, with every difference resolved. The planted calibration takes the gain at depth 3/3 as its threshold, and it was not repeated with the smaller gain at the default depth.} \corpus{Enron}'s own comparison therefore met the second and third criteria, since constant weights contributed less than half of the gain and permuting the graph's weights removed it.

Across OSLOM's settings on the planted suite, a stability difference at least as large as \corpus{Enron}'s, in either direction, pointed to the more accurate configuration in 420 of 425 comparisons, almost all of which favoured the unweighted configuration. When each graph carried its own weights, a gain of that size for the weighted configuration was rare, occurring in only 14 of the 4,234 graph--setting combinations whose two configurations differed in accuracy. In nine of those 14, the weighted configuration was the more accurate, against 12.6 percent of all 4,234. Similarly, when the graphs carried \corpus{Enron}'s weights, a gain of that size occurred in 20 of 14,390 such combinations, and in 11 of them the weighted configuration was the more accurate (Table~\ref{tab:weight-families}). When the graphs carried \corpus{Advogato}'s three trust levels, which barely distinguished the planted communities, gains of this size occurred in 1,054 of 20,971 such combinations, and the weighted configuration was the more accurate in 33.

\input{tables/weight-families}

Applying the second criterion reduced these to 219 comparisons across the three sources of weights. Of these, all but seven involved \corpus{Advogato}'s weights, and the weighted configuration was more accurate in only three of the 219. Of the 219, 84 met all three criteria. All 84 involved \corpus{Advogato}'s weights, and the weighted configuration was less accurate in every case.\footnote{The result did not depend on the form of the criteria. Three other versions of the second criterion were applied: a constant-weight gain below 0.03, no gain from constant weights, and a weighted gain exceeding the constant-weight gain by at least \corpus{Enron}'s gain. Under each, between 105 and 187 comparisons met the first two criteria, and the weighted configuration was the more accurate in at most two. When each of the five permutations, rather than their mean, was required to lower stability, 28 comparisons met all three criteria, and the weighted configuration was the more accurate in none.}

On \corpus{Enron}, stability favoured the weighted configuration, but the planted calibration did not support interpreting that gain as evidence of greater accuracy. In every planted comparison classified as \corpus{Enron}-like, the weighted configuration was less accurate, and each carried \corpus{Advogato}'s weights. Without a reference cover, the accuracy ranking on \corpus{Enron} remains undetermined.

\subsection{The Reporting Rule}
\label{sec:reporting}

The comparisons of the two preceding sections, on planted graphs and on \corpus{Enron}, both show why stability should not be reported alone. Higher agreement can accompany degenerate covers, and without a reference it cannot establish accuracy. The \emph{reporting rule} proposed here therefore has two parts. A stability figure is reported with the index under which it was computed, and it is preceded by four measures of the shape of the covers it describes. The four are the number of communities above the size floor, the share of very small communities, the fraction of covered nodes with more than one membership, and the share of nodes that the cover places. Together, they help distinguish agreement about recovered structure from agreement about a degenerate cover. With \(m_v\) and \(V_{\mathcal{K}}\) as defined in Section~\ref{sec:notation}, the measures are

\vspace{5pt}
\begin{equation}\label{eq:shape}
  \begin{aligned}
    k_{\ge 20} &= \bigl|\{C \in \mathcal{K} : |C| \ge 20\}\bigr|, & d &= \frac{\bigl|\{C \in \mathcal{K} : |C| \le 5\}\bigr|}{|\mathcal{K}|}, \\[1.3em]
    o &= \frac{\bigl|\{v \in V_{\mathcal{K}} : |m_v| > 1\}\bigr|}{|V_{\mathcal{K}}|}, & \phi &= \frac{|V_{\mathcal{K}}|}{n}.
  \end{aligned}
\end{equation}
\vspace{5pt}

\begin{wrapfigure}{R}{0pt}
  \includegraphics[width=\floatcolumn,alt={Scatter plot of stability, measured by the Omega index, against the fraction of communities with five or fewer members, one point per measured method configuration, with a vertical bar for each point's interval. Colour distinguishes the methods and marker shape the datasets, and a light segment joins the two weight configurations of each method on each dataset. The points with the highest stability are those with the largest shares of very small communities, while most segments are short or cut across that slope.}]{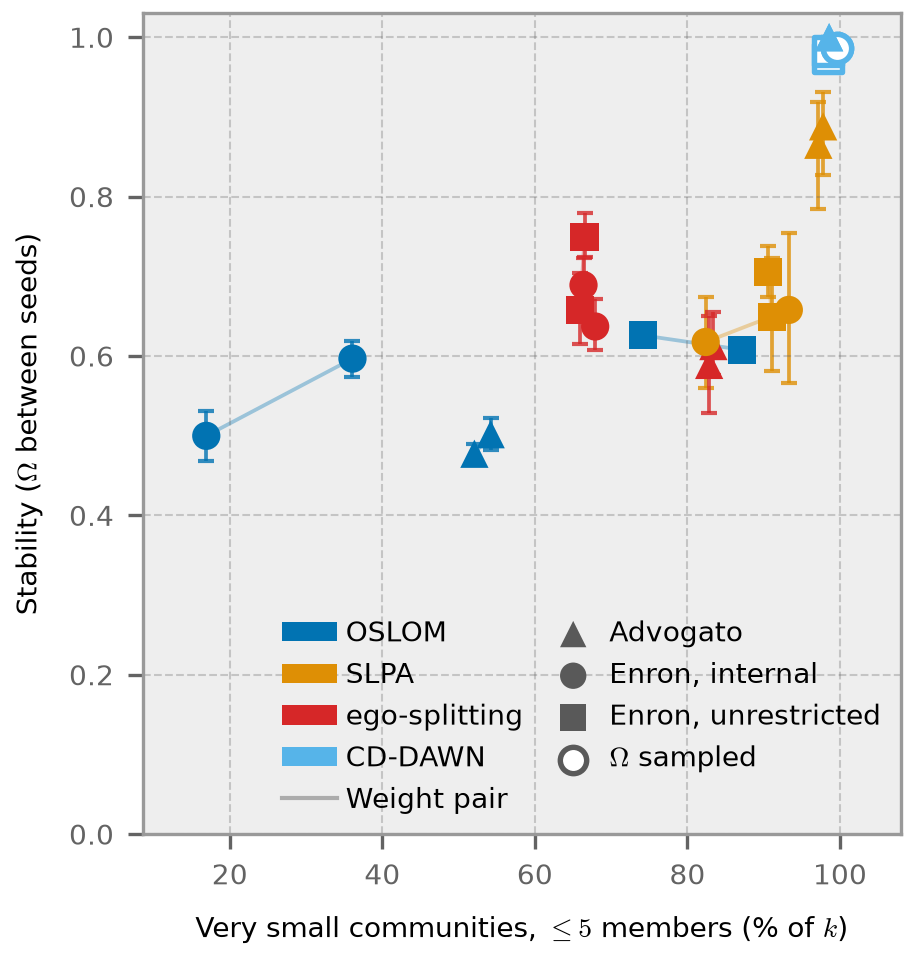}
  \caption{Stability (Ω) against the share of very small communities, those with five or fewer members, one point per measured configuration. Colour marks the method and marker shape marks the dataset. Each bar is the 95 percent interval from a bootstrap over the 20 seeds, and each light segment joins the two weight configurations of one method on one dataset.}
  \label{fig:stability-debris}
\end{wrapfigure}

\noindent The size floor of 20 members and the bound of five members on a very small community are those of Section~\ref{sec:notation}.\footnote{The count above the floor is reported in place of the total, which includes the very small communities, so that a cover can have more communities in all and fewer above the floor. Across the configurations of Table~\ref{tab:shape-drivers}, stability showed no association with the total count per 1,000 nodes (\(\rho = -0.05\)) but fell as the count above the floor per 1,000 nodes rose (\(\rho = -0.83\)).} The overlap fraction is calculated over nodes placed by the cover, not over all of \(V\), so a method is not credited with less overlap simply because it leaves nodes unassigned. The share placed, \(\phi\), states how much of the graph a stability figure describes. Ego-splitting placed between 27.4 and 67.6 percent of the nodes of the graphs in Table~\ref{tab:cover-shape}, and its stability there describes those nodes only. None of these measures requires a reference cover, and all are cheaper to compute than the stability figure they qualify. Figure~\ref{fig:stability-debris} and Table~\ref{tab:shape-drivers} show the association that motivates the rule.

\input{tables/shape-drivers}

The association was measured over 24 method configurations, each combining a method, a graph, and a choice of edge weights, with stability computed under both \(\Omega\) and \(\text{NMI}_{\max}\). Across the 24, the share of very small communities was positively associated with stability, with a rank correlation of 0.833 under \(\Omega\) and 0.769 under \(\text{NMI}_{\max}\). The association was slightly stronger under the chance-corrected \(\Omega\), so it was not specific to the reward that \(\text{NMI}_{\max}\) gives such communities when both covers share them, which is shown below.

The association reflected differences amongst methods and graphs rather than differences between the weighted and unweighted configurations within a method--graph pair. CD-DAWN illustrated the pattern, since its configurations were both the most stable and the most fragmented. On \corpus{Advogato}, each returned the same cover from every seed, so its agreement was exactly one. A permutation test that exchanged only the two weight configurations within each method--graph pair gave \(p = 0.33\), providing no evidence that these within-pair differences explained the overall association. When OSLOM and SLPA were rerun on the planted families at their real-graph settings and the results pooled (Section~\ref{sec:stability-planted}), larger shares of very small communities were associated with lower stability where detection succeeded (\(\mu \le 0.5\)). A positive association appeared only on the block model above that boundary (Table~\ref{tab:realdata-planted}).

The other part of the rule, naming the index, is needed because the two indices can rank the same results differently. Of the 276 pairs that the 24 configurations form, 27 were ranked in opposite orders by \(\Omega\) and \(\text{NMI}_{\max}\). Two of the 27 reversed pairs compared weighted and unweighted configurations of the same method on the same graph. This paper draws only this type of comparison from those figures. A bootstrap over the shared seeds resolved both differences in one of those pairs.\footnote{On the unrestricted \corpus{Enron} graph, \(\Omega\) preferred OSLOM's unweighted configuration. \(\text{NMI}_{\max}\) preferred the weighted configuration. In the weighted configuration, 87.2 percent of communities were very small, with five or fewer members, compared with 74.1 percent in the unweighted configuration.} The index must therefore be named whenever a stability figure is reported.

\(\text{NMI}_{\max}\) can prefer a cover for its very small communities, a preference that resembles the documented bias of normalized mutual information towards outputs with more communities (Section~\ref{sec:related-accuracy}). Its mechanism differs, however. The preference arises from the index definitions, regardless of which method produced the covers. To isolate this effect, a sequence of covers was constructed with fixed disagreement and an increasing number of very small communities shared by both covers. Across the nine partition graphs, the proportion of very small communities shared by both covers rose from none to 99.4 percent. Over that range, at the largest of the three levels of retained disagreement, \(\Omega\) fell from 0.8122 to 0.2065, while \(\text{NMI}_{\max}\) \emph{rose} from 0.7656 to 0.8669. The indices moved in opposite directions on all nine graphs at each of the three levels. A shared singleton is predicted perfectly by both covers, so it adds mutual information and raises \(\text{NMI}_{\max}\). Under \(\Omega\), the same singleton increases the share of node pairs with zero shared communities in both covers. The agreement expected by chance therefore rises, and the chance correction removes the corresponding observed gain. This effect matters for stability because the same method on the same graph tends to return the same singletons across seeds. When the reference cover does not share them, a cover containing very small communities scores lower under both indices, with a steeper decline under \(\text{NMI}_{\max}\). Both indices reached zero on all nine graphs.

\input{tables/collapse-rules}

The choice of index matters further when a partition index is used. Stability was also compared under the Adjusted Rand index, which is applied only after each cover has been collapsed to a partition. Table~\ref{tab:collapse-rules} compares stability under that index and \(\Omega\) for the same LFR covers collapsed by four rules. Where the covers were effectively partitions, all four rules agreed on average, and the mean difference between the indices was at most 0.0009. Where more than a tenth of nodes had multiple memberships, the relative scores depended on the collapse rule. The first-listed rule placed SLPA's Adjusted Rand stability score 0.0742 below \(\Omega\). The rule that retains the community with most of each node's edges placed that score 0.0876 above \(\Omega\). In this high-overlap band, changing the collapse rule reversed the sign of SLPA's difference between the two indices, while OSLOM's stayed positive under every rule.

The degenerate covers that the rule is meant to expose also made \(\Omega\) itself hard to compute. Ten of the 24 \(\Omega\) values could not be computed by enumeration, because the memory it needs grows quadratically with community size. The largest communities occurred in some of the most degenerate covers. In each of SLPA's four \corpus{Enron} configurations, more than 80 percent of communities were very small, while one community contained between 24.7 and 77.9 percent of the nodes. Enumerating these covers would have required 27 to 352 gigabytes. Six of the ten values were instead computed exactly from node membership counts by Equations~\eqref{eq:omega-moments} and~\eqref{eq:omega-inversion} of Section~\ref{sec:accuracy}, at a cost that does not depend on community size. These were the four SLPA configurations just described and ego-splitting's two on the unrestricted graph.

The remaining four values came from CD-DAWN's configurations on the two \corpus{Enron} graphs. The membership-count calculation was impractical for these covers because its cost grows exponentially with the number of memberships, and in the first-seed covers one node belonged to as many as 110 communities. Their \(\Omega\) values were therefore estimated from 2,000,000 sampled node pairs per seed pair, and a dagger in Table~\ref{tab:cover-shape} marks these estimates.

The order of reporting matters as well as its content. The four shape measures should precede the stability figure they qualify, not follow it as auxiliary detail, because they indicate whether the agreement reported reflects recovered structure or degenerate covers. The next subsection applies the rule to a procedure designed to raise agreement.

\subsection{Consensus Clustering}
\label{sec:consensus}

Consensus clustering is a procedure for obtaining stable results from a stochastic method \autocite[1]{lancichinettiConsensusClusteringComplex2012}. It therefore offers a direct test of the reporting rule, since the agreement it adds can be read beside the shape of the covers it returns. The procedure aggregates repeated runs by recording how often each node pair shares a community, thresholds the resulting co-association matrix at \(\tau\), and runs the method again on the thresholded graph \autocite[5--6]{lancichinettiConsensusClusteringComplex2012}. For \(T\) runs yielding covers \(\mathcal{K}_1, \ldots, \mathcal{K}_T\), the co-association frequency is

\vspace{-4pt}
\begin{equation}\label{eq:coassociation}
  D_{uv} = \frac{1}{T} \bigl|\{ t : u \in C \text{ and } v \in C \text{ for some } C \in \mathcal{K}_t \}\bigr|.
\end{equation}
\vspace{-5pt}

\noindent Entries below \(\tau\) are removed. A node left without edges is reconnected to its highest-weight partner, as in the published procedure. \citeauthor{lancichinettiConsensusClusteringComplex2012} defined the procedure for partitions, in which each node pair is either together or apart. For covers, it was adapted to count each pair once per run in which the nodes share at least one community, so that \(D\) remains a frequency in \([0,1]\), not a count of shared memberships. The input graphs are directed and the co-association graph is undirected, so OSLOM's directed binary was used for the base runs and its undirected binary for the co-association graph, in which each edge was listed once.

\input{tables/consensus}

One execution of the procedure, from its own seeds, is termed a \emph{stream}. Within a stream, the procedure was iterated at most three times, and convergence required every run to return exactly the same cover. If a stream did not converge, its reported cover was the most representative run of the final iteration, the one with the greatest co-association mass in common with the other final runs. This score sums the co-association weights of node pairs placed together in the candidate cover, so merging communities cannot lower it. Because the original sweep did not record convergence, the \corpus{Advogato} streams at eight to 128 runs were repeated with every ensemble retained. Six of the 80 streams, counted at every threshold and number of runs, converged, all at the third iteration and at \(\tau = 0.7\). The selection rule had little effect on these ensembles. Choosing the most representative final run rather than the first or a uniformly random run changed cross-stream agreement by at most 0.006 and detected overlap by 0.2 percentage points. The comparisons below use the most representative run.

The procedure did what it is designed to do. For OSLOM on the \corpus{Advogato} trust network, agreement amongst eight consensus streams increased with the number of input runs, from 0.637 to 0.857 at \(\tau = 0.5\) and from 0.696 to 0.931 at \(\tau = 0.7\) (Table~\ref{tab:consensus}).

Three qualifications limit what can be concluded from these curves. First, the sweep ended at 1,024 runs and did not show where the gains stopped. Each stream kept its seeds at every run count, so a doubling was judged on the paired difference between its two points, with a 95 percent interval from resampling the streams (Section~\ref{sec:matched}). At \(\tau = 0.7\), every doubling was resolved, including the last, to 1,024 runs, at 0.006 (0.001 to 0.011), so the gains had not stopped where the sweep ended. For \(\tau = 0.5\), every doubling through 512 runs was resolved, and the final point was lower than the preceding one by a difference that was not resolved (−0.015, −0.032 to 0.001). The curves therefore provided no stopping rule.

Second, the intervals in Table~\ref{tab:consensus} belong to single points, and read pair by pair they understate what the paired design resolved. The intervals of three of the seven adjacent pairs of points at the lower threshold overlapped, although only one paired difference was unresolved.\footnote{Eight streams yielded 28 pairwise comparisons per setting. Because the comparisons shared streams and were not independent, the intervals in Table~\ref{tab:consensus} were obtained by resampling the streams themselves. They were between 1.6 and 2.3 times as wide as Student's \(t\) intervals over the 28 comparisons, which treated the comparisons as independent and separated every doubling at both thresholds, including the fall at the top of the lower-threshold curve. By the overlap of its intervals, a leave-one-stream-out jackknife with seven degrees of freedom resolved four of the seven doublings at \(\tau = 0.7\) and three at \(\tau = 0.5\). At 1,024 runs, adjacent streams shared 24 of their base-run seeds, so the streams were not fully independent at the top of the curve.}

Third, the increase in agreement was accompanied by a loss of detected overlap. At \(\tau = 0.5\), the share of nodes with multiple memberships fell at six of the seven doublings, from 8.90 percent to 5.69 percent at 1,024 runs. Every point was also below the 9.18 percent that single runs of the same method at the same settings averaged over 20 seeds.\footnote{The single-run figure was measured at the settings of each base run of the sweep, with a significance threshold of 0.1, a search depth of 3/3, and the graph's own edge weights withheld. A base run of the consensus procedure and a single run were therefore the same computation. The only weights used were those of the co-association matrix, and only in the iterations.} The rise in agreement therefore coincided with fewer of the multiple memberships that OSLOM was designed to detect. At every run count, detected overlap was lower at \(\tau = 0.7\) than at \(\tau = 0.5\), and after the first doubling it stayed between 3.25 and 3.43 percent. The higher threshold also left between 6.7 and 7.9 percent of nodes unassigned, compared with 1.9 percent at \(\tau = 0.5\). Because agreement was computed over all nodes, two streams that left the same nodes unassigned also agreed on those omissions.

\input{tables/consensus-planted}

Accuracy could not be measured on the real graphs because they have no reference cover, so it was assessed on three planted LFR graphs with 30 percent overlap at \(\mu = 0.3\), \(0.5\), and \(0.6\). The same design was used, with eight streams, each swept from eight to 1,024 input runs at both thresholds (Table~\ref{tab:consensus-planted}). Agreement increased on every graph at both thresholds, by between 0.046 and 0.155, while accuracy stayed within 0.011 of its value at eight runs. At \(\tau = 0.5\), detected overlap fell from between 5.75 and 6.46 percent at eight runs to between 4.40 and 4.66 percent at 1,024, far below the 30 percent planted. Cut off at three iterations, consensus clustering thus increased agreement while detected overlap fell and accuracy remained effectively unchanged.

\input{tables/consensus-transfer}

Table~\ref{tab:consensus-transfer} compares agreement on \corpus{Advogato}, the 5,155-node developer trust network, and \corpus{Enron}, the 21,510-node corporate email network. At both thresholds, agreement rose on both graphs at every doubling up to 256 runs. With eight runs, the agreement values of the two graphs differed by 0.039 at \(\tau = 0.5\) and 0.008 at \(\tau = 0.7\), and with 256 runs by 0.024 and 0.032, respectively.\footnote{Reported differences are computed from unrounded values and rounded once. Subtracting the displayed values can therefore produce a result that differs in the last decimal place.} On \corpus{Enron} at \(\tau = 0.5\), the share of nodes with multiple memberships also fell at every doubling, from 7.56 to 4.91 percent at 256 runs. These results showed that rising agreement was not unique to \corpus{Advogato}, but two graphs did not establish that it was a general property of consensus clustering. On both, the overlap fraction of the reporting rule showed what the agreement figure alone did not, that the added agreement came with less detected overlap.

\section{Accuracy and the Selection Protocol}
\label{sec:comparison}

Accuracy against a reference cover is the third criterion by which overlapping methods are evaluated, and the one against which the other two were judged in Sections~\ref{sec:peak} and~\ref{sec:stability}. It is not free of choices of its own. A comparison of methods by accuracy depends on an index and on a selection protocol, the rule that reduces a method's many runs to one score per graph. This section tests the protocol, on a comparison of the four methods of Section~\ref{sec:methods} with the partition baseline on planted graphs, and Section~\ref{sec:discussion-index} takes up the index. Section~\ref{sec:matched} defines the protocols and measures how far the choice amongst them moves each method's score on the same runs. Section~\ref{sec:comparison-results} reports the comparison under the protocol used throughout this paper, and then shows which of its conclusions the protocol decides. Section~\ref{sec:competence} examines the covers behind the scores where detection failed. The companion paper reports the comparison in full, over seven methods \autocite{pizzornoCoversPartitionsOverlapping2026}.

Accuracy is measured against each planted cover with a chance-corrected index, so no internal quality score is needed. For graphs without a reference cover, rankings on planted graphs may inform expectations about the choice of method, but they cannot establish accuracy on real data, since the benchmark scores themselves do not transfer.

\subsection{Selection Protocols and Paired Comparisons}
\label{sec:matched}

Each method can return multiple covers for a graph, across parameter settings and random seeds. To compare methods, those runs must be reduced to one score per graph, and the rule by which they are reduced is the \emph{selection protocol}. The \emph{matched protocol} used throughout this paper first averages accuracy over seeds within each setting and then selects the highest of the setting-level means. The \emph{unmatched protocol} instead takes the best individual run, with settings and seeds pooled. For a graph, let \(\Omega_{s,r}\) denote the accuracy of the cover returned at setting \(s\) and seed \(r\), with \(R\) seeds at each setting. The matched and unmatched scores are

\vspace{-12pt}
\begin{equation}\label{eq:protocols}
  \text{matched} = \max_{s} \frac{1}{R} \sum_{r = 1}^{R} \Omega_{s,r}, \qquad \text{unmatched} = \max_{s,\, r} \Omega_{s,r}.
\end{equation}
\vspace{2pt}

\noindent The matched score gives each setting's seed-averaged value one chance to be selected, while the unmatched score gives each run its own chance. Under the unmatched protocol, a method run at five seeds per setting therefore has five times as many chances to produce a high score as a method run at a single seed.

\input{tables/selection-protocol}

Table~\ref{tab:selection-protocol} compares the protocols on the main suite (Section~\ref{sec:data}), on which every method was run at five seeds per setting. The matched and unmatched columns use identical runs and differ only in the selection rule, so the difference between them isolates the effect of selection, not of detection. Every method gained from the unmatched protocol, by between 0.0051 and 0.0132 for four of them and by 0.0487 for SLPA. These gains were large enough to matter to the comparison. The largest gain exceeded OSLOM's margin over the partition baseline under every protocol in the table (Section~\ref{sec:comparison-results}). Under the unmatched protocol, the deficit for SLPA relative to the partition baseline narrowed from 0.0835 to 0.0446. Such gains reflect selection bias in model evaluation. In machine learning, the effects of model selection have been described as \enquote{often of comparable magnitude to differences in performance between learning algorithms} \autocite[2079]{cawleyOverfittingModelSelection2010}. Because this bias varied by method, no common correction was appropriate.

The matched protocol removes only part of the bias. Averaging over seeds avoids selecting the best run, but the maximum over settings still uses the planted cover. The best setting for each graph was chosen using that graph's planted cover, which makes the matched score an oracle score, not one available without a reference. The practice is analogous to selecting hyperparameters on the test data, a protocol that \enquote{exhibits a consistent optimistic bias, potentially of a very substantial magnitude even when the number of hyper-parameters is small} \autocite[2103]{cawleyOverfittingModelSelection2010}. The remaining columns of Table~\ref{tab:selection-protocol} report alternatives that do not use each graph's planted cover to select its setting: the \emph{held-out} setting, selected by average accuracy on graphs at the suite's other parameter sets; each method's documented \emph{default}; and the \emph{median} setting. Held-out selection cost the partition baseline 0.0059 on average and OSLOM 0.0134. Each graph's best setting was chosen from 104 settings for OSLOM and from nine for the baseline, so selection on the scored graph had more settings to choose amongst for OSLOM.

A maximum over settings also depends on the grid of settings offered. The parameter grids were therefore extended to check whether each contained its maximum, and the continuation runs were pooled with the original results. For OSLOM, both parameters, the significance threshold and the coverage, were extended at both ends, and those runs contributed to every score reported in this section for the 1,000-node suites. For two of the methods, an edge of the grid still limited the best score, where the final step of the grid raised it by more than 0.01: CD-DAWN's on 76 of the 360 graphs, mostly at its lowest overlap threshold, and OSLOM's on eight, on six of which the edge was a coverage of 1.0, beyond which the parameter cannot be extended \autocite{pizzornoCoversPartitionsOverlapping2026}. The reported scores of CD-DAWN on those graphs, and of OSLOM on the other two, were therefore lower bounds. For every other method, the final step raised the score by more than 0.01 on at most one graph. No comparison in this paper placed CD-DAWN's bound within reach of another method.

The scale of a grid matters as well as its extent. A resolution grid designed for 1,000-node graphs is poorly scaled for larger ones. On the graphs of the 50,000-node suite with planted overlap at \(\mu \le 0.5\), ego-splitting scored 0.1961 at its documented default, whose resolution is the unscaled value, and 0.5507 at its best setting on the unscaled grid. At its best setting on a grid scaled by graph size, it scored 0.9006, enough to lead OSLOM there under held-out selection as well \autocite{pizzornoCoversPartitionsOverlapping2026}. OSLOM scored 0.8476 at its default and 0.8515 at its best setting there, on its original grid of 15 settings, which was not extended at this size. A resolution left unscaled can therefore decide which method wins. All other comparisons against planted covers used 1,000-node graphs, for which the grids were designed.

The scores that a protocol yields were then compared in pairs. Differences between methods, and between two configurations of one method, were tested on paired scores. When two configurations are evaluated on the same graphs, the uncertainty of their difference depends on whether it is calculated graph by graph or from separately aggregated means. The point estimate is the same either way, but the interval can differ substantially. This is why comparisons across multiple datasets use paired tests \autocite[6--7]{demsarStatisticalComparisonsClassifiers2006}. Across 120 settings on the five \corpus{wiki-topcats} subsamples, where two configurations of one method were evaluated on identical graphs, the paired interval on the difference had a median half-width of 0.027 against 0.229 when the configurations were treated as independent.\footnote{The subsamples come from different regions of the graph, and the share of their nodes in more than one reference community ranges from 26 to 66 percent. Scores vary with that share, so the spread between subsamples is several times the effect being compared. Pairing cancels the spread, because both configurations are scored on the same subsample, whereas an unpaired interval inherits all of it. The unpaired figure is the root of the sum of the squared intervals of the two configurations, which treats configurations that share a graph as independent.} The interval width, and therefore whether a difference of a given size can be distinguished from zero, depends on this choice. \Textcite[3--4]{demsarStatisticalComparisonsClassifiers2006} found that comparisons of algorithms over several datasets frequently included no statistical procedure or did not describe the one used. Accordingly, every comparison of two configurations in this paper was paired graph by graph, as the captions of the tables that report such differences note. Each parameter set of the main suite had ten realizations, scored separately. Margins were tested graph by graph with the two-sided Wilcoxon signed-rank test at the five percent level \autocite[7]{demsarStatisticalComparisonsClassifiers2006}.\footnote{The companion paper corrects such tests for multiple comparisons by Holm's step-down procedure, and every margin reported as significant in this paper remains significant under that correction \autocite{pizzornoCoversPartitionsOverlapping2026}.}

\subsection{Method Ranking and OSLOM's Margin}
\label{sec:comparison-results}

Of the four methods compared in this paper, only OSLOM scored significantly above the partition baseline across the suite under the matched protocol, by 0.0209, and its margin was positive in each of the ten realizations. The other three scored below the baseline under every protocol of Table~\ref{tab:selection-protocol}. OSLOM's margin combined three regimes. At \(\mu \le 0.4\), OSLOM led the partition baseline on every graph on which the two differed, by 0.0755 on average over all graphs at those levels. The baseline led at \(\mu = 0.6\) and \(0.7\), by 0.0656. From \(\mu = 0.75\), every method in the comparison scored near zero, below 0.06 on average, and there OSLOM kept a significant lead of 0.0064 over the baseline. The whole-suite margin described none of the three well.

\label{sec:where}
On the graphs with planted overlap at \(\mu \le 0.5\), OSLOM exceeded the partition baseline by 0.0817. Under the selection protocols of Table~\ref{tab:selection-protocol}, its margin ranged from 0.0807 to 0.1345 (\(p < 0.001\)). The margin opened only where overlap was planted, rising to 0.0533 at ten percent planted overlap and 0.1088 at 30 percent, then falling to 0.0831 at 50 percent. This downturn was specific to \(\Omega\), and Section~\ref{sec:discussion-index} traces it to the chance correction. When OSLOM's covers were collapsed to one membership per node, they scored below the partition baseline at every overlap level, so under \(\Omega\) the advantage came from the multiple memberships OSLOM recovered \autocite{pizzornoCoversPartitionsOverlapping2026}.

\label{sec:generalisation}
The margins above were measured on LFR graphs in which each overlapping node belongs to exactly two communities. The results could therefore depend on either the generator or the membership cap. Two controls tested these possibilities. The first replaced LFR with the directed, overlapping, degree-corrected block model of Section~\ref{sec:data}, while holding the parameter grid fixed. The second retained LFR but raised the membership cap from two to three, then to four and five.

The two controls gave different results. On the graphs with planted overlap at \(\mu \le 0.5\), as in Section~\ref{sec:where}, the margin under the block model widened as overlap increased, reaching 0.1955 at 50 percent planted overlap and remaining positive on all 50 graphs. OSLOM's collapsed covers also outperformed the baseline on all 50 graphs, so its advantage on this generator was not limited to recovering multiple memberships. With three memberships per overlapping node, the margin nearly disappeared at 50 percent overlap, and with four and five memberships it reversed there, while at ten percent overlap it remained positive but declined from 0.0533 with two memberships to 0.0227 with five \autocite{pizzornoCoversPartitionsOverlapping2026}. Under \(\Omega\), changing the generator did not remove the margin, but increasing the membership cap weakened and eventually reversed it, as OSLOM recovered fewer of the additional memberships \autocite{pizzornoCoversPartitionsOverlapping2026}. Section~\ref{sec:discussion-index} examines how far the latter pattern was specific to \(\Omega\).

The choice of protocol changed some of these conclusions but not all. OSLOM's advantage on graphs with planted overlap at \(\mu \le 0.5\) held under every selection protocol, as reported above. The whole-suite comparison was less robust. The mixing parameter at which OSLOM fell below the partition baseline moved from 0.6 to 0.5 under held-out selection, stayed at 0.6 under the defaults, and moved to 0.8 at the median setting. On the whole suite, OSLOM led the partition baseline significantly at its best setting on each graph, at the median setting, and at the defaults, where its margin was 0.0368. Under held-out selection, its margin of 0.0134 was not significant, and OSLOM scored above the baseline on only 138 of the 342 graphs on which the two differed, significantly fewer than half under a two-sided binomial test. On the suite with three memberships per overlapping node, held-out selection placed OSLOM significantly below the baseline, its best setting and its default did not separate the two, and the median setting placed it significantly above. Relative to each graph's best setting, the defaults and the median setting moved the comparison in OSLOM's favour, because a change of resolution affected the partition baseline's accuracy more than a change of settings affected OSLOM's. Held-out selection moved it the other way, because choosing the setting on each graph was worth more to OSLOM than to the baseline. On the same runs, the protocol could thus decide whether an overlapping method appeared to lead the partition baseline.

\subsection{Failure Modes and Their Diagnostics}
\label{sec:competence}

A score against the planted cover says how accurate a cover is, not how it fails. This subsection examines the covers behind the scores at high mixing, where detection failed. Under the matched protocol, OSLOM's accuracy crossed below the partition baseline's at \(\mu = 0.6\) (Section~\ref{sec:comparison-results}). Beyond that crossing, OSLOM no longer outperformed the partition baseline, apart from its lead of 0.0064 from \(\mu = 0.75\) (Section~\ref{sec:comparison-results}). At \(\mu = 0.8\), both methods scored near zero against the planted cover, at 0.0223 for OSLOM and 0.0197 for the baseline. Accuracy alone therefore made them appear equally unsuccessful, but their covers differed substantially.

On those graphs, the partition baseline returned between 11 and 71 communities at the resolution selected by the matched protocol, where between 16 and 30 were planted. Its covers were inaccurate, yet nothing in their size or shape gave a sign of failure. Modularity maximization can produce answers of this kind even when no structure is present, since \enquote{random graphs and scale-free networks have high modularity due to fluctuations in the establishment of links} \autocite[{4}]{guimeraModularityFluctuationsRandom2004}. OSLOM, at the setting selected for it, which was mostly its highest significance threshold and coverage, returned a median of 127 communities on a 1,000-node graph. That was more than four times as many as were planted, and almost none of them were very small. At its lowest thresholds, by contrast, it found no community that cleared its significance test and returned the all-singleton partition in most cases, reporting almost no structure.

These outputs show why the diagnostics in the reporting rule must be read together. The partition baseline's cover scored 0.0197 against the planted cover, yet neither the shape diagnostics nor any quality measure would expose its failure. Stability did. At the resolution selected by the matched protocol, the baseline's covers agreed at 0.073 on average at \(\mu = 0.8\), compared with at least 0.908 at every \(\mu \le 0.5\). OSLOM's selected covers were exposed in the same way, agreeing at 0.211. At its lowest thresholds, by contrast, OSLOM's share of very small communities exposed its covers, despite near-perfect stability (Section~\ref{sec:stability-planted}). Stability alone would make those covers appear reliable, while the shape diagnostics alone would not flag the partition baseline's cover. Read together, the diagnostics exposed each of these three failures. Of the three, OSLOM's fragmented output at its lowest thresholds was the easiest to expose, because the failure was apparent in a single run.

The diagnostics did not expose every failure on this suite. At \(\mu = 0.8\), averaged over its settings, SLPA scored 0.013 against the planted cover, yet its covers agreed across seeds at 0.525, against 0.444 at \(\mu = 0.4\), where it scored 0.527. They had 18 communities above the size floor on average, within the planted range, and no very small community, which its size shaping forbids (Figure~\ref{fig:stability-regime}).

\section{Reference-Free Diagnostics}
\label{sec:reference-free}

Sections~\ref{sec:peak} and~\ref{sec:stability} tested the quality measures and stability one at a time. On a real graph without a reference cover, where accuracy cannot be measured, they are read together, and their agreement has been taken to strengthen the claim that a detected structure is good \autocite[26]{chakrabortyMetricsCommunityAnalysis2017}. This section tests that reading. The \emph{reference-free diagnostics} it considers are four: the two quality measures defined for overlapping covers, \(Q_{ov}\) and \(Q_{dwo}\), and stability under each of its two indices, \(\Omega\) and \(\text{NMI}_{\max}\). All four are applied to one comparison, that of OSLOM run with and without a graph's edge weights (Section~\ref{sec:inversion}). Section~\ref{sec:weights} establishes which configuration was the more accurate on planted graphs and what the weights did to detected overlap on real graphs. Section~\ref{sec:battery} then compares the diagnostics' preferences with that accuracy and applies them to the two \corpus{Enron} graphs, where no reference exists.

\subsection{OSLOM With and Without Edge Weights}
\label{sec:weights}

\corpus{Enron} and \corpus{Advogato} are both weighted, and the ability to use edge weights is often treated as an advantage \autocites[4]{newmanAnalysisWeightedNetworks2004}[157]{fortunatoCommunityDetectionGraphs2010}[1]{lancichinettiConsensusClusteringComplex2012}[31]{xieOverlappingCommunityDetection2013}. The effect of supplying weights was tested for OSLOM on the weighted planted suite of Section~\ref{sec:data}, in a paired design that held the graphs, the parameter grid, and the seeds fixed and varied only whether the edge weights were supplied.

Supplying the weights lowered accuracy. The suite comprised 16 parameter sets, each with ten graph realizations, and every graph was run at 20 seeds. Withholding weights improved accuracy on 127 of the 160 realizations, by up to 0.0954, and reduced it on 12, by at most 0.0158. All of the latter had no more than ten percent planted overlap. The remaining 21 were ties, on graphs with no planted overlap at \(\mu \le 0.5\), where both configurations recovered the planted partition exactly. The accuracy gain from withholding weights increased to 0.0381 at 30 percent planted overlap and was 0.0340 at 50 percent. At both levels, the unweighted configuration was more accurate on every graph.

\input{tables/weight-arms}

The loss was not peculiar to OSLOM. Table~\ref{tab:weight-arms} compares the configurations with the partition baseline. Across the full suite, the unweighted configuration scored 0.7734, compared with 0.7506 for the weighted configuration. On the 90 overlapping realizations at \(\mu \le 0.5\), the unweighted configuration also outperformed the partition baseline (0.7477 against 0.7129), while the weighted configuration scored 0.7241. The partition baseline lost more from the same weights than OSLOM did, falling from 0.7688 to 0.6264 over the full suite. The mean contrast of the weights therefore does not measure how usable they are.

Nor was the loss explained by the weights carrying less structure than the topology. On every graph in the suite, the weights separated planted communities more clearly than the edges did (Section~\ref{sec:data}). Across the \(\mu = 0.6\) graphs, 61 to 64 percent of edges crossed community boundaries, while only 39 to 43 percent of the total weight crossed. Despite this stronger separation in the weights, the unweighted configuration scored 0.662, compared with 0.623 for the weighted configuration.

The loss of accuracy went with a loss of detected overlap. Both configurations substantially underdetected planted overlap. At 50 percent planted overlap, even the better configuration marked only nine percent of nodes as overlapping, averaged over its settings, and no method examined recovered the planted overlap in full (Section~\ref{sec:limits-overlap}). The weighted configuration detected less overlap than the unweighted one, and this gap widened as planted overlap increased.

The same suppression of overlap appeared more strongly on real data. On the internal \corpus{Enron} graph, 2.57 percent of nodes belonged to more than one weighted community, compared with 6.59 percent under the unweighted configuration. On the \corpus{Advogato} trust network, the corresponding shares were 4.19 and 9.18 percent (Table~\ref{tab:cover-shape}). Thus, the weighted covers were close to partitions despite the expected overlap in these networks.

One possible mechanism lies in OSLOM's significance test for weighted graphs. The test compares a node's attachment against a null based on the weight distribution, which has greater variance than the degree distribution, so a weaker second attachment, characteristic of an overlapping membership, might fail the significance threshold. Three controls challenged this account, two of them the constant and permuted configurations of Section~\ref{sec:inversion}. The first was \corpus{Advogato} itself, whose three declared trust levels make its strength distribution nearly a rescaled version of its degree distribution, yet under which detected overlap fell by more than half. The second assigned unit weight to every edge, under which OSLOM placed 4.37 percent of nodes in more than one community on \corpus{Advogato} and 3.33 percent on \corpus{Enron}, compared with 9.18 and 6.59 percent under the unweighted configuration. The third permuted the weights across edges, and the suppression persisted on both graphs. Taken together, the controls indicated that suppression arose mostly from OSLOM's weighted code path, not from the information encoded by the weights \autocite{pizzornoCoversPartitionsOverlapping2026}.

These results support a conditional recommendation for OSLOM, which is to withhold each graph's own weights by default, except when the topology has largely lost the structure carried by those weights. The exception was located on a second weighted suite, which varied \(\mu\) and the contrast between the mean weights of internal and external edges. There the weighted configuration was more accurate on average only within the region of \(\mu \ge 0.5\) and contrasts of at least four, and not everywhere in it, and at six of the eight combinations where it led, the partition baseline was more accurate than both of OSLOM's configurations \autocite{pizzornoCoversPartitionsOverlapping2026}.

The recommendation also held under two other kinds of weight. The same 16 parameter sets were rerun on the same ten graph realizations, with edge weights drawn from either \corpus{Enron}'s distribution or \corpus{Advogato}'s three trust levels and tilted towards the planted communities at several strengths (Table~\ref{tab:weight-families}). At the setting used on \corpus{Enron} (Section~\ref{sec:battery}), withholding weights was more accurate on between 158 and 160 of the 160 realizations at each \corpus{Enron}-weight tilt, and on between 123 and 127 with the \corpus{Advogato} weights.

Two qualifications limit this recommendation. First, when the weights alone carried the planted communities, OSLOM's weighted configuration recovered them only for large communities with two-valued weights at high contrast \autocite{pizzornoCoversPartitionsOverlapping2026}, and none of the weighted suites tested graphs in which the weights and topology encoded different planted covers.

Second, the recommendation depends on the role of the weights. In the consensus procedure of Section~\ref{sec:consensus}, edge weights are co-association frequencies, which record the accumulated evidence that the procedure is designed to retain. On \corpus{Advogato}, running the iterations without these weights reduced agreement by 0.026 to 0.044 at eight and 32 input runs under both thresholds, with each stream's base runs held fixed. All four reductions were distinguishable from zero.\footnote{The two conditions repeated the consensus procedure for the eight streams of Table~\ref{tab:consensus} from the same base runs, and differed only in whether the iterations used the co-association frequencies. Both were run on one machine, not the one used for Table~\ref{tab:consensus}, so their agreement levels are not compared with the table's. The intervals on the reduction come from resampling the streams, the same ones under both conditions. Agreement fell at both thresholds although the covers changed in opposite directions, becoming coarser without the weights at \(\tau = 0.5\) and finer at \(\tau = 0.7\).} The consensus procedure accordingly kept its co-association weights while the graph's own weights were withheld. The recommendation is thus conditional, not categorical. For OSLOM, a graph's own weights should be withheld unless its topology has largely lost the structure those weights carry, and where it has, a partition was mostly the more accurate choice.

\subsection{The Diagnostics With and Without a Reference Cover}
\label{sec:battery}

Section~\ref{sec:weights} established which configuration was the more accurate on each graph of the weighted planted suite, so each diagnostic's preference there can be compared with accuracy. A diagnostic is said to prefer the configuration that it scores higher. OSLOM was run with and without each graph's own edge weights at the setting used on \corpus{Enron} (Table~\ref{tab:planted-battery}). The quality measures were \(Q_{ov}\) and \(Q_{dwo}\), since the partition measures \(Q_{d}\) and \(Q_{lr}\) are not defined for overlapping covers. In both configurations, \(Q_{dwo}\) was scored on each graph's own weights, a choice fixed before scoring.

Of the 160 graph realizations, the unweighted configuration was more accurate on 124. On those, \(Q_{ov}\) and \(Q_{dwo}\) preferred the weighted configuration on 88 and 117, respectively, and stability under \(\Omega\) and under \(\text{NMI}_{\max}\) on 49 and 45. Scored on unit weights instead, \(Q_{dwo}\) preferred the weighted configuration on 65 of the 124. Amongst the 18 realizations where the weighted configuration was more accurate, \(Q_{ov}\) and \(Q_{dwo}\) preferred it on 12 and 17, respectively, while \(\Omega\) and \(\text{NMI}_{\max}\) did so on 12 and 13. Thus, \(Q_{ov}\) and \(Q_{dwo}\) often favoured the weighted configuration when it was less accurate, while \(\Omega\) and \(\text{NMI}_{\max}\) did so less often. Across the four indices, a majority favoured the weighted, less accurate configuration on 48 of the 124 realizations, and the unweighted configuration on 18, while the indices split two to two on 58. At the planted suites' depth of 10/50, and over all of the method's settings at either depth, the stability indices preferred the weighted configuration on between 30 and 40 percent of the realizations on which it was the less accurate, and the quality measures on between 68 and 97 percent.

Diagnostic preferences also varied with the source of the weights (Table~\ref{tab:weight-families}). With weights drawn from \corpus{Enron}'s distribution, \(Q_{dwo}\) favoured the weighted configuration in 154 to 160 of the realizations on which the unweighted configuration was more accurate, while \(\Omega\) did so in 18 to 28. With \corpus{Advogato}'s three trust levels, \(\Omega\) favoured the weighted configuration in 58 to 70 of the 123 to 127 realizations on which the unweighted configuration was more accurate, about half. Under \corpus{Advogato}'s weights, the preference of \(\Omega\) therefore did not reliably track accuracy.

For \corpus{Enron}, the planted results are evidence about the accuracy ranking but cannot confirm it, since there is no reference cover. Two of the findings might seem to transfer, and neither does. The controls traced overlap suppression to OSLOM's weighted code path, which also runs on real graphs, but planted graphs given \corpus{Enron}'s weights reproduced the real behaviour only in part \autocite{pizzornoCoversPartitionsOverlapping2026}, and the controls cannot establish which configuration was more accurate on \corpus{Enron}. The second weighted suite located the region in which the weighted configuration was more accurate on average, at \(\mu \ge 0.5\) and contrasts of at least four (Section~\ref{sec:weights}). That boundary cannot be applied to \corpus{Enron} either, since how clearly its weights separate its communities cannot be measured without a reference.

\input{tables/blind-battery}

On \corpus{Enron} itself, the diagnostics can be read but not checked. Table~\ref{tab:blind-battery} reports them for both \corpus{Enron} graphs. The internal graph receives closer attention because, on the unrestricted graph, most communities under either configuration were very small, which inflated \(\text{NMI}_{\max}\) (Section~\ref{sec:reporting}). On the internal graph, all four diagnostics preferred the weighted configuration.

\input{tables/enron-hubs}

The quality measures' preference for the weighted configuration is consistent with two mechanisms. The first is granularity (Section~\ref{sec:peak-results}), since the weighted covers were coarser, with 146 communities on average where the unweighted covers had 296, and 87 communities above the size floor where the unweighted covers had 222. The measures were therefore compared across covers differing by a factor of two in granularity and, as the merge ladder predicts, both rated the coarser cover higher. The second is specific to \(Q_{dwo}\), whose first factor falls as memberships per node rise and so penalizes overlap directly \autocite[17--18]{kumarOverlappingCommunityDetection2025}. Stability also favoured the weighted configuration on internal \corpus{Enron}, but neither coarseness nor overlap suppression explained that preference. On the planted suite, stability did not generally favour the coarser configuration and, on \corpus{Enron}, its advantage disappeared with constant weights and reversed with permuted weights (Section~\ref{sec:inversion}).\footnote{On the unrestricted \corpus{Enron} graph, the two stability indices gave opposite preferences on the same pair of covers (Section~\ref{sec:reporting}). \(\Omega\) favoured the unweighted configuration and \(\text{NMI}_{\max}\) the weighted one, whose cover had the larger share of very small communities. The difference is consistent with the mechanism described there, in which very small communities shared by two covers raise \(\text{NMI}_{\max}\) but not \(\Omega\). It is a reason to report the chance-corrected \(\Omega\) first.}

The two covers between which the diagnostics chose differed in substance as well as in score. Table~\ref{tab:enron-hubs} shows how each configuration placed the ten highest-degree nodes on internal \corpus{Enron}. Every configuration assigned all ten hubs to at least one community, but their groupings differed. In OSLOM's unweighted cover, the largest community holding each of the ten had between 127 and 702 members. All but two of the ten belonged to several communities, one to as many as six, and no more than three shared a community. In its weighted cover, five shared a single community of 1,303 members, and four of the ten held one membership each. The weighted cover thus grouped several of the graph's highest-degree nodes in the same large community, a consequential change for any interpretation of the graph's correspondence structure.

SLPA provides a more extreme example of a cover that the placement of the hubs exposes and stability does not. Without weights, its first-seed cover placed nine of the ten hubs in one community of 16,813 members, 78 percent of the graph, and eight of the nine held no other membership. This concentration made the cover's degeneracy apparent even without a reference. Yet SLPA's covers agreed across seeds at \(\Omega = 0.658\), exceeding the 0.500 to 0.626 range for OSLOM's configurations on the two \corpus{Enron} graphs (Table~\ref{tab:cover-shape}). Applying the published grouping and nesting rules to the same 20 seeds reproduced the concentration pattern. Nine of the ten hubs again shared a community of 16,811 members, and the covers agreed across seeds at \(\Omega = 0.664\).

The agreement of the diagnostics on \corpus{Enron} can now be weighed. Counted by index, the reference-free diagnostics favoured the weighted configuration in six of the eight comparisons, including all four on internal \corpus{Enron}, although on the internal graph the unweighted configuration they rejected detected 2.6 times as much overlap.\footnote{\(Q_{dwo}\) was also scored on both \corpus{Enron} graphs with unit weight on every edge. Scored in that way, it favoured the unweighted configuration on the unrestricted graph, and on the internal graph the difference between the configurations was not resolved.} The count is smaller when the indices are grouped by instrument. \(\Omega\) and \(\text{NMI}_{\max}\) are two readings of one instrument, stability, and the preference of \(Q_{dwo}\) depends on scoring the graph's own weights. Three instruments therefore favoured the weighted configuration on internal \corpus{Enron} with \(Q_{dwo}\) scored on those weights, and two did with unit weights, while the instruments split on unrestricted \corpus{Enron}. The granularity constraint of Section~\ref{sec:granularity-constraint} bars both quality measures because the two covers differ in granularity, so only stability remains, with one instrument preferring the weighted configuration on the internal graph, and a split on the unrestricted one. Amongst planted comparisons matching \corpus{Enron}'s weighted, constant, and permuted stability pattern, the weighted configuration was less accurate in every case (Section~\ref{sec:inversion}). What appeared to be the agreement of four diagnostics was thus the preference of one instrument, and the planted calibration did not support reading that preference as evidence of accuracy.

\section{Discussion}
\label{sec:discussion}

Sections~\ref{sec:peak} to~\ref{sec:reference-free} tested the three criteria in turn, and this section draws their results together. Two limitations of reference-free evaluation follow directly. The first belongs to the quality measures. Tests across cover granularities exposed failures in all four (Section~\ref{sec:peak}), although only \(Q_{ov}\) and \(Q_{dwo}\) are defined for overlapping covers. Because methods often return different numbers of communities, a mismatch of granularity is common when methods are compared, so this limitation affects most rankings of methods by quality score. The second belongs to run-to-run stability (Section~\ref{sec:stability}). It measures consistency, not correctness, and so cannot establish that detection succeeded without an independent reference cover. This limitation is narrower, but it is especially consequential for the directed, weighted graphs that the methods target, where reference covers are scarce \autocite[131]{malliarosClusteringCommunityDetection2013}. It matters there because stability is used precisely where no reference is available. It is a popular criterion for choosing amongst clusterings without ground truth \autocite[236]{luxburgClusteringStabilityOverview2010}, and it has been proposed as a measure of the significance of community structure \autocite[1]{karrerRobustnessCommunityStructure2008}.

Beyond these two limitations, the results support four further conclusions, each of which draws on more than one of the preceding sections. First, the two criteria can agree for reasons unrelated to accuracy (Section~\ref{sec:discussion-composition}). Second, a paired design cancels a shared limitation only when its effect is the same across the conditions being compared (Section~\ref{sec:discussion-shared}). Third, the choice amongst admissible accuracy indices can change the verdict (Section~\ref{sec:discussion-index}). Fourth, a gap in available datasets may help explain the adoption of these criteria without empirical validation (Section~\ref{sec:discussion-gap}).

\subsection{Agreement Between \texorpdfstring{\textit{Q}\tmsubx{\textit{ov}}}{Q\_ov} and Stability}
\label{sec:discussion-composition}

Section~\ref{sec:reference-free} found the quality measures and stability agreeing on internal \corpus{Enron}, and Sections~\ref{sec:peak} and~\ref{sec:stability} found each responding to the shape of a cover, \(Q_{ov}\) to its coarseness and stability to its degeneracy. If the two share a preference, their agreement follows from the shape of a cover, whatever its accuracy. The analysis that follows therefore tests whether \(Q_{ov}\) and stability share a preference for covers that resolve little of the graph. Such a cover is termed \emph{under-resolved}, and how much a cover resolves is measured by its \emph{effective community count}. Communities of size 20 to \(n/2\) are treated as resolved. Let \(\mathcal{K}_{\text{res}} = \{C \in \mathcal{K} : 20 \le |C| \le n/2\}\) denote those communities, and let \(n_0 = n - |\bigcup_{C \in \mathcal{K}_{\text{res}}} C|\) count the nodes in none of them. One weight \(w = |C|\) is assigned to each \(C \in \mathcal{K}_{\text{res}}\) and, when \(n_0 > 0\), an additional weight \(w = n_0\), with \(W = \sum_w w\). The effective community count is then

\vspace{-10pt}
\begin{equation}\label{eq:effective-count}
  k_{\text{eff}}(\mathcal{K}) = \exp\left( - \sum_{w} \frac{w}{W} \log \frac{w}{W} \right), \qquad w \in \{|C| : C \in \mathcal{K}_{\text{res}}\} \cup \{n_0\},
\end{equation}
\vspace{0pt}

\noindent and it is reported per 1,000 nodes as \(1000\,k_{\text{eff}}(\mathcal{K})/n\). The weights represent community memberships rather than distinct nodes, so a node in two resolved communities contributes to both. The count equals one both for a single community containing every node and for a cover of singletons, so complete concentration and complete fragmentation are both under-resolved. Twenty equal resolved communities covering every node yield \(k_{\text{eff}} = 20\).

If a criterion favours under-resolved covers, its values should correlate negatively with the effective community count across the settings of one method on one graph. Table~\ref{tab:coarseness} tests this prediction on the planted families.

\input{tables/coarseness}

For \(Q_{ov}\), the prediction held for SLPA on every family but not for OSLOM. Across SLPA's settings, its median correlation with the effective community count lay between −0.95 and −0.91 on the three families at \(\mu \le 0.5\). Across OSLOM's settings, it was positive on all three, between 0.31 and 0.39.

For stability, the prediction held only in part. Stability also favoured SLPA's less resolved settings, with median correlations between −0.49 and −0.30. However, at \(\mu \le 0.5\) SLPA resolved slightly more communities than were planted on the median graph, so its less resolved settings were also the more accurate, and this association did not indicate a bias towards under-resolved covers. The same caution applies to \(Q_{ov}\) for SLPA, although its correlation was far stronger than that of accuracy itself (Table~\ref{tab:coarseness}). Across OSLOM's settings, the correlations were positive, between 0.64 and 0.75 at \(\mu \le 0.5\), and stability did not favour the less resolved settings.

The same question can be asked of the two weight configurations. For OSLOM on the real graphs, \(Q_{ov}\) and stability both preferred the coarser configuration on internal \corpus{Enron} but disagreed on the unrestricted graph (Table~\ref{tab:cover-shape}). On the weighted planted suite, the weighted configuration was both less resolved and less overlapping on 88 of the 124 realizations on which it was less accurate. \(Q_{ov}\) preferred it on 76 of those, and stability under \(\Omega\) on 27 (Table~\ref{tab:planted-battery}). The preference for coarser, less resolved covers was therefore consistent for \(Q_{ov}\) across configurations and across SLPA's settings, but not across OSLOM's. Stability favoured OSLOM's less resolved settings only above \(\mu = 0.5\) (Table~\ref{tab:coarseness}), and on the LFR suites only from \(\mu = 0.75\), where its lowest thresholds returned the all-singleton partition (Section~\ref{sec:stability-planted}). Resolution alone did not explain when stability favoured a configuration.

A third view comes from the real graphs taken together. Across the 24 real-data configurations in Table~\ref{tab:shape-drivers}, stability fell as the density of communities above the size floor increased, both pooled (\(\rho = -0.83\)) and within each method. The association was not resolved within pairs of weight configurations (\(p = 0.29\) under a permutation that exchanges only the two members of each pair). \(Q_{ov}\) showed no consistent relationship with this density. It fell within OSLOM and ego-splitting (\(\rho = -0.71\) and \(-0.60\)), rose within SLPA and CD-DAWN (\(\rho = 0.60\) and \(0.74\)), and was positive when pooled across the four methods (0.28).\footnote{Within each method, stability was negatively associated with overlap (\(\rho\) between −0.83 and −0.71), and the pooled association was also negative (\(\rho = -0.41\)). The association between stability and \(Q_{ov}\) was positive within OSLOM (\(\rho = 0.60\)), negative within SLPA and CD-DAWN (−0.60 and −0.75), and near zero within ego-splitting (0.09). Pooled, it was −0.34. The six configurations per method are not a sample, so these coefficients describe the observed covers only.}

The two criteria thus share no single preference for under-resolved covers, but both respond to the shape of a cover, and that is enough for them to agree for reasons unrelated to accuracy. Without a reference cover, agreement amongst reference-free criteria may appear to corroborate a result. However, the quality measures responded to cover shape regardless of accuracy, and stability did so in some regimes, so their agreement is not independent evidence. On internal \corpus{Enron}, all four reference-free diagnostics favoured the weighted configuration, which returned half as many communities, even though the unweighted configuration detected 2.6 times as much overlap. This agreement did not independently establish the accuracy of the weighted cover. On the planted suite, the quality measures favoured the weighted configuration at similar rates regardless of its accuracy. The planted calibration, likewise, did not support reading the stability preference on \corpus{Enron} as an accuracy signal, since the weighted configuration was less accurate in every planted comparison classified as \corpus{Enron}-like, all of them under \corpus{Advogato}'s weights (Section~\ref{sec:inversion}).

This dependence on cover shape motivates the reporting rule (Section~\ref{sec:reporting}). The rule requires four direct descriptions of each cover, namely the number of communities above the size floor, the share of very small communities, the fraction of covered nodes with more than one membership, and the share of nodes that the cover places. These are not quality scores but descriptions of the cover's resolution, overlap, and extent. Because both failure mechanisms respond to cover shape, a criterion with the same preference cannot expose its own failure. The direct descriptions make that shape visible.

\subsection{Limits of Paired Designs}
\label{sec:discussion-shared}

Most comparisons in this paper are paired. Two configurations, or two methods, are scored on the same graphs at the same seeds, on the reasoning that a limitation they share cancels from their difference. That reasoning holds only when the limitation affects all compared conditions equally. In that case, it remains in each condition's result but cancels from their comparison. The comparison of weighted and unweighted configurations in Section~\ref{sec:weights} relies on this reasoning for the incomplete recovery of planted overlap, which both configurations share, but not equally (Section~\ref{sec:limits-overlap}). A second comparison tests whether the same logic applies to a shared procedural setting.

In that comparison, a connector hub was removed from the internal \corpus{Enron} graph before detection, and the cover then detected was set against the one detected with the hub in place. A connector hub is a node adjacent to at least half of all communities but placed in only one. The node removed, \texttt{\footnotesize{jeff.skilling}}, is one of two such nodes amongst the ten of Table~\ref{tab:enron-hubs}. Both conditions used the same method and settings, so any effect of those choices should have cancelled from the comparison. The search depth was one such setting, a choice rather than a limitation of the benchmark, and it was set to 3/3 rather than the default 10/50 (Section~\ref{sec:methods}).

The comparison was made at four search depths, from 3/3 to 20/50, with the graph, hub, and 20 seeds held fixed. Although stability with the hub in place changed from 0.500 at 3/3 to 0.526 at 10/50, the paired effect of hub removal was similar at all four depths. It lowered stability by 0.066 to 0.073 and increased the share of very small communities. No differences in overlap or \(Q_{ov}\) were resolved at any depth (Table~\ref{tab:hub-depth}). The shared setting therefore cancelled from this comparison, as the reasoning requires.

A second case shows the reasoning failing. Pairing does not cancel a shared limitation when its effect depends on the factor compared, and OSLOM's directed binary illustrates such an interaction. When an undirected graph is represented as pairs of reciprocated arcs, the binary creates spurious overlap without edge weights and suppresses it when they are supplied, so weighting can appear beneficial. \Textcite[664]{demeoEnhancingCommunityDetection2013a} evaluated a network weighting scheme with OSLOM on an undirected benchmark. Through the directed binary, across 1,440 such graphs, each detected ten times, assigning unit weights outperformed omitting weights on 239 or 240 of the 240 graphs at each of six mixing parameters.

\input{tables/hub-depth}

With OSLOM's undirected binary, however, the comparison reversed at \(\mu = 0.6\). At \(\mu \le 0.4\), every configuration averaged at least 0.999. At \(\mu = 0.6\), constant weights significantly reduced accuracy on 107 of 240 graphs, scoring 0.848 against 0.912 without weights.\footnote{Accuracy is the normalized mutual information of \textcite[4--5]{danonComparingCommunityStructure2005}. A difference is significant at \(p < 0.001\) over the ten runs, as in the original study \autocite[665]{demeoEnhancingCommunityDetection2013a}. The weighting scheme itself and its replication under the original conditions are reported in \textcite{pizzornoCoversPartitionsOverlapping2026}.} The defect affected both conditions but interacted with weighting, so it did not cancel from their comparison. The comparison in Section~\ref{sec:weights} was less exposed to this interaction. Its planted suites had between one and 14 percent reciprocated arcs, rather than the full reciprocity of the undirected benchmark.

As the two cases show, a shared limitation cancels only when its effect is equal across the compared conditions. An adjustable parameter of the procedure can affect conditions differently, so a comparison at one value should be checked at other plausible values, as in the hub analysis. The incomplete recovery of planted overlap is an outcome of each method, not a property of the benchmark, and it differed between the configurations compared. It is therefore reported with each result and is not assumed to cancel.

\subsection{Index Choice}
\label{sec:discussion-index}

A comparison by accuracy depends on a selection protocol and on an index. Section~\ref{sec:comparison} tested the protocol, and this subsection takes up the index. The method comparison used \(\Omega\) to measure accuracy and the matched protocol to select settings (Section~\ref{sec:matched}). \(\Omega\) was chosen because it corrects for chance agreement and does not reward degenerate covers (Section~\ref{sec:accuracy}). The analysis below tests whether \(\text{NMI}_{\max}\) and \(F_1\) change the conclusions when applied to the same covers.

Applying the other indices changed two conclusions materially. Table~\ref{tab:index-margin} compares the margins over the partition baseline under all three indices, on the same covers and at the settings selected by \(\Omega\). Holding these settings fixed isolates the effect of the index.\footnote{The margins were also computed with each index selecting its own best setting. The table's entries then changed by up to 0.1002, and on the three-membership suite the \(\text{NMI}_{\max}\) and \(F_1\) margins, which rise and then fall in the table, rose steadily. Three statements of this subsection held under both selections. On the main suite, the \(\Omega\) margin turned downward and the other two rose monotonically, and those two were positive where the \(\Omega\) margin collapsed. The disagreement between the indices is therefore not an effect of how the settings were selected.} On the main suite, only the \(\Omega\) margin turned downward at 50 percent planted overlap, while the other two margins continued to rise. With three memberships, the margin of \(\Omega\) at 50 percent was 0.0011 and positive on 26 of the 50 graphs, whereas the \(\text{NMI}_{\max}\) and \(F_1\) margins were positive on 40 and 30 graphs, respectively. With five memberships, \(F_1\) also placed OSLOM below the partition baseline at 50 percent overlap, and \(\text{NMI}_{\max}\) placed its margin within 0.01 of zero. At 30 percent, both remained positive while the \(\Omega\) margin was −0.0205 \autocite{pizzornoCoversPartitionsOverlapping2026}.

\input{tables/index-margin}

The disagreement between indices does not undermine the role of \(\Omega\) as the primary index (Section~\ref{sec:accuracy}). Only \(\Omega\) corrects observed pairwise agreement for agreement expected by chance. \(F_1\) has no chance correction, and \(\text{NMI}_{\max}\) normalizes by entropy. Before chance correction, OSLOM's margin over the partition baseline rose from 0.0054 at ten percent planted overlap to 0.0127 at 30 percent and 0.0145 at 50 percent. Between 30 and 50 percent, the agreement expected by chance fell for both methods, but further for the partition baseline, from 0.880 to 0.857, than for OSLOM, from 0.877 to 0.866. Because the \(\Omega\) correction divides each method's disagreement by one minus its expected agreement, the baseline's lower expected agreement at 50 percent discounted more of its disagreement. That outweighed the increase in the observed-agreement margin and turned the \(\Omega\) margin downward. The downturn therefore reflects chance correction, not a decline in pairwise agreement. \(\Omega\) remains the conservative index to report.

The remaining method-comparison conclusions were checked at both the \(\Omega\)-selected settings and each index's own best settings. Two further conclusions changed with the index. On the graphs with planted overlap at \(\mu \le 0.5\), SLPA ranked above ego-splitting under \(\Omega\), on the strength of the graphs at \(\mu = 0.5\), but below it under \(\text{NMI}_{\max}\) and \(F_1\), though not significantly. At \(\mu = 0.8\), OSLOM and the partition baseline both scored near zero under \(\Omega\) and \(\text{NMI}_{\max}\), but not under \(F_1\), which gave the partition baseline 0.1580 and OSLOM 0.1544 (Section~\ref{sec:competence}). All other conclusions remained unchanged.\footnote{In every evaluation, the crossing remained at \(\mu = 0.6\), and OSLOM's margin over the partition baseline on the full suite was significantly positive under all three indices. The ranking of the methods on all graphs at \(\mu \le 0.5\) was also unchanged, and OSLOM remained significantly above the partition baseline on both subsets at \(\mu \le 0.5\).}

Index choice can thus change a conclusion, most materially the one about the effect of the membership cap. Unlike the preference for coarse covers, the disagreement arises even when the reference cover is known, and it reflects differences in how the indices treat chance agreement. The conclusions that change with the index should be qualified, not withdrawn, and each should name the index under which it holds, as Section~\ref{sec:reporting} requires of a stability figure. The analysis in this subsection applies that requirement to the paper itself. The finding about weak recovery of additional memberships \autocite{pizzornoCoversPartitionsOverlapping2026} is unchanged, because it rests on the number of communities assigned per node, not on an accuracy index.

\subsection{The Validation Gap}
\label{sec:discussion-gap}

The criteria tested in this paper came into use without the kind of test applied to them here, and the datasets available help to explain why. Empirical validation of the quality measures is constrained by those datasets. Table~\ref{tab:corpora} summarizes the mismatch between the directed, weighted networks that the methods target and the datasets with reference covers. Directed, weighted graphs with planted overlapping covers can be generated, and several are used in this paper. Their weights, however, are either synthetic or sampled from a real network and placed on a synthetic topology (Section~\ref{sec:data}). Real graphs with reference covers are either unweighted or use node metadata as the cover. Across the three collections examined, no directed, weighted graph was found with an overlapping reference cover independent of node attributes. The gap lies in the datasets that exist, not in the selection made for this study.

Of the 286 entries in the \emph{Netzschleuder} catalogue \autocite{peixotoNetzschleuderNetworkCatalogue2020}, only one combines directed edges, numeric edge weights, node metadata, and at least 1,000 nodes.\footnote{The count was made through the catalogue's API on 20 September 2026. An entry was counted if it was directed, carried a non-boolean edge property named \texttt{\footnotesize{weight}}, and was tagged as having metadata. The minimum of 1,000 nodes excludes the many small networks that the catalogue also holds.} It is a connectome of 21,739 neurons with a mean degree of 196, annotated by neuron type. Neuron type is a taxonomy, not an overlapping community cover. Treating it as a reference cover would substitute metadata for detected structure. \Textcite[1--2]{peelGroundTruthMetadata2017} showed that metadata need not correspond to the structure recovered by community detection, and that whether they do is itself a question to be tested \autocite[4]{peelGroundTruthMetadata2017}.

The SNAP collection \autocite{leskovecSNAPDatasetsStanford2014} presents a similar mismatch. Its networks with ground-truth communities are undirected, except two that are directed and unweighted. These are \corpus{wiki-topcats}, used in this paper, and an email network whose reference is a partition. A third collection comprises communication graphs amongst computing hosts. These graphs have directed edges \autocite[100]{madaniDatasetNetworksComputing2022}, and their attributes can be used to derive weights \autocite[103--104]{madaniDatasetNetworksComputing2022}. Yet the only reference available, for one graph, groups hosts by function, assigning each to one of 21 groups of at most seven hosts \autocite[104]{madaniDatasetNetworksComputing2022}. This is a partition by node function, not an overlapping community cover.

The divide between real-network applications and reference-cover benchmarks helps explain why quality measures may be adopted without empirical validation. On real graphs, reference covers are generally unavailable, so quality measures are often used without direct accuracy measurements \autocites[26]{chakrabortyMetricsCommunityAnalysis2017}[656]{jebabliCommunityDetectionAlgorithm2018}[24]{rossettiCommunityDiscoveryDynamic2018}. On benchmarks with known covers, accuracy can be measured directly, so an internal quality measure is unnecessary for judging whether a detected cover is correct. Because these settings rarely overlap, a measure may be proposed with an axiomatic rationale and a worked example without testing whether it ranks a trusted reference above systematically degraded covers.

The protocol proposed in Section~\ref{sec:peak} addresses this validation problem, though it cannot supply the missing real-world datasets. Applying it requires only a graph with a trusted reference cover and the measure under test. Three one-parameter ladders test whether a measure ranks the reference above systematically degraded covers. A separate ring-of-cliques test checks whether it detects structure at small scales. No community-detection run is needed, whereas the cluster jobs behind this work total 172,197 core-hours. The protocol therefore makes it feasible to test a measure before adoption. None of the four measures examined appears to have undergone such a test before this paper.

Two external measures illustrate what the protocol can reveal. The measure proposed by \textcite[{3}]{lazarModularityMeasureNetworks2010} weights each node's contribution by the inverse of its membership count and has no null term.\footnote{On an undirected graph, the measure is \(M^{ov} = |\mathcal{K}|^{-1} \sum_{C} \rho_C\, |C|^{-1} \sum_{i \in C} (x_{i,C} - y_{i,C}) / (k^{u}_i\,|m_i|)\), where \(x_{i,C}\) and \(y_{i,C}\) count the edges from node \(i\) to nodes inside and outside \(C\), \(k^{u}_i = x_{i,C} + y_{i,C}\) is its degree, and \(\rho_C\) is the density of \(C\), its number of internal edges divided by \(\binom{|C|}{2}\). A community of one node scores zero. The measure was computed on each graph's undirected view.} From \(\mu = 0.4\) onwards, the single all-node community scored above the planted cover on every graph. Across all 46 graphs in both families, the mean score also rose from one level of the noise ladder to the next. The four measures examined showed neither pattern on any of the 720 planted graphs (Section~\ref{sec:peak-results}). A second example is \(Q_{ds}^{ov}\), the overlapping modularity density of \textcite{chenExtensionModularityDensity2014} discussed in Section~\ref{sec:resolution}. At low mixing, it resisted the merge ladder but favoured finer covers, and at higher mixing it also favoured a coarse cover. Together, these examples show that the protocol can expose preferences for covers that are too fine or too coarse.

The protocol depends on graphs with trusted reference covers. Synthetic benchmarks supply them, and so partly close the gap. The LFR construction has been extended to directed, weighted graphs with overlapping communities \autocite[{1}]{lancichinettiBenchmarksTestingCommunity2009}, and benchmark construction has continued along other axes, most recently to hierarchical communities with heterogeneous connectivity between them \autocite[1]{crossHierarchicalCommunityDetection2025}. The weighted suites used in this paper also plant their weights. In one suite, weights are sampled from the distributions of two real networks but assigned to synthetic topologies. That suite reflects the weight distributions of the source networks, but not the association between weights and topology in them. The remaining need is a real directed, weighted graph with an overlapping reference cover defined independently of node attributes. Until such a dataset is available, accuracy on the graphs that the methods target cannot be measured directly, and reference-free diagnostics may agree for reasons unrelated to accuracy (Section~\ref{sec:discussion-composition}).

\section{Limitations}
\label{sec:limitations}

The preceding section concerned the limits of the criteria, and this one concerns the limits of the evidence behind those findings. Four limitations qualify the conclusions of this paper. Two pertain to how far the results generalize, namely the scope of the datasets and the incomplete recovery of planted overlap by every method. The other two are confined to particular experiments, specifically the computational cost of the consensus experiment and the reproducibility of one OSLOM configuration across processor architectures. Each subsection identifies the conclusions it affects.

\subsection{Scope of the Datasets}
\label{sec:limits-corpora}

The findings rest on bases of different breadth. The quality-measure results have the broadest, since they span the LFR, block-model, and \corpus{DBLP} families, which differ in construction and in how communities are defined (Section~\ref{sec:data}). The failure of \(Q_{ov}\) also recurred on LFR graphs of 10,000 and 50,000 nodes, so it is not limited to the 1,000-node suite.

The method comparison has a narrower empirical basis. The overall ranking and the point at which detection failed were measured in this paper only on LFR, whose planted covers are recoverable at low mixing, and the companion paper measured both under the block model as well \autocite{pizzornoCoversPartitionsOverlapping2026}. Parts of the comparison were nevertheless checked beyond the main suite. OSLOM's margin was tested under a second generator, where it persisted after its covers were collapsed to one membership per node (Section~\ref{sec:generalisation}), and separate LFR controls showed that the margin declined as the membership cap rose. The relationship between stability and accuracy was likewise tested under a second generator (Section~\ref{sec:stability-planted}).

One limit applies to every dataset. Accuracy in this paper is agreement with a reference cover, planted or drawn from node metadata, and no experiment tests whether a more accurate cover improves a downstream use, such as graph-based retrieval-augmented generation, the application that first motivated this work (Section~\ref{sec:related-detection}).

\subsection{Recovery of Planted Overlap}
\label{sec:limits-overlap}

The planted comparisons are limited in two respects, the number of memberships planted for each overlapping node and the share of the planted overlap that the methods recovered. The main suite assigns each overlapping node to exactly two communities. The controls in Section~\ref{sec:generalisation} raised that cap to three, four, and five, and OSLOM's margin declined as the cap rose. No experiment in this paper tests more than five memberships, although the methods impose no such cap on covers returned for real data.\footnote{Three memberships is also the point at which the two candidate forms of \(F\) in Section~\ref{sec:measures} cease to agree, making the unresolved choice consequential.}

Recovery was limited as well. Every method examined in this paper recovered only part of the planted overlap, and OSLOM identified on average at most 54 percent of the nodes planted in two communities at \(\mu \le 0.5\) \autocite{pizzornoCoversPartitionsOverlapping2026}. The detected-overlap column in Table~\ref{tab:cover-shape} is therefore useful for comparing configurations, but not for estimating a graph's actual overlap.

\subsection{Cost of Consensus}
\label{sec:limits-cost}

Computational cost limited the consensus experiment in two ways. First, it shortened the \corpus{Enron} sweep. Section~\ref{sec:consensus} measures agreement to 1,024 input runs on \corpus{Advogato} but only to 256 on \corpus{Enron}, where a single consensus stream at 256 runs took up to 25.6 hours. The \corpus{Enron} results therefore do not establish whether agreement would continue to increase beyond 256 runs, and Table~\ref{tab:consensus-transfer} compares the two networks only over the run counts they share. Second, the consensus procedure was cut off at three iterations, before most streams had converged, so its results describe the procedure as truncated.

\subsection{Architecture-Dependent Reproducibility}
\label{sec:limits-repro}

The last limitation concerns exact reproducibility. With a fixed seed, three of the four configurations on the internal \corpus{Enron} graph gave identical results when rerun on a second machine, of a different processor architecture, but OSLOM's unweighted configuration, rerun at three seeds, differed in most of the quantities derived from its covers. Its output was bit-identical across the nodes of the cluster on which the experiments were run. Both ran the same versions of the Python dependencies, so a difference of library versions does not explain the discrepancy. Because OSLOM was compiled separately for each architecture, this comparison establishes cross-architecture variation but cannot distinguish processor effects from architecture-specific compilation. OSLOM's unweighted configuration was also the least stable across seeds, which is consistent with sensitivity to small numerical perturbations, although the cause was not tested. The discrepancy was small beside the variation across seeds. Stability differed by 0.005 across architectures, compared with a spread of 0.047 across seeds on the same machine. No conclusion in this paper depends on a difference of that size, and no table compares values from the two machines directly.

\section{Conclusion}
\label{sec:conclusion}

This paper tested the three criteria by which overlapping community detection is evaluated, on graphs whose planted cover is known, and each proved to need qualification. The first criterion, internal quality, was tested by asking whether each of four measures attained its maximum at the planted cover. The overlapping modularity \(Q_{ov}\) of \textcite{nicosiaExtendingDefinitionModularity2009} failed on every graph in both planted families, favouring covers with far fewer communities than were planted. The second criterion, stability, measures agreement, not accuracy. On planted graphs, its relationship with accuracy varied by method and regime, and it sometimes remained high after detection failed. The third criterion, accuracy against a reference, depended on the selection protocol. OSLOM's lead over the partition baseline was significant when each graph's setting was chosen with its planted cover, but not when the setting was chosen on other graphs, and on the suite with three memberships the protocol alone decided the comparison.

Where no reference exists, the first two criteria are all that remain, and the practical consequence is clearest on internal \corpus{Enron}. All four reference-free diagnostics favoured OSLOM's weighted configuration, although that configuration detected less overlap than the unweighted one. Their agreement did not establish accuracy. On planted graphs, the quality measures did not reliably track accuracy, and in every planted comparison classified as \corpus{Enron}-like the unweighted configuration was the more accurate, all under weights that barely distinguished the planted communities (Section~\ref{sec:inversion}). Because \corpus{Enron} has no reference cover, its accuracy ranking remains undetermined.

Taken together, these findings motivate four practical safeguards. First, a quality measure should be tested with the protocol of Section~\ref{sec:peak} before adoption, which requires only a graph with a trusted reference cover and the measure itself. Second, scores from the four measures examined should not be compared across covers that differ in granularity or shape. Third, stability should be reported alongside the number of communities above the size floor, the share of very small communities, the fraction of covered nodes with more than one membership, and the share of nodes that the cover places. Fourth, method comparisons should use the matched selection protocol. Margins obtained by selecting settings on each graph's planted cover should be identified as oracle scores and reported beside scores from a selection that does not use the scored graph's reference.

What the safeguards cannot do is settle accuracy on the graphs that these methods target. The results do not establish which method is most accurate on real directed, weighted graphs, because no available real dataset combines direction and weights with an overlapping reference cover defined independently of node attributes. Until such a dataset is assembled, direct evidence of accuracy for such graphs comes from planted graphs, while agreement amongst reference-free diagnostics on real graphs remains inconclusive. The validation protocol and the reporting rule can expose the limitations of the reference-free criteria, but they cannot replace that missing benchmark.

\section*{Data and Code Availability}
\label{sec:availability}
\addcontentsline{toc}{section}{Data and Code Availability}

Everything required to reproduce the analyses, tables, and figures reported in this paper is publicly available. The released Python package implements the methods and the measures, and it provides the harness that ran the parameter sweeps and the code that generates the tables and figures.\footnote{\url{https://github.com/gpizzorno/coverchecks}} The accompanying deposit contains the manifest and the per-run result table of every sweep, the covers retained for every scored run, and the scheduler logs.\footnote{\url{https://doi.org/10.5281/zenodo.24541807}} The scheduler recorded 172,197 core-hours over 47,801 allocations for the cluster jobs behind this paper and the companion paper \autocite{pizzornoCoversPartitionsOverlapping2026}. The planted suites are deposited as well, each graph file carrying its planted cover beside its edge list, and the \corpus{wiki-topcats} subsamples can be rebuilt by the package from the published network. Every accuracy reported in this paper can therefore be recomputed from the deposited files and the rebuilt subsamples. The source datasets are public and are not redistributed. Where a required file is absent, the package's loaders report how to retrieve or build it.

The package also implements the computation of the \(\Omega\) index from membership counts described in Section~\ref{sec:accuracy}. It recovers the exact index without enumerating node pairs, at a cost independent of community size, whereas pair enumeration on degenerate covers can require tens to hundreds of gigabytes of memory.

The companion paper, \emph{Covers and Partitions} \autocite{pizzornoCoversPartitionsOverlapping2026}, reports the method comparison, implementation validation, and OSLOM configuration study in full. The tables and figures in both papers are generated from the same results using the same process.

\printbibliography{}

\end{document}

%% file: tables/corpora.tex
\begin{table}[tbp]
 \sffamily\footnotesize%
 \begin{tabularx}{\textwidth}{>{\raggedright\arraybackslash}Xcrrrccll}
  \toprule
  \scriptsize\textbf{Dataset} & \scriptsize\textbf{Memb.} & \scriptsize\textbf{Graphs} & \scriptsize\textbf{Nodes} & \scriptsize\textbf{Edges} & \scriptsize\textbf{Dir.} & \scriptsize\textbf{Wgt.} & \scriptsize\textbf{Reference cover} & \scriptsize\textbf{Use} \\
  \midrule
  \corpus{DBLP} co-authorship &  & 10 & 10,996 & 83,934 & no & no & venues & \S\ref{sec:peak} \\
  \corpus{wiki-topcats} &  & 5 & 5,746 & 20,173 & yes & no & categories & \S\ref{sec:matched} \\
  \corpus{Advogato} &  & 1 & 5,155 & 47,135 & yes & yes & none & \S\ref{sec:stability}, \S\ref{sec:weights} \\
  \corpus{Enron}, internal &  & 1 & 21,510 & 233,276 & yes & yes & none & \S\ref{sec:stability}, \S\ref{sec:weights}, \S\ref{sec:battery} \\
  \corpus{Enron}, unrestricted &  & 1 & 85,741 & 360,121 & yes & yes & none & \S\ref{sec:stability}, \S\ref{sec:battery} \\
  Block-model suite & 2 & 36 × 10 & 1,000 & 14,303 & yes & no & planted cover & \S\ref{sec:stability}, \S\ref{sec:generalisation} \\
  Main suite & 2 & 36 × 10 & 1,000 & 13,996 & yes & no & planted cover & \S\ref{sec:peak}, \S\ref{sec:stability}, \S\ref{sec:comparison} \\
  10,000-node suite & 2 & 36 × 5 & 10,000 & 139,296 & yes & no & planted cover & \S\ref{sec:peak} \\
  50,000-node suite & 2 & 24 × 3 & 50,000 & 681,162 & yes & no & planted cover & \S\ref{sec:peak}, \S\ref{sec:matched} \\
  Three-membership suite & 3 & 36 × 10 & 1,000 & 13,798 & yes & no & planted cover & \S\ref{sec:stability}, \S\ref{sec:generalisation}, \S\ref{sec:discussion-index} \\
  Four-membership suite & 4 & 36 × 10 & 1,000 & 13,542 & yes & no & planted cover & \S\ref{sec:generalisation} \\
  Five-membership suite & 5 & 36 × 10 & 1,000 & 13,312 & yes & no & planted cover & \S\ref{sec:generalisation} \\
  Weighted suite & 2 & 16 × 10 & 1,000 & 13,790 & yes & yes & planted cover & \S\ref{sec:inversion}, \S\ref{sec:weights}, \S\ref{sec:battery} \\
  Contrast suite & 2 & 140 × 10 & 1,000 & 13,996 & yes & yes & planted cover & \S\ref{sec:weights} \\
  Weights-only suite & 2 & 128 × 5 & 1,000 & 13,829 & yes & yes & planted cover & \S\ref{sec:weights} \\
  Real-weight suite & 2 & 128 × 10 & 1,000 & 13,790 & yes & yes & planted cover & \S\ref{sec:inversion}, \S\ref{sec:weights}, \S\ref{sec:battery} \\
  \bottomrule
  \end{tabularx}
  \caption{Datasets used in this paper. The planted suites were generated by LFR, except the block-model suite, and \tabterm{Memb.} is the number of memberships of each overlapping node. \tabterm{Graphs} is the number of graphs, which for a planted suite is the number in one realization times the number of realizations, each a fresh draw from the same parameters. Node and edge counts are medians for datasets with more than one graph. \tabterm{Dir.} and \tabterm{Wgt.} mark directed and weighted graphs, and \tabterm{Use} gives the sections in which each dataset is mainly used.}
  \label{tab:corpora}  
\end{table}

%% file: tables/peaks-at-truth.tex
\begin{wraptable}{R}{0pt}
 \sffamily\footnotesize%
 \begin{tabular}{lccc}
  \toprule
  \scriptsize\textbf{Measure} & \scriptsize\textbf{Peaks} & \scriptsize\textbf{Gap} & \scriptsize\textbf{Peak \textit{k}} \\
  \midrule
  \textit{Q}\tmsubx{\textit{ov}} & 0/9 & +0.1197 & 2 \\
  \textit{Q}\tmsubx{\textit{dwo}} & 2/9 & +0.0008 & 17 \\
  \textit{Q}\tmsubx{\textit{d}} & 7/9 & −0.0007 & 21 \\
  \textit{Q}\tmsubx{\textit{lr}} & 7/9 & −0.0006 & 21 \\
  \bottomrule
  \end{tabular}
  \caption{Whether each quality measure attained its maximum at the planted cover on the merge and random-split ladders, over the nine partition graphs of the main suite. \tabterm{Peaks} counts the graphs on which it did. \tabterm{Gap} is the median amount by which the best rival on either ladder outscored the planted cover, negative where the planted cover scored highest. \tabterm{Peak \textit{k}} is the median number of communities at the maximum over the planted cover and the merge ladder, against a planted median of 21.}
  \label{tab:peaks-at-truth}  
\end{wraptable}

%% file: tables/granularity-dose.tex
\begin{table}[tbp]
 \sffamily\footnotesize%
 \begin{tabular*}{\textwidth}{@{\extracolsep{\fill}}lccccccccc}
  \toprule
  \scriptsize\textbf{Reference} & \scriptsize\textbf{Target \textit{k}} & \scriptsize\textbf{\textit{n}} & \scriptsize\textbf{Null ratio} & \scriptsize\textbf{\textit{Q}\tmsubx{\textit{ov}} at ref.} & \scriptsize\textbf{\textit{Q}\tmsubx{\textit{ov}} gap} & \scriptsize\textbf{\textit{Q}\tmsubx{\textit{ov}} peak \textit{k}} & \scriptsize\textbf{\textit{Q}\tmsubx{\textit{d}} gap} & \scriptsize\textbf{\textit{Q}\tmsubx{\textit{d}} at ref.} & \scriptsize\textbf{\textit{Q}\tmsubx{\textit{dwo}} gap} \\
  \midrule
  \multirow{7}{*}{venues} & 5 & 10 & 3.0\tmmed{}×\tmmed{}10\tmsup{−2} & 0 & +0.0299 & 3 & --- & --- & −0.0191 \\
   & 10 & 10 & 5.5\tmmed{}×\tmmed{}10\tmsup{−3} & 0 & +0.0891 & 3 & --- & --- & +0.0015 \\
   & 20 & 10 & 9.8\tmmed{}×\tmmed{}10\tmsup{−4} & 0 & +0.1239 & 2 & --- & --- & +0.0052 \\
   & 40 & 10 & 1.8\tmmed{}×\tmmed{}10\tmsup{−4} & 0 & +0.1576 & 2 & --- & --- & +0.0129 \\
   & 60 & 10 & 7.8\tmmed{}×\tmmed{}10\tmsup{−5} & 0 & +0.1716 & 2 & --- & --- & +0.0145 \\
   & 100 & 10 & 2.5\tmmed{}×\tmmed{}10\tmsup{−5} & 0 & +0.1983 & 2 & --- & --- & +0.0312 \\
   & 150 & 10 & 9.7\tmmed{}×\tmmed{}10\tmsup{−6} & 0 & +0.2098 & 2 & --- & --- & +0.0397 \\
  \cmidrule(lr){1-10}
  \multirow{7}{*}{partition} & 5 & 10 & 6.4\tmmed{}×\tmmed{}10\tmsup{−2} & 1 & +0.0197 & 3 & −0.0080 & 8 & −0.0085 \\
   & 10 & 10 & 2.0\tmmed{}×\tmmed{}10\tmsup{−2} & 0 & +0.0419 & 4 & −0.0006 & 6 & +0.0041 \\
   & 20 & 10 & 4.4\tmmed{}×\tmmed{}10\tmsup{−3} & 0 & +0.0767 & 3 & +0.0001 & 5 & +0.0217 \\
   & 40 & 10 & 1.2\tmmed{}×\tmmed{}10\tmsup{−3} & 0 & +0.1114 & 3 & +0.0006 & 1 & +0.0385 \\
   & 60 & 10 & 6.3\tmmed{}×\tmmed{}10\tmsup{−4} & 0 & +0.1292 & 3 & +0.0022 & 0 & +0.0562 \\
   & 100 & 10 & 2.4\tmmed{}×\tmmed{}10\tmsup{−4} & 0 & +0.1396 & 3 & +0.0005 & 0 & +0.0635 \\
   & 150 & 10 & 1.1\tmmed{}×\tmmed{}10\tmsup{−4} & 0 & +0.1459 & 2.5 & +0.0002 & 0 & +0.0571 \\
  \bottomrule
  \end{tabular*}
  \caption{The granularity sweep, in which ten distinct \corpus{DBLP} subsamples at each of seven target numbers of reference communities (\tabterm{Target \textit{k}}) were scored against the venue cover and against the partition that keeps each author in one venue. \textit{n} is the number of subsamples. \tabterm{Null ratio} is the median ratio of the damped null term to Newman's at the reference cover. \tabterm{At ref.} counts the graphs on which the reference scored at least as well as every rung of either ladder, \tabterm{gap} is the median amount by which the best rung outscored the reference, and \tabterm{peak \textit{k}} is the median number of communities at the maximum of \textit{Q}\tmsubx{\textit{ov}}. \textit{Q}\tmsubx{\textit{d}} is shown only against the partition, where it is defined.}
  \label{tab:granularity-dose}  
\end{table}

%% file: tables/eq-checks.tex
\begin{table}[tbp]
 \sffamily\footnotesize%
 \begin{tabular*}{\textwidth}{@{\extracolsep{\fill}}llccccccc}
  \toprule
  \scriptsize\textbf{Dataset} & \scriptsize\textbf{Measure} & \scriptsize\textbf{\textit{n}} & \scriptsize\textbf{At ref.} & \scriptsize\textbf{Random merges} & \scriptsize\textbf{Structural split} & \scriptsize\textbf{Noise} & \scriptsize\textbf{Peak \textit{k}} & \scriptsize\textbf{Gap} \\
  \midrule
  \multirow{3}{*}{LFR, partition} & \textit{EQ} & 9 & 7/9 & 9/9 & 9/9 & 9/9 & 21 (19--21) & −0.0006 \\
   & \textit{Q}\tmsubx{\textit{ov}} & 9 & 0/9 & 0/9 & 9/9 & 9/9 & 2 (2--5) & +0.1197 \\
   & \textit{Q}\tmsubx{\textit{dwo}} & 9 & 2/9 & 2/9 & 9/9 & 9/9 & 17 (2--21) & +0.0008 \\
  \cmidrule(lr){1-9}
  \multirow{3}{*}{LFR, overlapping} & \textit{EQ} & 27 & 25/27 & 16/27 & 18/27 & 27/27 & 26 (22--30) & −0.0005 \\
   & \textit{Q}\tmsubx{\textit{ov}} & 27 & 0/27 & 0/27 & 19/27 & 27/27 & 2 (2--4) & +0.1877 \\
   & \textit{Q}\tmsubx{\textit{dwo}} & 27 & 13/27 & 2/27 & 27/27 & 27/27 & 7 (3--30) & +0.0023 \\
  \cmidrule(lr){1-9}
  \multirow{3}{*}{DBLP} & \textit{EQ} & 10 & 3/10 & 1/10 & 1/10 & 10/10 & 58.5 (53--60) & +0.0004 \\
   & \textit{Q}\tmsubx{\textit{ov}} & 10 & 0/10 & 0/10 & 10/10 & 10/10 & 2 & +0.1716 \\
   & \textit{Q}\tmsubx{\textit{dwo}} & 10 & 0/10 & 0/10 & 2/10 & 10/10 & 37.5 (11--56) & +0.0145 \\
  \bottomrule
  \end{tabular*}
  \caption{The overlapping modularity \textit{EQ} of Shen et al.\ on the validation ladders, beside \textit{Q}\tmsubx{\textit{ov}} and \textit{Q}\tmsubx{\textit{dwo}}, on the main suite and on ten distinct \corpus{DBLP} subsamples scored against the venue cover. \textit{EQ} is computed on each graph's undirected view, and \textit{n} is the number of graphs. The count columns give the graphs on which the reference cover, planted on the main suite and the venue cover on \corpus{DBLP}, scored at least as well as every cover on the named ladder. \tabterm{At ref.} combines the merge and random-split ladders, and \tabterm{Random merges} requires it on the ladders of all three seeds. \tabterm{Noise} counts the graphs on which the mean score over 30 draws at each level never rose as memberships were randomized. \tabterm{Peak \textit{k}} is the median number of communities at the maximum over the reference cover and the merge ladder, with its range. \tabterm{Gap} is the median amount by which the best rival on either ladder outscored the reference cover, negative where the reference cover scored highest.}
  \label{tab:eq-checks}  
\end{table}

%% file: tables/resolution-limits.tex
\begin{wraptable}{R}{0pt}
 \sffamily\footnotesize%
 \begin{tabular}{cccccc}
  \toprule
  \scriptsize\textbf{Cliques} & \scriptsize\textbf{Nodes} & \scriptsize\textbf{\textit{Q}\tmsubx{\textit{ov}}} & \scriptsize\textbf{\textit{Q}\tmsubx{\textit{dwo}}} & \scriptsize\textbf{\textit{Q}\tmsubx{\textit{d}}} & \scriptsize\textbf{\textit{Q}\tmsubx{\textit{lr}}} \\
  \midrule
  4 & 16 & −0.0379 & −0.1786 & −0.1786 & −0.1547 \\
  6 & 24 & \textbf{+0.0390} & −0.0952 & −0.0952 & −0.0838 \\
  8 & 32 & \textbf{+0.0578} & −0.0536 & −0.0536 & −0.0484 \\
  10 & 40 & \textbf{+0.0644} & −0.0286 & −0.0286 & −0.0272 \\
  12 & 48 & \textbf{+0.0674} & −0.0119 & −0.0119 & −0.0130 \\
  14 & 56 & \textbf{+0.0689} & --- & --- & −0.0029 \\
  16 & 64 & \textbf{+0.0697} & \textbf{+0.0089} & \textbf{+0.0089} & \textbf{+0.0047} \\
  20 & 80 & \textbf{+0.0706} & \textbf{+0.0214} & \textbf{+0.0214} & \textbf{+0.0153} \\
  24 & 96 & \textbf{+0.0709} & \textbf{+0.0298} & \textbf{+0.0298} & \textbf{+0.0224} \\
  30 & 120 & \textbf{+0.0712} & \textbf{+0.0381} & \textbf{+0.0381} & \textbf{+0.0295} \\
  \bottomrule
  \end{tabular}
  \caption{Resolution-limit behaviour on a ring of four-node cliques joined by single reciprocal edges. Each cell is the score of the cover that fuses adjacent cliques in pairs minus the planted cover's, set in bold where positive. A dash marks an exact tie.}
  \label{tab:resolution-limits}  
\end{wraptable}

%% file: tables/cover-shape.tex
\begin{table}[tbp]
 \sffamily\footnotesize%
 \setlength{\tabcolsep}{2.2pt}%
 \begin{tabular*}{\textwidth}{@{\extracolsep{\fill}}lllcccccccccc}
  \toprule
  \scriptsize\textbf{Dataset} & \scriptsize\textbf{Method} & \scriptsize\textbf{Weights} & \scriptsize\textbf{\textit{k}} & \scriptsize\textbf{\textit{k}\tmsubx{≥20}} & \scriptsize\textbf{Very small} & \scriptsize\textbf{Med.} & \scriptsize\textbf{Largest} & \scriptsize\textbf{Overlap} & \scriptsize\textbf{Placed} & \scriptsize\textbf{Ω} & \scriptsize\textbf{Interval} & \scriptsize\textbf{\textit{Q}\tmsubx{\textit{ov}}} \\
  \midrule
  \multirow{8}{*}{\corpus{Advogato}} & OSLOM & \texttt{-uw} & 310 & 104 & 52.1\% & 5 & 3.9\% & 9.18\% & 100.0\% & 0.477 & 0.466--0.489 & 0.300 \\
   & OSLOM & \texttt{-w} & 223 & 62 & 54.2\% & 3 & 7.4\% & 4.19\% & 100.0\% & 0.502 & 0.482--0.522 & 0.381 \\
   & SLPA & \texttt{-uw} & 965 & 7 & 97.8\% & 1 & 80.1\% & 14.09\% & 100.0\% & 0.888 & 0.827--0.932 & 0.415 \\
   & SLPA & \texttt{-w} & 1,015 & 8 & 97.2\% & 1 & 77.3\% & 18.02\% & 100.0\% & 0.865 & 0.784--0.919 & 0.454 \\
   & ego-splitting & \texttt{-uw} & 92 & 13 & 82.9\% & 3 & 16.3\% & 10.77\% & 67.6\% & 0.589 & 0.529--0.650 & 0.490 \\
   & ego-splitting & \texttt{-w} & 91 & 12 & 83.4\% & 3 & 16.6\% & 10.69\% & 67.6\% & 0.612 & 0.576--0.656 & 0.494 \\
   & CD-DAWN & \texttt{-uw} & 70 & 1 & 98.6\% & 2 & 97.1\% & 0.00\% & 99.6\% & 1.000 & 1.000--1.000 & 0.061 \\
   & CD-DAWN & \texttt{-w} & 71 & 1 & 98.6\% & 2 & 97.1\% & 0.02\% & 99.8\% & 1.000 & 1.000--1.000 & 0.060 \\
  \cmidrule(lr){1-13}
  \multirow{8}{*}{\shortstack[l]{\corpus{Enron},\\internal}} & OSLOM & \texttt{-uw} & 296 & 222 & 16.9\% & 60 & 4.2\% & 6.59\% & 100.0\% & 0.500 & 0.468--0.531 & 0.239 \\
   & OSLOM & \texttt{-w} & 146 & 87 & 36.1\% & 67 & 6.2\% & 2.57\% & 100.0\% & 0.597 & 0.574--0.619 & 0.403 \\
   & SLPA & \texttt{-uw} & 902 & 24 & 93.3\% & 1 & 77.9\% & 17.49\% & 100.0\% & 0.658 & 0.566--0.754 & 0.510 \\
   & SLPA & \texttt{-w} & 1,071 & 103 & 82.4\% & 1 & 48.6\% & 48.46\% & 100.0\% & 0.618 & 0.560--0.675 & 0.472 \\
   & ego-splitting & \texttt{-uw} & 38 & 11 & 66.4\% & 3 & 11.4\% & 7.36\% & 56.0\% & 0.689 & 0.659--0.723 & 0.578 \\
   & ego-splitting & \texttt{-w} & 37 & 10 & 67.9\% & 3 & 16.5\% & 7.23\% & 56.0\% & 0.637 & 0.607--0.671 & 0.561 \\
   & CD-DAWN & \texttt{-uw} & 337 & 1 & 99.6\% & 1 & 94.4\% & 0.01\% & 96.1\% & 0.986\textsuperscript{\dag} & 0.984--0.988 & 0.109 \\
   & CD-DAWN & \texttt{-w} & 300 & 1 & 99.7\% & 1 & 98.2\% & 0.00\% & 99.8\% & 0.987\textsuperscript{\dag} & 0.986--0.988 & 0.037 \\
  \cmidrule(lr){1-13}
  \multirow{8}{*}{\shortstack[l]{\corpus{Enron},\\unrestricted}} & OSLOM & \texttt{-uw} & 2,845 & 622 & 74.1\% & 1 & 2.0\% & 3.87\% & 100.0\% & 0.626 & 0.614--0.637 & 0.359 \\
   & OSLOM & \texttt{-w} & 2,893 & 318 & 87.2\% & 1 & 2.3\% & 0.90\% & 100.0\% & 0.608 & 0.598--0.616 & 0.477 \\
   & SLPA & \texttt{-uw} & 10,025 & 474 & 91.1\% & 1 & 43.1\% & 29.02\% & 100.0\% & 0.649 & 0.582--0.724 & 0.697 \\
   & SLPA & \texttt{-w} & 10,143 & 529 & 90.6\% & 1 & 24.7\% & 37.72\% & 100.0\% & 0.706 & 0.674--0.738 & 0.510 \\
   & ego-splitting & \texttt{-uw} & 340 & 62 & 65.9\% & 4 & 5.2\% & 5.39\% & 27.4\% & 0.658 & 0.615--0.704 & 0.510 \\
   & ego-splitting & \texttt{-w} & 339 & 61 & 66.5\% & 4 & 5.5\% & 5.49\% & 27.4\% & 0.750 & 0.724--0.780 & 0.481 \\
   & CD-DAWN & \texttt{-uw} & 4,112 & 43 & 98.4\% & 1 & 79.7\% & 0.15\% & 88.8\% & 0.982\textsuperscript{\dag} & 0.981--0.984 & 0.354 \\
   & CD-DAWN & \texttt{-w} & 3,339 & 27 & 98.4\% & 1 & 88.8\% & 0.43\% & 97.1\% & 0.974\textsuperscript{\dag} & 0.970--0.978 & 0.227 \\
  \bottomrule
  \end{tabular*}
  \caption{Cover shape on the two networks without a reference cover, with \corpus{Enron} as both the internal and the unrestricted graph. \textit{k} is the number of communities and \textit{k}\tmsubx{≥20} the number with 20 or more members. \tabterm{Very small} is the share of communities with five or fewer members, \tabterm{Med.} is the median community size, \tabterm{Largest} is the share of the graph's nodes in the largest community, \tabterm{Overlap} is the share of placed nodes in more than one community, and \tabterm{Placed} is the share of the graph's nodes in at least one community, each a mean over the seeds' covers. Ω is stability, the mean agreement over every pair of the 20 seeds on the nodes that at least one of their covers places, with a 95 percent interval from a bootstrap over seeds. \textit{Q}\tmsubx{\textit{ov}} is the overlapping modularity, averaged over the seeds' covers. A dagger (\textsuperscript{\dag}) marks a sampled estimate of Ω (Section~\ref{sec:reporting}).}
  \label{tab:cover-shape}  
\end{table}

%% file: tables/realdata-planted.tex
\begin{table}[tbp]
 \sffamily\footnotesize%
 \setlength{\tabcolsep}{3.5pt}%
 \begin{tabular*}{\textwidth}{@{\extracolsep{\fill}}lcccccccccc}
  \toprule
   &  &  & \multicolumn{2}{c}{\scriptsize\textbf{Very small}} & \multicolumn{2}{c}{\scriptsize\textbf{Stability}} & \multicolumn{2}{c}{\scriptsize\textbf{Accuracy}} &  &  \\
  \cmidrule(lr){4-5}\cmidrule(lr){6-7}\cmidrule(lr){8-9}
  \scriptsize\textbf{Suite} & \scriptsize\textbf{\textit{μ}} & \scriptsize\textbf{Graphs} & \scriptsize\textbf{OSLOM} & \scriptsize\textbf{SLPA} & \scriptsize\textbf{OSLOM} & \scriptsize\textbf{SLPA} & \scriptsize\textbf{OSLOM} & \scriptsize\textbf{SLPA} & \scriptsize\textbf{Concordant} & \scriptsize\textbf{\textit{ρ}} \\
  \midrule
  \multirow{2}{*}{Main} & ≤\tmthick{}0.5 & 200 & 0.0\% & 18.9\% & 0.914 & 0.460 & 0.822 & 0.453 & 200/200 & −0.82 \\
   & >\tmthick{}0.5 & 160 & 33.2\% & 17.6\% & 0.783 & 0.393 & 0.200 & 0.000 & 113/156 & −0.03 \\
  \cmidrule(lr){1-11}
  \multirow{2}{*}{Three-membership} & ≤\tmthick{}0.5 & 20 & 0.0\% & 20.8\% & 0.877 & 0.409 & 0.692 & 0.397 & 20/20 & −0.80 \\
   & >\tmthick{}0.5 & 16 & 24.4\% & 19.1\% & 0.728 & 0.290 & 0.185 & 0.000 & 14/16 & −0.22 \\
  \cmidrule(lr){1-11}
  \multirow{2}{*}{Block-model} & ≤\tmthick{}0.5 & 20 & 0.2\% & 72.4\% & 0.931 & 0.562 & 0.856 & 0.160 & 17/20 & −0.55 \\
   & >\tmthick{}0.5 & 16 & 56.9\% & 87.6\% & 0.861 & 0.880 & 0.150 & 0.000 & 4/15 & +0.74 \\
  \bottomrule
  \end{tabular*}
  \caption{The two configurations used on the real graphs (Table~\ref{tab:cover-shape}), run on the planted suites at the same settings: OSLOM at search depth 3/3 with threshold 0.1 and coverage 0.5, and SLPA without size shaping at \textit{r}\tmthick{}=\tmthick{}0.15, each at ten seeds. The suites are named as in Table~\ref{tab:corpora}, and the one with two memberships is read in ten realizations, the other two in one. \tabterm{Very small} communities have five or fewer members, \tabterm{Stability} is Ω between seeds, and \tabterm{Accuracy} is Ω against the planted cover, each a mean over graphs. \tabterm{Concordant} counts the graphs on which the more stable configuration was the more accurate, amongst those on which the two differed in both, and \textit{ρ} is the rank correlation between the share of very small communities and stability over both configurations' graphs together.}
  \label{tab:realdata-planted}  
\end{table}

%% file: tables/planted-battery.tex
\begin{table}[b!]
 \sffamily\footnotesize%
 \begin{tabular*}{\textwidth}{@{\extracolsep{\fill}}cccccccc}
  \toprule
   &  &  &  & \multicolumn{4}{c}{\scriptsize\textbf{Prefer \texttt{-w}}} \\
  \cmidrule(lr){5-8}
  \scriptsize\textbf{Planted overlap} & \scriptsize\textbf{Graphs} & \scriptsize\textbf{\texttt{-uw} more accurate} & \scriptsize\textbf{Of which less resolved} & \scriptsize\textbf{Ω} & \scriptsize\textbf{\textup{NMI}\tmsubx{\textup{max}}} & \scriptsize\textbf{\textit{Q}\tmsubx{\textit{ov}}} & \scriptsize\textbf{\textit{Q}\tmsubx{\textit{dwo}}} \\
  \midrule
  0\% & 40 & 14 & 1 & 2 & 2 & 2 & 7 \\
  10\% & 40 & 35 & 14 & 24 & 25 & 14 & 35 \\
  30\% & 40 & 38 & 37 & 8 & 7 & 36 & 38 \\
  50\% & 40 & 37 & 36 & 15 & 11 & 36 & 37 \\
  All & 160 & 124 & 88 & 49 & 45 & 88 & 117 \\
  \bottomrule
  \end{tabular*}
  \caption{The reference-free diagnostics of Table~\ref{tab:blind-battery} on the weighted planted suite, for OSLOM with and without the graph's edge weights at the setting used on \corpus{Enron} (threshold 0.1, coverage 0.5, search depth 3/3) and 20 seeds, on the 16 graphs in ten realizations each. At each level of planted overlap, the table counts the realizations on which the unweighted configuration was the more accurate. Of those, it counts the ones on which the weighted configuration was also the less resolved, by the effective community count of Section~\ref{sec:discussion-composition}, and the less overlapping, and the ones on which each diagnostic preferred the weighted configuration. \textit{Q}\tmsubx{\textit{dwo}} is read on the graph's own weights, as on \corpus{Enron}.}
  \label{tab:planted-battery}  
\end{table}

%% file: tables/weight-families.tex
\begin{table}[tbp]
 \sffamily\footnotesize%
 \setlength{\tabcolsep}{3.5pt}%
 \begin{tabular*}{\textwidth}{@{\extracolsep{\fill}}lclccccccccccc}
  \toprule
   &  &  & \multicolumn{2}{c}{\scriptsize\textbf{More accurate}} &  & \multicolumn{3}{c}{\scriptsize\textbf{Prefer \texttt{-w}}} & \multicolumn{3}{c}{\scriptsize\textbf{Change in Ω}} & \multicolumn{2}{c}{\scriptsize\textbf{Gains ≥\tmthick{}0.097}} \\
  \cmidrule(lr){4-5}\cmidrule(lr){7-9}\cmidrule(lr){10-12}\cmidrule(lr){13-14}
  \scriptsize\textbf{Weights from} & \scriptsize\textbf{Tilt} & \scriptsize\textbf{Contrast} & \scriptsize\textbf{\texttt{-uw}} & \scriptsize\textbf{\texttt{-w}} & \scriptsize\textbf{Gain} & \scriptsize\textbf{Ω} & \scriptsize\textbf{\textit{Q}\tmsubx{\textit{ov}}} & \scriptsize\textbf{\textit{Q}\tmsubx{\textit{dwo}}} & \scriptsize\textbf{Own} & \scriptsize\textbf{Const.} & \scriptsize\textbf{Perm.} & \scriptsize\textbf{Cells} & \scriptsize\textbf{\texttt{-w} better} \\
  \midrule
  Power law & --- & 2.17--2.52 & 124 & 18 & 0.0195 & 49 & 88 & 117 & −0.016 & −0.002 & −0.065 & 14/4,234 & 9 \\
  \cmidrule(lr){1-14}
  \multirow{3}{*}{\corpus{Enron}} & 0.2 & 1.65--2.86 & 158 & 2 & 0.0487 & 28 & 54 & 154 & −0.046 & −0.002 & −0.103 & 8/4,790 & 6 \\
   & 0.3 & 1.29--2.14 & 158 & 2 & 0.0741 & 19 & 43 & 156 & −0.062 & −0.002 & −0.103 & 5/4,800 & 4 \\
   & 0.5 & 0.78--1.29 & 160 & 0 & 0.1319 & 18 & 35 & 160 & −0.104 & −0.002 & −0.102 & 7/4,800 & 1 \\
  \cmidrule(lr){1-14}
  \multirow{5}{*}{\corpus{Advogato}} & 0.0 & 1.19--1.21 & 125 & 14 & 0.0742 & 70 & 79 & 56 & +0.003 & −0.002 & −0.001 & 204/4,191 & 5 \\
   & 0.1 & 1.15--1.17 & 127 & 12 & 0.0777 & 70 & 80 & 54 & +0.005 & −0.002 & −0.001 & 223/4,193 & 7 \\
   & 0.2 & 1.11--1.12 & 126 & 13 & 0.0798 & 69 & 76 & 55 & +0.005 & −0.002 & 0.000 & 203/4,194 & 6 \\
   & 0.3 & 1.07--1.08 & 126 & 13 & 0.0828 & 61 & 81 & 55 & +0.002 & −0.002 & −0.001 & 240/4,193 & 9 \\
   & 0.5 & 0.99--1.01 & 123 & 15 & 0.0888 & 58 & 75 & 53 & −0.002 & −0.002 & −0.001 & 184/4,200 & 6 \\
  \bottomrule
  \end{tabular*}
  \caption{The diagnostics of Table~\ref{tab:planted-battery} under three kinds of edge weight, on the same 16 planted graphs in ten realizations each, at 20 seeds: the power law of the weighted planted suite, and weights drawn from \corpus{Enron}'s and \corpus{Advogato}'s own. \tabterm{Tilt} is the probability that an edge is given the weight that runs against its community, from 0 (every weight follows the planted communities) to 0.5 (none does), and the power law has none. \tabterm{Contrast} is the realized ratio of the mean internal to the mean external edge weight. At the setting used on \corpus{Enron}, \tabterm{More accurate} counts the realizations on which each configuration was the more accurate, and \tabterm{Gain} is the mean accuracy gained by withholding the weights. \tabterm{Prefer \texttt{-w}} counts, amongst the realizations on which the unweighted configuration was the more accurate, those on which each diagnostic preferred the weighted one. \tabterm{Change in Ω} is the mean change in stability from the unweighted configuration when the weighted code path read the graph's own weights (\tabterm{Own}), constant weights (\tabterm{Const.}), and weights permuted across edges (\tabterm{Perm.}, the mean of five permutations). Over every setting at both search depths, \tabterm{Cells} counts those in which the weighted configuration was the more stable by at least 0.097, the gain on \corpus{Enron}, amongst the cells whose configurations differed in accuracy, and \tabterm{\texttt{-w} better} counts those of them in which it was also the more accurate.}
  \label{tab:weight-families}  
\end{table}

%% file: tables/shape-drivers.tex
\begin{table}[tbp]
 \sffamily\footnotesize%
 \begin{tabularx}{\textwidth}{@{\hspace{10pt}}l@{\hspace{18pt}}Xccccc}
  \toprule
   & \scriptsize\textbf{Shape of the cover} & \scriptsize\textbf{Pooled (24)} & \scriptsize\textbf{OSLOM (6)} & \scriptsize\textbf{SLPA (6)} & \scriptsize\textbf{Ego-splitting (6)} & \scriptsize\textbf{CD-DAWN (6)} \\
  \midrule
  \multirow{6}{*}{\rotatebox[origin=c]{90}{Stability, Ω}} & \textit{k}\tmsubx{≥20} per 1,000 nodes & −0.83 & −0.77 & −0.31 & −0.60 & −0.45 \\
   & \textit{k} per 1,000 nodes & −0.05 & −0.37 & +0.77 & −0.60 & −0.93 \\
   & Very small communities (\textit{d}) & +0.83 & +0.66 & +0.83 & −0.71 & +0.46 \\
   & Overlap (\textit{o}) & −0.41 & −0.83 & −0.71 & −0.71 & −0.75 \\
   & Largest community's share of nodes & +0.81 & −0.49 & +0.49 & −0.66 & +0.75 \\
   & \textit{Q}\tmsubx{\textit{ov}} & −0.34 & +0.60 & −0.60 & +0.09 & −0.75 \\
  \cmidrule(lr){1-7}
  \multirow{5}{*}{\rotatebox[origin=c]{90}{\textit{Q}\tmsubx{\textit{ov}}}} & \textit{k}\tmsubx{≥20} per 1,000 nodes & +0.28 & −0.71 & +0.60 & −0.60 & +0.74 \\
   & \textit{k} per 1,000 nodes & +0.01 & −0.09 & −0.49 & −0.60 & +0.77 \\
   & Very small communities (\textit{d}) & −0.39 & +0.60 & −0.60 & −0.37 & −0.83 \\
   & Overlap (\textit{o}) & +0.71 & −0.89 & +0.49 & −0.09 & +0.66 \\
   & Largest community's share of nodes & −0.38 & +0.03 & −0.77 & +0.09 & −1.00 \\
  \bottomrule
  \end{tabularx}
  \caption{Spearman rank correlations between the shape of each cover and two reference-free diagnostics, stability and \textit{Q}\tmsubx{\textit{ov}}, over the 24 configurations of Table~\ref{tab:cover-shape}, pooled and within each method (six configurations each). The last row of the first panel is the correlation between the two diagnostics.}
  \label{tab:shape-drivers}  
\end{table}

%% file: tables/collapse-rules.tex
\begin{table}[b!]
 \sffamily\footnotesize%
 \begin{tabular*}{\textwidth}{@{\extracolsep{\fill}}llcccccc}
  \toprule
  \scriptsize\textbf{Method} & \scriptsize\textbf{Detected overlap} & \scriptsize\textbf{Cells} & \scriptsize\textbf{Mean} & \scriptsize\textbf{First} & \scriptsize\textbf{Largest} & \scriptsize\textbf{Most edges} & \scriptsize\textbf{Random} \\
  \midrule
  \multirow{3}{*}{OSLOM} & at most 1\% & 9,386 & 0.1\% & +0.0006 & +0.0005 & +0.0002 & +0.0009 \\
   & 1\% to 10\% & 16,046 & 5.4\% & +0.0250 & +0.0193 & +0.0056 & +0.0359 \\
   & more than 10\% & 12,008 & 17.0\% & +0.0715 & +0.0532 & +0.0304 & +0.0944 \\
  \cmidrule(lr){1-8}
  \multirow{3}{*}{SLPA} & at most 1\% & 4,677 & 0.2\% & +0.0008 & +0.0006 & +0.0000 & +0.0007 \\
   & 1\% to 10\% & 3,796 & 4.3\% & +0.0131 & +0.0063 & −0.0063 & +0.0113 \\
   & more than 10\% & 4,487 & 35.7\% & +0.0742 & +0.0244 & −0.0876 & +0.0862 \\
  \bottomrule
  \end{tabular*}
  \caption{Stability under Ω minus stability under the Adjusted Rand index on the same covers of the main suite, by range of detected overlap. The Adjusted Rand index is computed after collapsing each node to one of its communities by four rules: the first membership listed, its largest community, the community holding most of its edges, and a uniform choice (mean of 20 draws). Every setting of each method is included, at five seeds, on ten realizations of each graph. \tabterm{Cells} counts the graph-setting combinations in each range, \tabterm{Mean} is their mean detected overlap, and each difference is the mean over those combinations, positive where the Adjusted Rand index agreed less than Ω.}
  \label{tab:collapse-rules}  
\end{table}

%% file: tables/consensus.tex
\begin{table}[tbp]
 \sffamily\footnotesize%
 \begin{tabular*}{\textwidth}{@{\extracolsep{\fill}}ccccccc}
  \toprule
   & \multicolumn{3}{c}{\scriptsize\textbf{\textit{τ}\tmthick{}=\tmthick{}0.5}} & \multicolumn{3}{c}{\scriptsize\textbf{\textit{τ}\tmthick{}=\tmthick{}0.7}} \\
  \cmidrule(lr){2-4}\cmidrule(lr){5-7}
  \scriptsize\textbf{Runs} & \scriptsize\textbf{Ω} & \scriptsize\textbf{95\% interval} & \scriptsize\textbf{Overlap} & \scriptsize\textbf{Ω} & \scriptsize\textbf{95\% interval} & \scriptsize\textbf{Overlap} \\
  \midrule
  8 & 0.6367 & 0.6243--0.6488 & 8.90\% & 0.6961 & 0.6848--0.7081 & 4.19\% \\
  16 & 0.7098 & 0.6959--0.7224 & 7.03\% & 0.7527 & 0.7416--0.7657 & 3.43\% \\
  32 & 0.7505 & 0.7367--0.7659 & 6.42\% & 0.8024 & 0.7960--0.8090 & 3.43\% \\
  64 & 0.7822 & 0.7701--0.7971 & 6.28\% & 0.8416 & 0.8272--0.8545 & 3.43\% \\
  128 & 0.8114 & 0.7901--0.8328 & 6.19\% & 0.8717 & 0.8597--0.8829 & 3.37\% \\
  256 & 0.8399 & 0.8322--0.8485 & 5.92\% & 0.9044 & 0.8950--0.9121 & 3.25\% \\
  512 & 0.8720 & 0.8653--0.8803 & 5.99\% & 0.9245 & 0.9186--0.9308 & 3.34\% \\
  1,024 & 0.8566 & 0.8384--0.8723 & 5.69\% & 0.9305 & 0.9254--0.9350 & 3.41\% \\
  \bottomrule
  \end{tabular*}
  \caption{Agreement between eight consensus streams of OSLOM on the \corpus{Advogato} trust network by number of input runs, at two co-association thresholds (\textit{τ}\tmthick{}=\tmthick{}0.5 left, \textit{τ}\tmthick{}=\tmthick{}0.7 right). Ω is the mean over the 28 pairs of streams, with a 95 percent percentile interval from resampling the streams, and \tabterm{Overlap} is the share of nodes placed in more than one community, averaged over the streams.}
  \label{tab:consensus}  
\end{table}

%% file: tables/consensus-planted.tex
\begin{table}[tbp]
 \sffamily\footnotesize%
 \begin{tabular*}{\textwidth}{@{\extracolsep{\fill}}ccccc@{\hspace{10pt}}cccc@{\hspace{10pt}}cccc}
  \toprule
   & \multicolumn{4}{c}{\scriptsize\textbf{\textit{μ}\tmthick{}=\tmthick{}0.3}} & \multicolumn{4}{c}{\scriptsize\textbf{\textit{μ}\tmthick{}=\tmthick{}0.5}} & \multicolumn{4}{c}{\scriptsize\textbf{\textit{μ}\tmthick{}=\tmthick{}0.6}} \\
  \cmidrule(lr){2-5}\cmidrule(lr){6-9}\cmidrule(lr){10-13}
   & \multicolumn{2}{c}{\scriptsize\textbf{\textit{τ}\tmthick{}=\tmthick{}0.5}} & \multicolumn{2}{c}{\scriptsize\textbf{\textit{τ}\tmthick{}=\tmthick{}0.7}} & \multicolumn{2}{c}{\scriptsize\textbf{\textit{τ}\tmthick{}=\tmthick{}0.5}} & \multicolumn{2}{c}{\scriptsize\textbf{\textit{τ}\tmthick{}=\tmthick{}0.7}} & \multicolumn{2}{c}{\scriptsize\textbf{\textit{τ}\tmthick{}=\tmthick{}0.5}} & \multicolumn{2}{c}{\scriptsize\textbf{\textit{τ}\tmthick{}=\tmthick{}0.7}} \\
  \cmidrule(lr){2-3}\cmidrule(lr){4-5}\cmidrule(lr){6-7}\cmidrule(lr){8-9}\cmidrule(lr){10-11}\cmidrule(lr){12-13}\scriptsize\textbf{Runs} & \scriptsize\textbf{Agr.} & \scriptsize\textbf{Acc.} & \scriptsize\textbf{Agr.} & \scriptsize\textbf{Acc.} & \scriptsize\textbf{Agr.} & \scriptsize\textbf{Acc.} & \scriptsize\textbf{Agr.} & \scriptsize\textbf{Acc.} & \scriptsize\textbf{Agr.} & \scriptsize\textbf{Acc.} & \scriptsize\textbf{Agr.} & \scriptsize\textbf{Acc.} \\
  \midrule
  8 & 0.821 & 0.700 & 0.932 & 0.699 & 0.867 & 0.634 & 0.944 & 0.630 & 0.882 & 0.538 & 0.886 & 0.527 \\
  16 & 0.843 & 0.698 & 0.939 & 0.696 & 0.895 & 0.633 & 0.956 & 0.628 & 0.905 & 0.540 & 0.903 & 0.523 \\
  32 & 0.872 & 0.702 & 0.963 & 0.700 & 0.928 & 0.637 & 0.961 & 0.627 & 0.932 & 0.542 & 0.932 & 0.527 \\
  64 & 0.931 & 0.710 & 0.974 & 0.698 & 0.948 & 0.639 & 0.965 & 0.623 & 0.931 & 0.544 & 0.945 & 0.529 \\
  128 & 0.923 & 0.696 & 0.979 & 0.698 & 0.970 & 0.637 & 0.976 & 0.623 & 0.951 & 0.544 & 0.957 & 0.527 \\
  256 & 0.897 & 0.690 & 0.986 & 0.698 & 0.984 & 0.638 & 0.973 & 0.622 & 0.966 & 0.544 & 0.966 & 0.527 \\
  512 & 0.938 & 0.700 & 0.988 & 0.698 & 0.990 & 0.638 & 0.983 & 0.622 & 0.971 & 0.542 & 0.963 & 0.529 \\
  1,024 & 0.976 & 0.708 & 0.989 & 0.697 & 0.980 & 0.637 & 0.990 & 0.624 & 0.985 & 0.543 & 0.968 & 0.528 \\
  \bottomrule
  \end{tabular*}
  \caption{Consensus clustering of OSLOM on three planted LFR graphs with 30 percent overlap, by number of input runs, at two co-association thresholds. \tabterm{Agr.} is the agreement between eight consensus streams, their mean pairwise Ω, and \tabterm{Acc.} is the accuracy of each stream, its Ω against the planted cover, averaged over the streams.}
  \label{tab:consensus-planted}  
\end{table}

%% file: tables/consensus-transfer.tex
\begin{wraptable}{R}{0pt}
 \sffamily\footnotesize%
 \begin{tabular}{ccccc}
  \toprule
   & \multicolumn{2}{c}{\scriptsize\textbf{\textit{τ}\tmthick{}=\tmthick{}0.5}} & \multicolumn{2}{c}{\scriptsize\textbf{\textit{τ}\tmthick{}=\tmthick{}0.7}} \\
  \cmidrule(lr){2-3}\cmidrule(lr){4-5}
  \scriptsize\textbf{Runs} & \scriptsize\textbf{\corpus{Advogato}} & \scriptsize\textbf{\corpus{Enron}} & \scriptsize\textbf{\corpus{Advogato}} & \scriptsize\textbf{\corpus{Enron}} \\
  \midrule
  8 & 0.637 & 0.598 & 0.696 & 0.688 \\
  16 & 0.710 & 0.677 & 0.753 & 0.753 \\
  32 & 0.751 & 0.720 & 0.802 & 0.793 \\
  64 & 0.782 & 0.754 & 0.842 & 0.833 \\
  128 & 0.811 & 0.791 & 0.872 & 0.848 \\
  256 & 0.840 & 0.816 & 0.904 & 0.873 \\
  \bottomrule
  \end{tabular}
  \caption{Agreement between eight consensus streams of OSLOM on \corpus{Advogato} and \corpus{Enron} at two co-association thresholds, over the run counts at which both were swept. Each cell is the mean Ω over the 28 pairs of streams.}
  \label{tab:consensus-transfer}  
\end{wraptable}

%% file: tables/selection-protocol.tex
\begin{table}[b!]
 \sffamily\footnotesize%
 \begin{tabular*}{\textwidth}{@{\extracolsep{\fill}}lccccccc}
  \toprule
  \scriptsize\textbf{Method} & \scriptsize\textbf{Runs per graph} & \scriptsize\textbf{Matched} & \scriptsize\textbf{Held out} & \scriptsize\textbf{Default} & \scriptsize\textbf{Median} & \scriptsize\textbf{Unmatched} & \scriptsize\textbf{Inflation} \\
  \midrule
  OSLOM & 520 & 0.5922 & 0.5789 & 0.5577 & 0.5609 & 0.6054 & +0.0132 \\
  SLPA & 180 & 0.4878 & 0.4714 & 0.4509 & 0.4533 & 0.5365 & +0.0487 \\
  ego-splitting & 160 & 0.4180 & 0.3898 & 0.3109 & 0.3173 & 0.4231 & +0.0051 \\
  CD-DAWN & 125 & 0.2834 & 0.2377 & 0.2493 & 0.1482 & 0.2939 & +0.0104 \\
  Leiden (baseline) & 45 & 0.5714 & 0.5655 & 0.5209 & 0.5158 & 0.5811 & +0.0098 \\
  \bottomrule
  \end{tabular*}
  \caption{The same runs reduced to one score per graph in several ways, as mean Ω against the planted cover over the whole main suite. The matched protocol, used throughout, takes the mean over seeds first and the maximum over settings second. \tabterm{Held out} scores each graph at the setting that is best on average over the suite's graphs at other parameter sets, \tabterm{Default} is the method's documented setting (for SLPA, whose size-shaping stage has no published default, a threshold of 0.1 and a lower size threshold of ten), and \tabterm{Median} is the median over settings. The unmatched protocol takes the maximum over every single run, seeds and settings pooled, and \tabterm{Inflation} is the unmatched score minus the matched one.}
  \label{tab:selection-protocol}  
\end{table}

%% file: tables/weight-arms.tex
\begin{wraptable}{R}{0pt}
 \sffamily\footnotesize%
 \begin{tabular}{llcc}
  \toprule
   &  & \multicolumn{2}{c}{\scriptsize\textbf{Mean Ω}} \\
  \cmidrule(lr){3-4}
  \scriptsize\textbf{Method} & \scriptsize\textbf{Weights} & \scriptsize\textbf{All 160} & \scriptsize\textbf{Overlapping} \\
  \midrule
  \multirow{2}{*}{OSLOM} & \texttt{-uw} & 0.7734 & 0.7477 \\
   & \texttt{-w} & 0.7506 & 0.7241 \\
  \cmidrule(lr){1-4}
  \multirow{2}{*}{Leiden} & \texttt{-uw} & 0.7688 & 0.7129 \\
   & \texttt{-w} & 0.6264 & 0.5958 \\
  \bottomrule
  \end{tabular}
  \caption{Mean Ω accuracy under the matched protocol for OSLOM and the partition baseline on the weighted planted suite, each with and without the edge weights. \tabterm{All 160} covers every graph realization, and \tabterm{Overlapping} the 90 with planted overlap at \textit{μ}\tmthick{}≤\tmthick{}0.5. The suite has 16 parameter sets, each in ten realizations at 20 seeds, and both configurations of a method use the same graphs, grid, and seeds.}
  \label{tab:weight-arms}  
\end{wraptable}

%% file: tables/blind-battery.tex
\begin{table}[b!]
 \sffamily\footnotesize%
 \begin{tabular*}{\textwidth}{@{\extracolsep{\fill}}llccccc}
  \toprule
  \scriptsize\textbf{Dataset} & \scriptsize\textbf{Weights} & \scriptsize\textbf{Ω} & \scriptsize\textbf{\textup{NMI}\tmsubx{\textup{max}}} & \scriptsize\textbf{\textit{Q}\tmsubx{\textit{ov}}} & \scriptsize\textbf{\textit{Q}\tmsubx{\textit{dwo}}} & \scriptsize\textbf{Overlap} \\
  \midrule
  \multirow{2}{*}{\corpus{Enron}, internal} & \texttt{-uw} & 0.500 & 0.372 & 0.239 & 0.238 & 6.59\% \\
   & \texttt{-w} & \textbf{0.597} & \textbf{0.473} & \textbf{0.403} & \textbf{0.257} & 2.57\% \\
  \cmidrule(lr){1-7}
  \multirow{2}{*}{\corpus{Enron}, unrestricted} & \texttt{-uw} & \textbf{0.626} & 0.535 & 0.359 & \textbf{0.135} & 3.87\% \\
   & \texttt{-w} & 0.608 & \textbf{0.573} & \textbf{0.477} & 0.080 & 0.90\% \\
  \bottomrule
  \end{tabular*}
  \caption{The four reference-free diagnostics for OSLOM with and without the graph's edge weights, on the internal and the unrestricted \corpus{Enron} graph, each the mean over 20 seeds. \tabterm{Overlap} is the fraction of nodes placed in more than one community. The better value of each pair is set in bold where a bootstrap that resamples the seeds the two configurations share resolved the difference at 95 percent. \textit{Q}\tmsubx{\textit{dwo}} is scored on the graph's own weights for both configurations.}
  \label{tab:blind-battery}  
\end{table}

%% file: tables/enron-hubs.tex
\begin{table}[b!]
 \sffamily\footnotesize%
 \begin{tabular*}{\textwidth}{@{\extracolsep{\fill}}lccccc}
  \toprule
   &  & \multicolumn{2}{c}{\scriptsize\textbf{OSLOM}} & \multicolumn{2}{c}{\scriptsize\textbf{SLPA}} \\
  \cmidrule(lr){3-4}\cmidrule(lr){5-6}
  \scriptsize\textbf{Node} & \scriptsize\textbf{Degree} & \scriptsize\textbf{\texttt{-uw}} & \scriptsize\textbf{\texttt{-w}} & \scriptsize\textbf{\texttt{-uw}} & \scriptsize\textbf{\texttt{-w}} \\
  \midrule
  outlook.team & 1,569 & 269 (3) & 291 (2) & 16,813 (1) & 12,456 (2) \\
  sally.beck & 1,502 & 162 (4) & 1,303 (3) & 16,813 (1) & 12,456 (1) \\
  david.forster & 1,495 & 290 (5) & 1,303 (2) & 16,813 (1) & 4,158 (1) \\
  kenneth.lay & 1,425 & 702 (4) & 1,184 (1) & 16,813 (1) & 12,456 (1) \\
  technology.enron & 1,255 & 260 (4) & 215 (2) & 128 (1) & 112 (1) \\
  jeff.skilling & 1,088 & 702 (1) & 1,303 (1) & 16,813 (1) & 12,456 (1) \\
  julie.clyatt & 941 & 343 (3) & 1,303 (1) & 16,813 (1) & 12,456 (1) \\
  louise.kitchen & 923 & 127 (1) & 1,303 (1) & 16,813 (1) & 12,456 (1) \\
  tracey.kozadinos & 921 & 401 (6) & 539 (4) & 16,813 (2) & 12,456 (3) \\
  bodyshop & 871 & 327 (4) & 539 (2) & 16,813 (1) & 12,456 (3) \\
  \bottomrule
  \end{tabular*}
  \caption{Placement of the ten highest-degree nodes of the internal \corpus{Enron} graph, by method and weight configuration, in each configuration's first-seed cover. \tabterm{Degree} is the number of distinct neighbours in either direction. Each cell gives the size of the largest community holding the node and, in parentheses, the number of communities in which it was placed.}
  \label{tab:enron-hubs}  
\end{table}

%% file: tables/coarseness.tex
\begin{table}[b!]
 \sffamily\footnotesize%
 \begin{tabular*}{\textwidth}{@{\extracolsep{\fill}}llcccccc}
  \toprule
  \scriptsize\textbf{Suite} & \scriptsize\textbf{Method} & \scriptsize\textbf{\textit{μ}} & \scriptsize\textbf{Graphs} & \scriptsize\textbf{Stability} & \scriptsize\textbf{\textit{Q}\tmsubx{\textit{ov}}} & \scriptsize\textbf{Accuracy} & \scriptsize\textbf{Over planted} \\
  \midrule
  \multirow{4}{*}{Main} & \multirow{2}{*}{OSLOM} & \textit{μ}\tmthick{}≤\tmthick{}0.5 & 200 & +0.72 (0) & +0.39 (2) & +0.69 (0) & 0.82 \\
   &  & \textit{μ}\tmthick{}>\tmthick{}0.5 & 160 & −0.32 (103) & +0.39 (46) & +0.81 (0) & 0.54 \\
   & \multirow{2}{*}{SLPA} & \textit{μ}\tmthick{}≤\tmthick{}0.5 & 200 & −0.37 (181) & −0.93 (200) & −0.35 (147) & 1.04 \\
   &  & \textit{μ}\tmthick{}>\tmthick{}0.5 & 160 & −0.32 (124) & −0.91 (160) & −0.33 (129) & 0.85 \\
  \cmidrule(lr){1-8}
  \multirow{4}{*}{Three-membership} & \multirow{2}{*}{OSLOM} & \textit{μ}\tmthick{}≤\tmthick{}0.5 & 200 & +0.75 (0) & +0.39 (5) & +0.68 (0) & 0.80 \\
   &  & \textit{μ}\tmthick{}>\tmthick{}0.5 & 160 & −0.19 (96) & +0.42 (44) & +0.77 (0) & 0.51 \\
   & \multirow{2}{*}{SLPA} & \textit{μ}\tmthick{}≤\tmthick{}0.5 & 200 & −0.30 (154) & −0.91 (200) & −0.15 (121) & 1.03 \\
   &  & \textit{μ}\tmthick{}>\tmthick{}0.5 & 160 & −0.30 (117) & −0.91 (160) & −0.33 (130) & 0.74 \\
  \cmidrule(lr){1-8}
  \multirow{4}{*}{Block-model} & \multirow{2}{*}{OSLOM} & \textit{μ}\tmthick{}≤\tmthick{}0.5 & 200 & +0.64 (17) & +0.31 (32) & +0.72 (7) & 0.87 \\
   &  & \textit{μ}\tmthick{}>\tmthick{}0.5 & 160 & −0.70 (139) & +0.61 (52) & +0.81 (0) & 0.59 \\
   & \multirow{2}{*}{SLPA} & \textit{μ}\tmthick{}≤\tmthick{}0.5 & 200 & −0.49 (184) & −0.95 (200) & −0.50 (183) & 1.13 \\
   &  & \textit{μ}\tmthick{}>\tmthick{}0.5 & 160 & −0.33 (137) & −0.89 (160) & −0.09 (91) & 1.06 \\
  \bottomrule
  \end{tabular*}
  \caption{Whether a criterion preferred the less resolved of a method's settings, over ten realizations of each planted suite. Each cell is the median over graphs of the rank correlation, across a method's settings on one graph, between the effective community count of Equation~\eqref{eq:effective-count} and stability, \textit{Q}\tmsubx{\textit{ov}}, or accuracy (Ω against the planted cover), with the number of graphs on which it was negative in parentheses. A negative coefficient means that the criterion preferred the less resolved setting. \tabterm{Over planted} is the median over graphs of the settings' mean effective community count divided by the planted cover's. \tabterm{Graphs} are counted where the settings can be ranked, and \textit{μ} is the requested value. The suites are named as in Table~\ref{tab:corpora}.}
  \label{tab:coarseness}  
\end{table}

%% file: tables/hub-depth.tex
\begin{table}[tbp]
 \sffamily\footnotesize%
 \begin{tabular*}{\textwidth}{@{\extracolsep{\fill}}lccccccc}
  \toprule
   & \multicolumn{2}{c}{\scriptsize\textbf{Hub in place}} & \scriptsize\textbf{Hub removed} & \multicolumn{4}{c}{\scriptsize\textbf{Removing the hub, paired difference}} \\
  \cmidrule(lr){2-3}\cmidrule(lr){4-4}\cmidrule(lr){5-8}
  \scriptsize\textbf{Search depth} & \scriptsize\textbf{Ω} & \scriptsize\textbf{Very small} & \scriptsize\textbf{Ω} & \scriptsize\textbf{Ω} & \scriptsize\textbf{Overlap \%} & \scriptsize\textbf{Very small \%} & \scriptsize\textbf{\textit{Q}\tmsubx{\textit{ov}}} \\
  \midrule
  3/3 & 0.500 & 16.9\% & 0.434 & \textbf{−0.066} & +0.37 & \textbf{+1.4} & 0.000 \\
  5/5 & 0.522 & 16.6\% & 0.451 & \textbf{−0.071} & −0.02 & \textbf{+1.1} & +0.003 \\
  10/50 & 0.526 & 16.1\% & 0.454 & \textbf{−0.071} & +0.15 & \textbf{+1.1} & +0.002 \\
  20/50 & 0.520 & 15.7\% & 0.447 & \textbf{−0.073} & +0.31 & \textbf{+1.1} & +0.001 \\
  \bottomrule
  \end{tabular*}
  \caption{Stability and cover shape on the internal \corpus{Enron} graph with a connector hub in place and with it removed before detection, at four of OSLOM's search depths (\texttt{-r}/\texttt{-hr}), on the same node at the same 20 seeds. A connector hub is a node adjacent to at least half of all communities but placed in only one, and \tabterm{Very small} communities have five or fewer members. Each difference is paired over a bootstrap of the shared seeds, in percentage points for the two shares, and set in bold where its 95 percent interval excludes zero.}
  \label{tab:hub-depth}  
\end{table}

%% file: tables/index-margin.tex
\begin{table}[b!]
 \sffamily\footnotesize%
 \begin{tabular*}{\textwidth}{@{\extracolsep{\fill}}lcccc}
  \toprule
  \scriptsize\textbf{Suite} & \scriptsize\textbf{Planted overlap} & \scriptsize\textbf{Ω} & \scriptsize\textbf{\textup{NMI}\tmsubx{\textup{max}}} & \scriptsize\textbf{\textit{F}\tmsub{1}} \\
  \midrule
  \multirow{3}{*}{Main} & 10\% & +0.0533 (45/50) & +0.0513 (43/50) & +0.0147 (41/50) \\
   & 30\% & +0.1088 (50/50) & +0.1433 (49/50) & +0.0567 (49/50) \\
   & 50\% & +0.0831 (43/50) & +0.2092 (48/50) & +0.1042 (48/50) \\
  \cmidrule(lr){1-5}
  \multirow{3}{*}{Three-membership} & 10\% & +0.0431 (36/50) & +0.0400 (33/50) & +0.0133 (33/50) \\
   & 30\% & +0.0419 (34/50) & +0.1626 (49/50) & +0.0798 (47/50) \\
   & 50\% & +0.0011 (26/50) & +0.0661 (40/50) & +0.0387 (30/50) \\
  \bottomrule
  \end{tabular*}
  \caption{OSLOM's margin over the partition baseline under each of the three indices of Section~\ref{sec:accuracy}, on the graphs at \textit{μ}\tmthick{}≤\tmthick{}0.5, paired graph by graph over ten realizations of each suite, at the setting Ω selects. Parentheses give the number of graphs on which the margin was positive. Margins may be compared down a column and counts across a row, but margins may not be compared across columns, since the three indices do not share a scale.}
  \label{tab:index-margin}  
\end{table}